\documentclass[11pt,preprint]{aastex}
\def\lsim{\raise0.3ex\hbox{$<$}\kern-0.75em{\lower0.65ex\hbox{$\sim$}}}
\def\gsim{\raise0.3ex\hbox{$>$}\kern-0.75em{\lower0.65ex\hbox{$\sim$}}}

\begin{document}

\title{The Dynamical State of The Serpens South Filamentary Infrared Dark Cloud}
\author{Tomohiro Tanaka\altaffilmark{1}, 
Fumitaka Nakamura\altaffilmark{2,3}, Yuya Awazu\altaffilmark{1}, 
Yoshito Shimajiri\altaffilmark{3}, Koji Sugitani\altaffilmark{4}, 
Toshikazu Onishi\altaffilmark{1},
Ryohei Kawabe\altaffilmark{2,3,5}, Hiroshige Yoshida\altaffilmark{6}, 
Aya E. Higuchi\altaffilmark{5}
}
\altaffiltext{1}{Department of Physical Science, Osaka Prefecture
University, Gakuen f1-1, Sakai, Osaka 599-8531, Japan}
\altaffiltext{2}{National Astronomical Observatory, Mitaka, Tokyo 181-8588, 
Japan; fumitaka.nakamura@nao.ac.jp}
\altaffiltext{3}{Nobeyama Radio Observatory, Minamimaki, Minamisaku, 
Nagano 384-1805, Japan}
\altaffiltext{4}{Graduate School of Natural Sciences, 
Nagoya City University, Mizuho-ku, Nagoya 467-8501, Japan}
\altaffiltext{5}{Joint ALMA Observatory, Alonso de Cordova 3107 OFC 129, 
Vitacura, Chile}
\altaffiltext{6}{Caltech Submillimeter Observatory, 111 Nowelo St. Hilo HI 96720 U.S.A.}
\begin{abstract}
We present the results of N$_2$H$^+$ ($J=1-0$) observations
toward Serpens South, the nearest cluster-forming, infrared dark cloud. 
The physical quantities are derived by fitting the hyperfine structure
of N$_2$H$^+$. The Herschel and 1.1-mm continuum maps show 
that a pc-scale filament fragments into 
three clumps with radii of $0.1-0.2$ pc and masses of $40-230M_\odot$. 
We find that 
the clumps contain smaller-scale ($\sim 0.04$ pc) structures, i.e., 
dense cores. We identify 70 cores by applying CLUMPFIND 
to the N$_2$H$^+$ data cube.
In the central cluster-forming clump, the excitation temperature and
line-width tend to be large, presumably due to 
protostellar outflow feedback and stellar radiation.
However, for all the clumps, the virial ratios 
are evaluated to be $0.1-0.3$, indicating that the internal motions 
play only a minor role in the clump support. 
The clumps exhibit no free-fall, but
low-velocity infall, and thus the clumps should be supported 
by additional forces. 
The most promising force is the globally-ordered magnetic field observed
toward this region. 
We propose that the Serpens South 
filament was close to magnetically-critical  
and ambipolar diffusion triggered the cluster formation.
We find that the northern clump, which shows
no active star formation,
has a mass and radius comparable to the central cluster-forming clump, 
and therefore, it is a likely candidate of a {\it pre-protocluster clump}.
The initial condition for cluster formation is likely to
be a magnetically-supported clump of cold, quiescent gas.
This appears to contradict the accretion-driven turbulence scenario, 
for which the turbulence in the clumps is maintained by the accretion flow.
\end{abstract}

\section{Introduction}
\label{sec:intro}

Infrared Dark Clouds (IRDCs) are cold, dense regions of molecular clouds
seen as extinction features against bright mid-infrared galactic
backgound \citep[e.g.,][]{rathborne06,peretto09,pillai06}. 
They were first identified on the basis of 
observations of ISO and MSX satellites \citep{perault96,egan98}.
Since the abundant discovery by the Spitzer Space telescope,
IRDCs have been paid attention as promising sites to study the earliest
phases of star cluster formation because their volume and surface 
densities resemble those of the nearby active cluster-forming clumps like NGC
1333 and $\rho$ Oph \citep{kauffmann10,hernandez11}.
Recent observations have revealed that the IRDCs often show elongated or 
filamentary shapes and appear to fragment into dense clumps
\citep[e.g.,][]{miettinen12}.  The observations also suggest that 
the self-gravity is likely to play an important role in the 
dynamics of the clumps in IRDCs 
\citep[e.g.,][]{pillai06, pillai11, rygl10, miettinen12, busquet13}.
Global inflow motions toward the IRDC filaments have also been 
reported \citep{schneider10}.  
These observations indicate that kinematics and dynamics of 
IRDCs and their substructures can be well characterized by
the combination between the dust continuum and molecular line
observations.
Although extensive observations towards IRDCs have just started 
recently, our current knowledge on the physical properties of IRDCs 
remain very limitted. 

In order to further characterize the physical properties of
IRDCs and to constrain how clusters form in IRDCs, we carried out 
mapping observations toward a nearby filamentary IRDC, 
Serpens South, which was 
discovered by the Spitzer observations \citep{gutermuth08}. 
The IRDC associated with the Serpens
South cluster is the nearest IRDC that shows a sign of active 
cluster formation.  
An interesting characteristic of the cluster is its 
extremely-high fraction of protostars. In the central region, 
the number fraction of protostars (Class I) relative to the 
Young Stellar Objects (YSOs) 
detected by the Spitzer telescope (Class I+Class II) reaches about 80 \%
\citep{gutermuth08}.
This fraction is largest among the cluster-forming regions known 
within the nearest 400 pc  \citep[see][]{evans09}. 
For example, the Serpens Main Cloud, 
a nearby embedded cluster having the size comparable to Serpens South,
have the protostar fraction of about 60 \%, 
about two thirds of Serpens South. Another nearby embedded cluster, 
NGC 1333, has the protostar fraction of about 30 \%, less than a half
of Serpens South.
Very recently, \citet{bontemps10} and \citet{konyves10} discovered 
7 Class 0 protostars in the Serpens South IRDC on the basis of 
the Herschel observations,
all of which haven't been identified by the Spitzer observations.
These observations strongly suggest that Serpens South is in 
the very early phase of star cluster formation. 
We note that the distance to Serpens South is somewhat uncertain
in the range from 260 and 700 pc \citep[see the discussion of][]{bontemps10}. 
In this paper, we adopt the distance of 415 pc \citep{dzib10}
because our preliminary analysis on the basis of the near-infrared
observations implies a relatively larger distance of $\gtrsim 400$ pc.

Recently, \citet{nakamura11} performed $^{12}$CO ($J=3-2$) mapping 
observations toward the Serpens South IRDC using the ASTE 10-m telescope
and discovered that a number of powerfull outflows 
are blowing out of the central dense clump that 
is located near the southern edge of a long filamentary cloud
\citep{andre10,bontemps10,arzoumanian13}.
The CO ($J=3-2$) images have revealed that several collimated outflow 
lobes are overlapping and interacting with one another, 
indicating that the very active star formation is ongoing.
From the H$_2$ $v=1-0$ S(1) 2.122 $\mu$m image, \citet{teixeira12} 
also identified 10 outflow knots in Serpens South.
The main filament appears to be penetrated by a globally-ordered
magnetic field, implying that the magnetic field plays a role 
in the filament formation \citep{sugitani11}.
Several less-dense filamentary structures also appear to converge toward 
the main filament and the dense clump. 
These less-dense filaments may give us clues to uncover the formation process
of the cluster \citep{myers09}.
Recently, \citet{kirk13} performed a Mopra observation toward 
Serpens South using several dense gas tracers 
such as N$_2$H$^+$ ($J=1-0$), HCO$^+$ ($J=1-0$), 
and H$^{13}$CO$^+$ ($J=1-0$).
The optically-thick lines like HCO$^+$ ($J=1-0$) showed
blue-skewed profiles, indicative of the infall motions
along the line-of-sight.
They also found a significant velocity gradient in the N$_2$H$^+$ emission 
along the southern part of the main filament.
They interpreted that the velocity gradient is caused by a mass infall 
along the filament toward the central cluster.

In this paper, we present the results of the 
N$_2$H$^+$ ($J=1-0$) observations toward 
the filamentary IRDC, Serpens South, 
using the Nobeyama 45-m radio telescope, 
and study the dynamical state of 
the dense clumps in Serpens South. 
The details of the observations are described in
Section \ref{sec:obs}. The results of the observations are presented in
Section \ref{sec:results}.
Our N$_2$H$^+$ ($J=1-0$) data have a finer angular resolution 
and higher sensitivity than those of \citet{kirk13}, and therefore 
allow us to estimate the quantities by fitting the hyperfine structure
of N$_2$H$^+$.
Applying the hyperfine fitting to the N$_2$H$^+$ data cube, we estimate
the physical quantities of the dense gas in this region in Section
\ref{sec:fitting}, and assess in Section \ref{sec:filament} 
the dynamical state of the main filament
that appears to fragment into a few dense clumps. 
Then, we apply in Section \ref{sec:virial} 
the virial analysis to the dense clumps,  
and clarify the clump dynamical state.
Then, in Section \ref{sec:core}, we identify the dense cores using the
clumpfind method and derive the physical quantities of the cores.
Finally, we summarize our main conclusion in Section \ref{sec:conclusion}.

\section{Observations and Data}
\label{sec:obs}

\subsection{Nobeyama 45-m Radio Telescope}

\begin{figure}[h]
\plotone{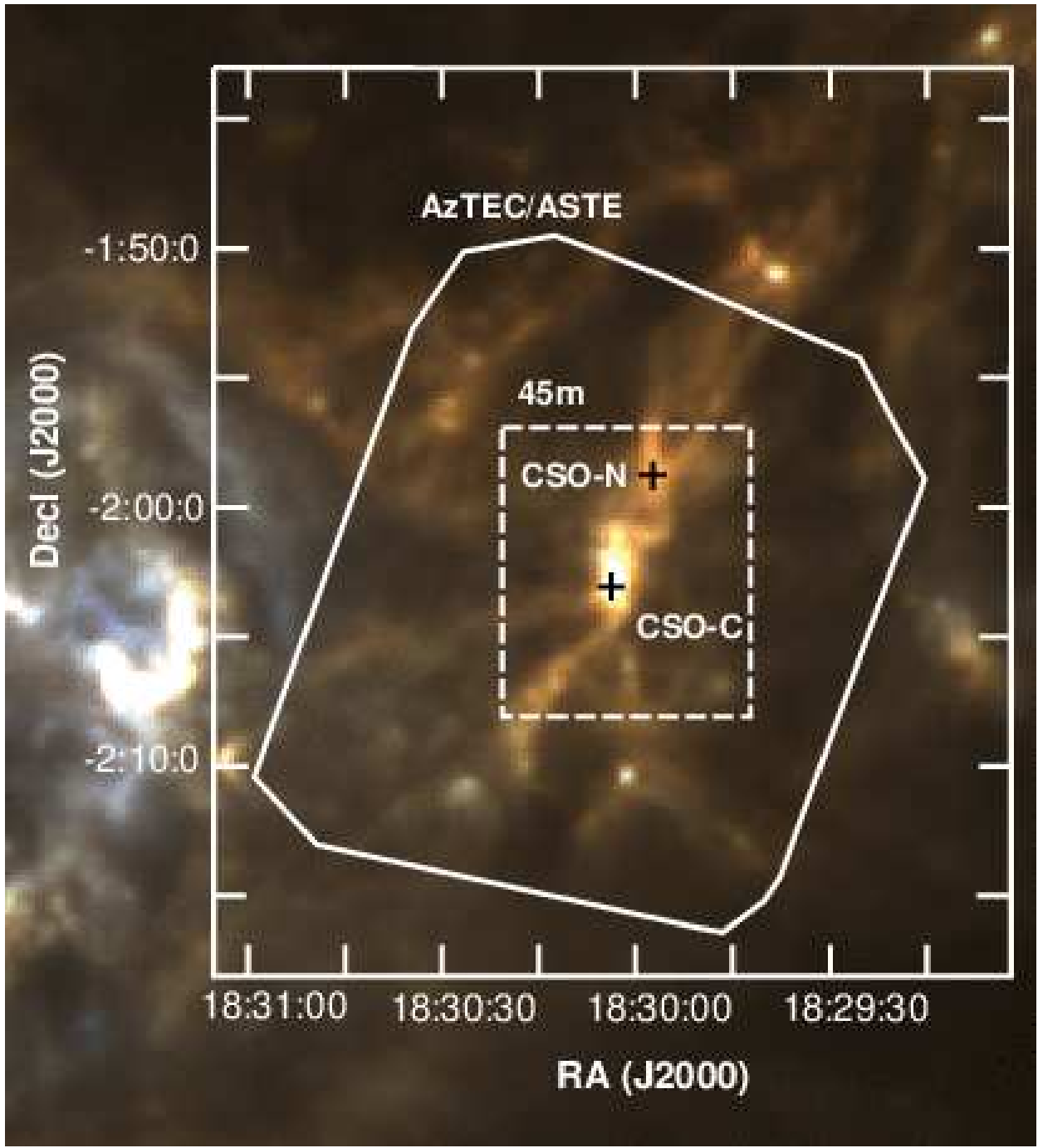}
\caption{Three-color Herschel image of Serpens South
with SPIRE 250 $\mu$m ({\it blue}), 
350 $\mu$m ({\it green}), and 
500 $\mu$m ({\it red}). 
The observed areas 
are the dashed box and the area enclosed by solid lines
for the NRO 45-m N$_2$H$^+$ observations and AzTEC/ASTE 1.1 mm observations, 
respectively.
Two points of the CSO observations are indicated with the crosses.
The eastern bright area outside the panel is the dense gas around 
the W40 HII region.
}  
\label{fig:obsbox}
\end{figure}

In January and March 2012, we carried out N$_2$H$^+$ 
[$J=1-0$; the rest frequency of the main component
($F_1F \rightarrow F'_1 F' = 23 \rightarrow 12$) = 93.173777 GHz] 
observations toward  
the Serpens South cluster 
with the 25-element focal plane receiver
BEARS equipped in the 45-m telescope at NRO. 
In Figure \ref{fig:obsbox}, the observed area is indicated 
by a dashed box on the three-color {\it Herschel} image.
The observation box size is $9.'5 \times 11'$ with the map center
of (R.A. [J2000], Decl. [J2000]) $=$ (18:30:1.85, $-$02:02:28.6).
At 93 GHz band, the telescope has a Full-Width-at-Half-Maximum 
(FWHM) beam size of 18$''$ and 
a main beam efficiency of $\eta=0.51$. 
The beam separation of the BEARS is 41.1$''$ on the plane of the sky 
\citep{sunada00,yamaguchi00}. 
At the back end, we used 25 sets of 1024 channel Auto Correlators 
(ACs), which have a bandwidth of 8 MHz and a frequency 
resolution of 7.8 kHz, corresponding to a velocity resolution 
of 0.025 km s$^{-1}$ at 93 GHz \citep{sorai00}. 
The final velocity resolution was binned to 0.05 km s$^{-1}$.
The N$_2$H$^+$ ($J=1-0$) consists of seven hyperfine components 
over 4.6 MHz. We determined the observed frequency range
such that all the 7 hyperfine components are distributed around 
the central frequency.
During the observations, the system noise temperatures 
were in the range between 300 and 450 K in a double sideband. 
The standard chopper wheel method \citep{kutner81}  
was used to convert the output signal into the antenna 
temperature $T_{A}$*(K), corrected for the atmospheric attenuation.

The telescope pointing was checked every 1 hour by observing a 
SiO maser source, IRC+00363, and the typical pointing offset were 
1$''$ $\sim$ 3$''$ during the whole observing period. 
Our mapping observations were made with the OTF mapping technique 
\citep{sawada08}.
We obtained an OTF map with xscan and yscan over 
27 boxes and combined them into a single map, 
in order to reduce the scanning effects. 
We adopted a convolutional scheme with a spheroidal function 
\citep{sawada08}
to calculate the intensity at each grid point 
of the final data with a spatial grid size 
of 7.5$''$, about a half of the beam size. 
The resultant effective angular resolution was 24$''$, 
corresponding to 0.05 pc at a distance of 415 pc. 
The rms noise level of the final maps was 0.2 K 
in $T_{A}^*$ (1$\sigma$) at a velocity resolution of 0.05 km s$^{-1}$.
We adopted the rest frequency of the main hyperfine component to make 
the data cube. To make the channel map and apply the clumpfind, 
we used the data cube whose rest frequency is set to that of 
the isolated hyperfine component.


\subsection{CSO 10.4-m Radio Telescope}

We carried out single-point observations toward
the Serpens South cluster in HCO$^+$ ($J=3-2$) and   
H$^{13}$CO$^+$ ($J=3-2$) on May 15, 2011 and August 27, 2011, 
using the 10.4-m telescope of the Caltech Submillimeter Observatory
(CSO) at Mauna Kea, Hawaii.
The observed points have the following coordinates:
(R.A. [J2000], Decl. [J2000]) $=$ (18:30:03.80, $-$02:03:03.9)
and (18:29:57.40, $-$01:58:45.9).
The former (hereafter, CSO-C) and latter (hereafter, CSO-N) are pointing toward 
the 1.1mm dust continuum 
emission peaks of the central and northern clumps, respectively,
as discussed in Section \ref{sec:results}.
The observed points are indicated by the crosses in Figure \ref{fig:obsbox}.
At 270 GHz band, the telescope has a main beam efficiency of $\eta=0.72$
and a FWHM beam size of $30''$,
corresponding to 0.06 pc at a distance of 415 pc. 
The observations were made with a position-switch mode,
using a 230 GHz receiver. 
At the back end, we used a FFTS1 spectrometer with
8192 channels that covers the 1 GHz bandwidth. 
The frequency resolution of 122 kHz corresponds to the velocity resolution 
of 0.14 km s$^{-1}$ at 270 GHz. 
During the observations, the system noise temperatures 
were about 500 K and 200 K on May 15 and August 27, respectively, 
in a double sideband. 
The rms noise levels of the HCO$^+$ ($J=3-2$) and  
H$^{13}$CO$^+$ ($J=3-2$) data were 
$\Delta T_{\rm rms} \sim $ 0.05 K
in $T_{A}^*$ (1$\sigma$) at a velocity resolution 
of about 0.3 km s$^{-1}$.
The two molecular lines were detected towards CSO-C, 
whereas none of the molecular lines were 
detected towards CSO-N.
For CSO-N, we observed additional two molecular lines, CS ($J=5-4$) and HCN
($J=3-2$). We could not detected these two line emissions 
with the same rms noise as 
that of HCO$^+$ ($J=3-2$) and H$^{13}$CO$^+$ ($J=3-2$).

\subsection{1.1mm AzTEC/ASTE Data}
\label{subsec:aztecdata}

The 1.1 mm continuum data were taken toward a $\sim 20' \times 20'$
area centered on the Serpens South cluster with 
the AzTEC camera mounted on the Atacama Submillimeter
Telescope Experiment (ASTE) 10-m telescope. 
The observed area is the area enclosed by solid lines
in Figure \ref{fig:obsbox}.
The rms noise level was about 10 mJy beam$^{-1}$ 
and the effective beam size was 40$''$ after the FRUIT 
imaging, which is an iterative mapping method to recover 
the spatially extended component.
The FRUIT method is based on an iterative mapping method applied
to the BOLOCAM data reduction pipeline \citep{enoch06}. 
The FRUIT data reduction pipeline is written in IDL and developed 
by the University of Massachusetts. The FRUIT
algorithm is described in \citet{liu10} and \citet{shimajiri11}.
The detail of the AzTEC data will be presented in 
Gutermuth et al. (2013, in preparation).

\subsection{Herschel Data}

We used the version 2.5 fits files of the Herschel archival data 
of the PACS 160 $\mu$m and SPIRE 250, 350, and 500 $\mu$m 
towards Serpens South.
The detail of the data will be described in 
Sugitani et al. (2013, in preparation).
In brief, all the data were convolved to the ones with 
the angular resolution of 36.3$''$, and then 
adopting the grey-body approximation, we performed 
the spectral energy distribution (SED) fitting at each grid point
to derive the physical quantities such as 
the column density and dust temperature.

\section{Results}
\label{sec:results}

\subsection{Spatial Distribution of the 1.1 mm Continuum Emission}
\label{subsec:1.1mmresults}

Previous observations have demonstrated that 
the optically-thin dust continuum emission at submillimeter wavelengths 
provides a powerful diagnostic to constrain the density and temperature 
structure of molecular clouds \citep[e.g.,][]{andre00}. 
Our 1.1mm dust continuum data also 
indicate that the dust continuum emission traces well several
YSO envelopes and the spatial distribution of the dense and/or warm
molecular gas in this region.

\begin{figure}[h]
\plottwo{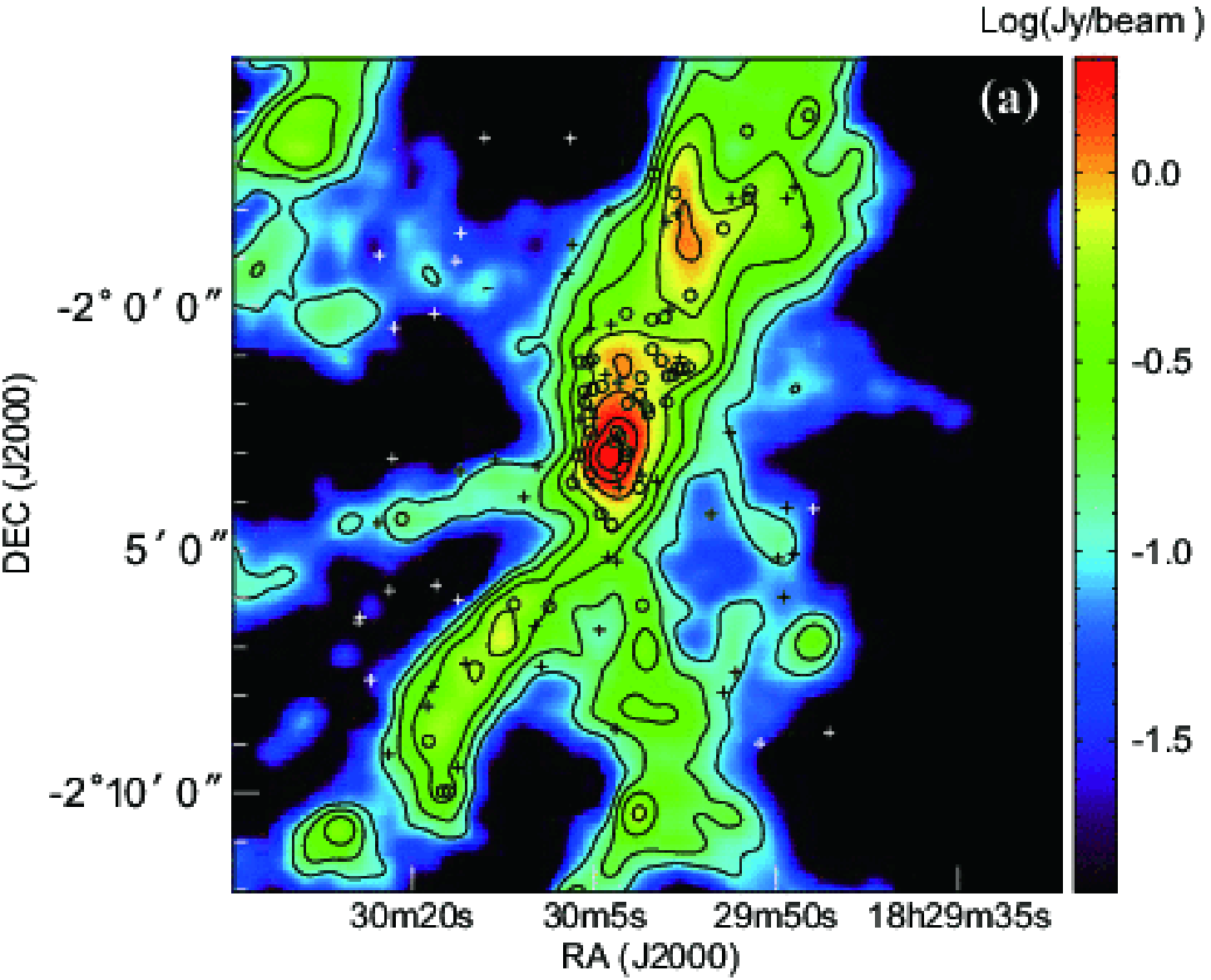}{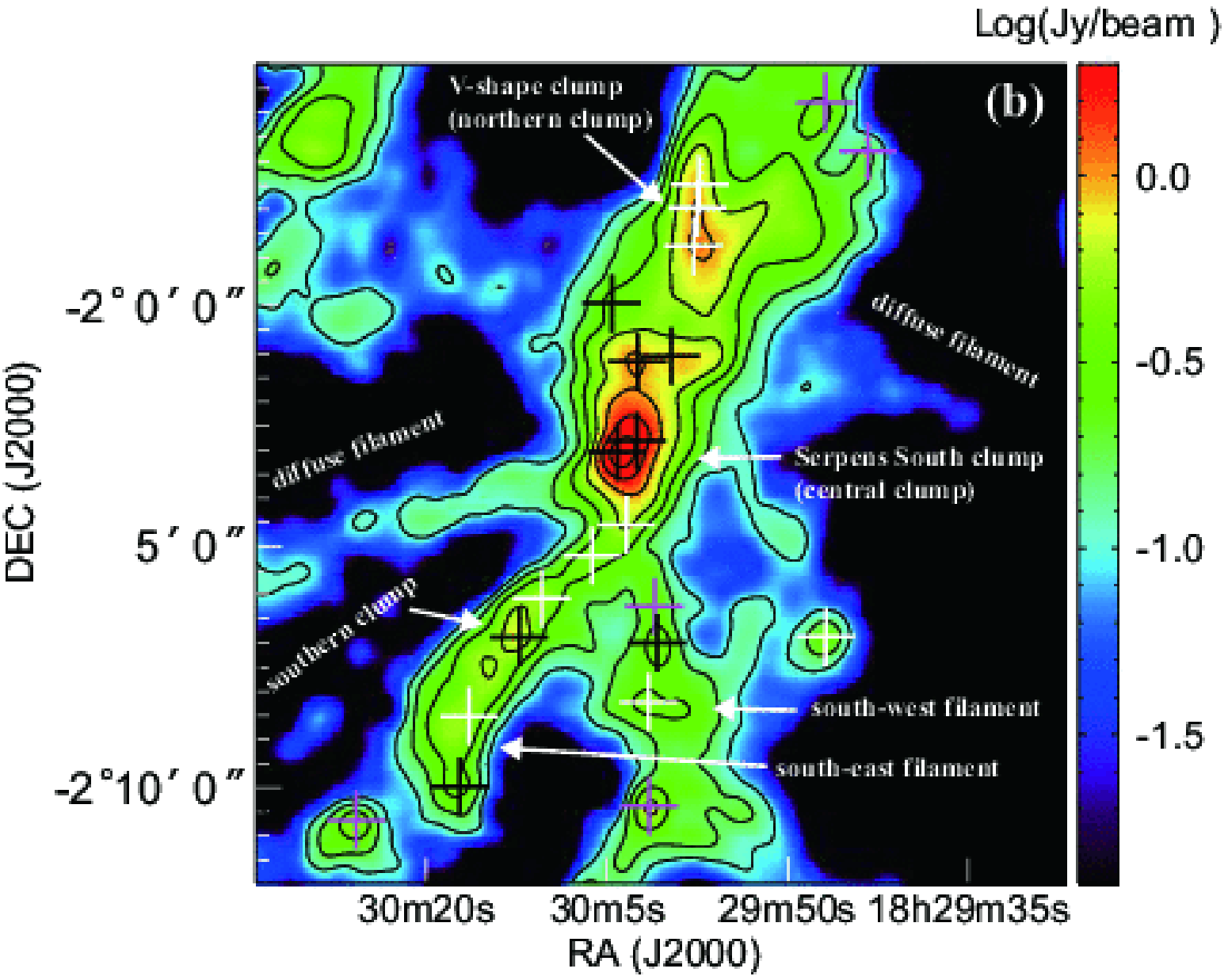}
\caption{(a) 1.1 mm continuum emission map taken by
the AzTEC camera on the ASTE telescope.
The circles and crosses indicate the positions of the Spitzer 
Class I and II YSOs, respectively.
(b) Same as panel (a) but with the positions 
of the starless cores, Class 0, and Class I sources
newly identified by the Herschel observations \citep{bontemps10}.
The white, black, and magenta crosses are 
the starless cores, Class 0, and Class I sources, respectively.
The names of several features discussed in the main text are designated
in the panel.
For both the panels, the contours start at -1.0 ($=10\sigma$)
with intervals 
of 0.25 on a logarithmic scale.
}  
\label{fig:aztec}
\end{figure}

In Figure \ref{fig:aztec}, we present the 1.1 mm continuum emission map 
taken with the AzTEC camera mounted on the ASTE 10-m telescope.
This is essentially the same as Figure 3 of \citet{nakamura11}.  
For comparison, the positions of the YSOs identified by the Spitzer 
observations \citep{gutermuth08} and the positions of the YSOs 
and prestellar cores identified from the MAMBO 1.3 mm 
observations \citep{maury11} are indicated in Figures \ref{fig:aztec}a 
and \ref{fig:aztec}b, respectively.
The dust image was obtained by applying the FRUIT method 
to recover the spatially-extended component.
The resultant effective angular resolution is $\sim 40''$.

The dust continuum emission has its maximum towards the Serpens 
South cluster at the position of 
(R.A. [J2000], Dec. [J2000]) = (18:30:03.81, $-$02:03:04), 
which is embedded in a dusty clump (hereafter referred to as 
the Serpens South clump or central clump).
The Serpens South cluster is located near the southern edge 
of a long filamentary cloud that was revealed by the Herschel 
observations \citep{andre10}.
The Serpens South clump is also located at the intersection 
between the two filaments 
(south-east and south-west filaments) seen in the southern part of the image.
At the southern part of the main filament (south-east filament), 
the main filament appears to fragment into an elongated clump
(hereafter, southern clump), 
at the peak of which a Herschel Class 0 object is located.
In the northern part of the map, we found a V-shape clump 
(hereafter, northern or V-shape clump).
Several starless cores and Class 0 sources identified by the Herschel 
observations can be recognized as compact structures even in our 
FRUIT map. 
Asides from the main filamentary structure, 
there are a couple of faint filamentary structures that appear to 
converge toward the Serpens South clump. 
These faint filaments can also be recognized in the Spitzer image
\citep[see ][]{gutermuth08,nakamura11}.
The detail of the 1.1-mm continuum image will be presented in 
Gutermuth et al. (2013, in preparation).

\subsection{Spatial Distribution of  N$_2$H$^+$ ($J=1-0$) Emission}

\begin{figure}[h]
\plottwo{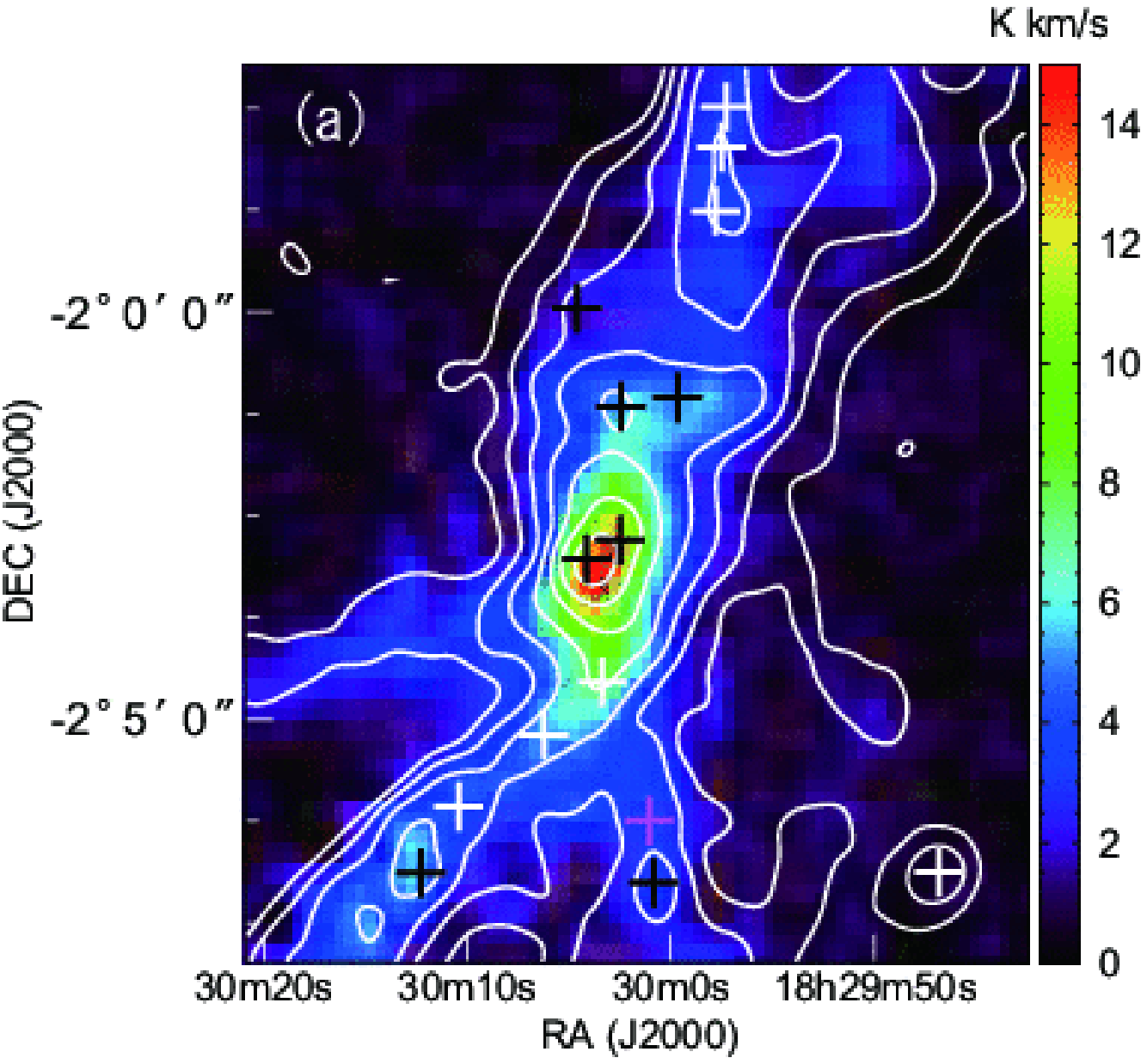}{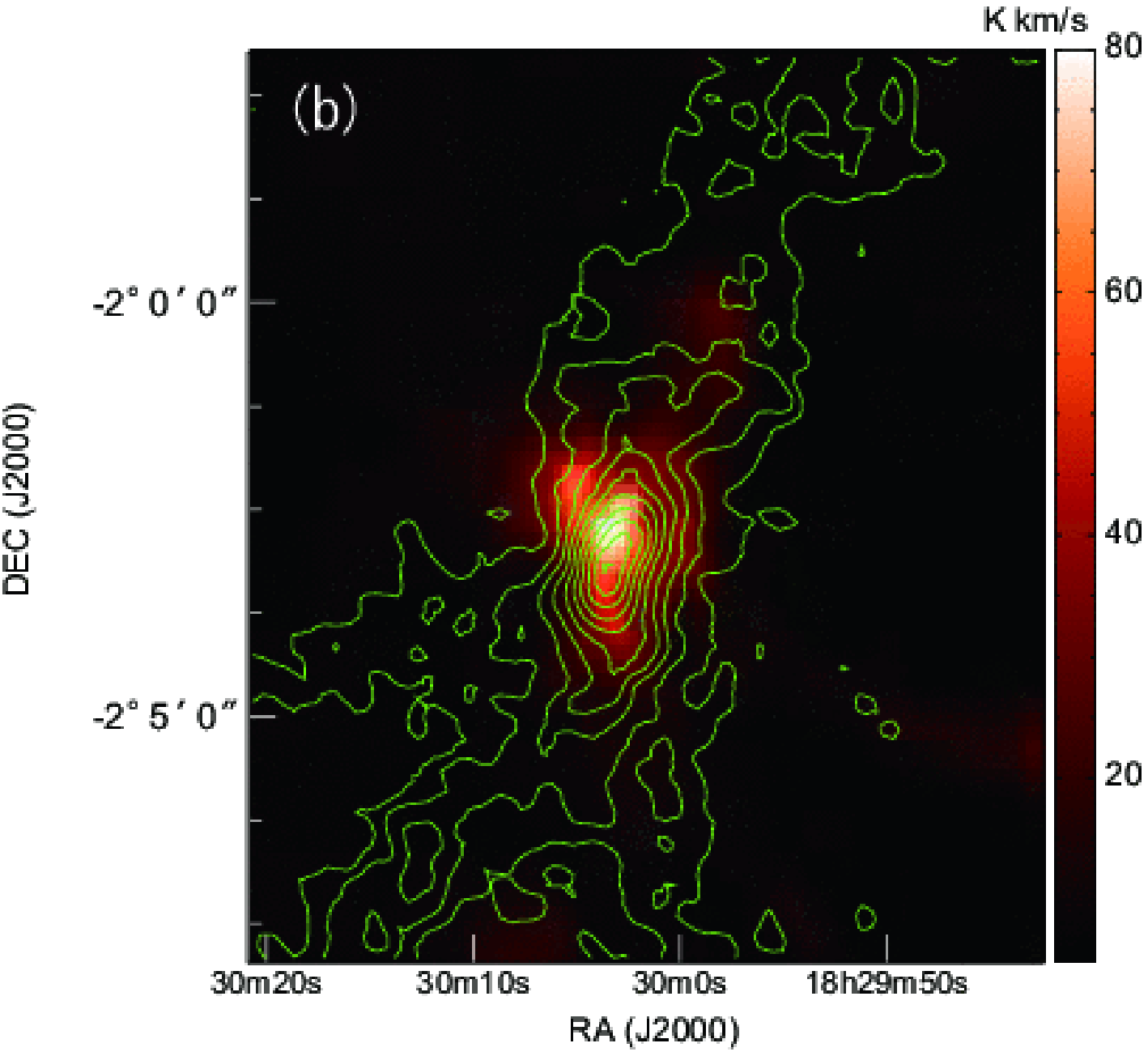}
\caption{(a) N$_2$H$^+$ ($J=1-0$) intensity map
integrated over the seven components of the 
hyperfine multiplet, taken with
the Nobeyama 45-m telescope toward Serpens South.
The contours of the AzTEC 1.1 mm dust continuum emission 
are overlaid on the image. The contours start
from $-$1.0 with an interval of 0.25 in a logarithmic scale.
The white, black, and magenta crosses indicate
the positions of the starless cores, Class 0, and Class I sources
newly identified by the Herschel observations \citep{bontemps10}.
(b) N$_2$H$^+$ ($J=1-0$) intensity contour map
overlaid on the $^{12}$CO $(J=3-2)$ integrated intensity map
in the range of $v_{\rm LSR} = -2$ km s$^{-1}$ to $+15$ km s$^{-1}$.
The contour level starts from 2 K km s$^{-1}$ with an interval 
of 1.5 K km s$^{-1}$.
}  
\label{fig:n2h+}
\end{figure}

Figure \ref{fig:n2h+}a shows the N$_2$H$^+$ ($J=1-0$) intensity map 
toward Serpens South, integrated over the seven components of 
the hyperfine multiplet, taken with the Nobeyama 45-m telescope. 
Our N$_2$H$^+$ ($J=1-0$) intensity distribution 
is in good agreement with that obtained by \citet{kirk13} 
and has a finer angular resolution and higher sensitivity.
On average, the integrated N$_2$H$^+$ intensity closely follows
the 1.1 mm continuum emission, 
suggesting that the N$_2$H$^+$ emisson is an excellent tracer 
of cold and dense emission, as is known from previous studies 
\citep[e.g.,][]{walsh07}. 
However, there are a couple of exceptions.
For example, the V-shape clump identified by the 1.1 mm continuum emission
is not so prominent in the N$_2$H$^+$ intensity map. 
In addition, the N$_2$H$^+$ integrated intensity 
takes its maximum at 
(R.A. [J2000], Decl. [J2000]) = (18:30:03.7, $-$02:03:10),
which is slightly offset from the continuum maximum by $\sim 10''$, 
about a half the NRO 45-m beam size.
Since the pointing errors of both the 1.1 mm dust continuum 
map and N$_2$H$^+$ are better than a few arcseconds, we consider
that the offset is real. However, the offset may be due to
the different angular resolutions of both maps.

In addition, the positions of the dust continuum sources 
identified by the Herschel observations  \citep{bontemps10}
often show deviations from the peaks of 
the N$_2$H$^+$ emission.
The N$_2$H$^+$ integrated intensity tends to be weaker 
toward the Class 0 and Class 0/I candidates 
(the black crosses in Fig. \ref{fig:n2h+}a), in particular
toward the three sources located in the northern part of the image.
The N$_2$H$^+$ integrated intensity is very weak 
toward the Herschel starless cores
(the white crosses in Fig. \ref{fig:n2h+}a). 
The starless core located at the southwest edge of the image 
appears almost devoid of N$_2$H$^+$ emission.
There are two possibilities to explain these descrepancies.
One possibility is that these dust cores are chemically young
and that the abundance of a late-stage molecule, N$_2$H$^+$, 
is still relatively low.  Another possibility is that N$_2$H$^+$
molecules tend to be destroyed by CO and the N$_2$H$^+$
abundance has become small.

Very recently, we carried out CCS ($J_N=2_1-1_0$), CCS ($J_N=4_3-3_2$), 
and CCS ($J_N=7_7-6_5$) observations toward a 20$' \times 10'$
area including the N$_2$H$^+$ mapping area, and found that
the CCS emissions tend to be stronger towards the starless 
dust cores identified by Herschel. 
Therefore, we interpret that the weak N$_2$H$^+$ emission 
towards the starless cores implies that the cores are chemically young.
We will present the results of the CCS observations in a forthcoming paper.

In Figure \ref{fig:n2h+}b, we compare the N$_2$H$^+$ integrated
intensity map with the $^{12}$CO ($J=3-2$) integrated intensity map
obtained by \citet{nakamura11}.
The strong CO emission is concentrated in the central clump
that has the strongest N$_2$H$^+$ emission.
As shown by \citet{nakamura11}, the CO emission represents
the protostellar outflows blowing out of the central clump.
The CO emission is more extended than the dense gas traced  by
N$_2$H$^+$, implying that the outflows have broken through
the dense clump.

\begin{figure}[h]
\epsscale{1.0}
\plotone{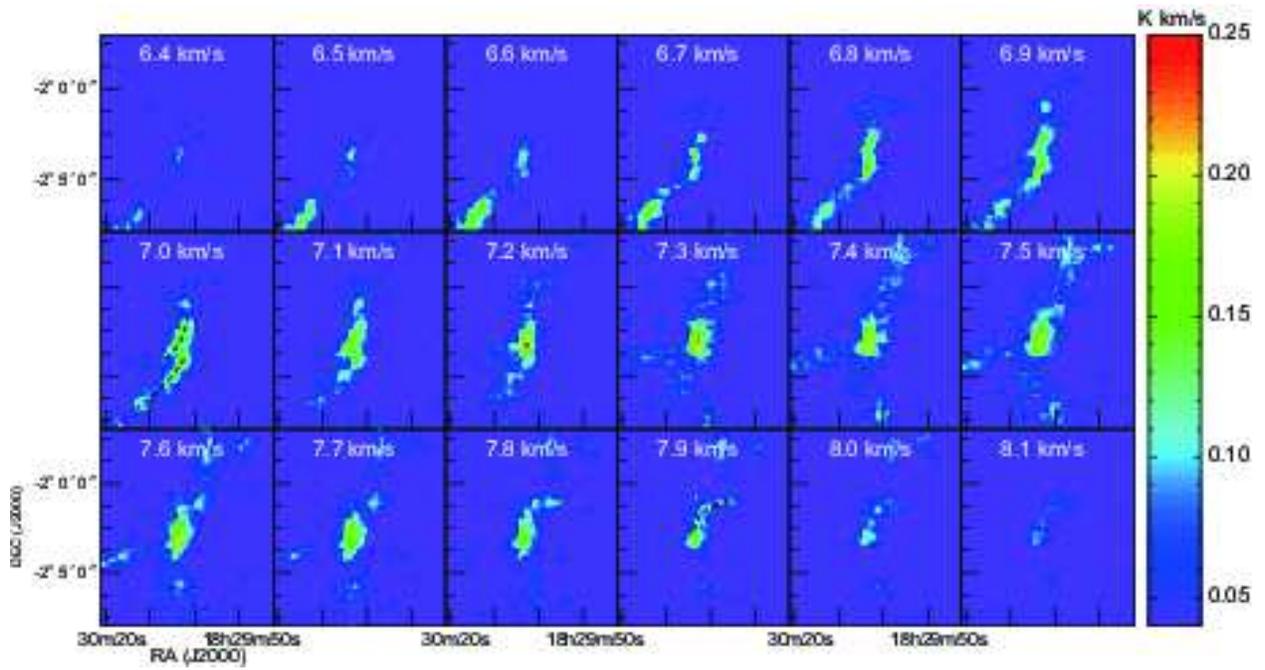}
\caption{N$_2$H$^+$ ($J=1-0$) velocity channel maps
of the isolated component of the hyperfine multiplet, taken with
the Nobeyama 45-m telescope toward Serpens South.
The reference frequency is set to the rest frequency 
of the isolated component to make the maps.
Some arc-like or filamentary structures are highlighted by the dashed curves.
}  
\label{fig:n2h+channel}
\end{figure}

\begin{figure}[h]
\epsscale{1.0}
\plotone{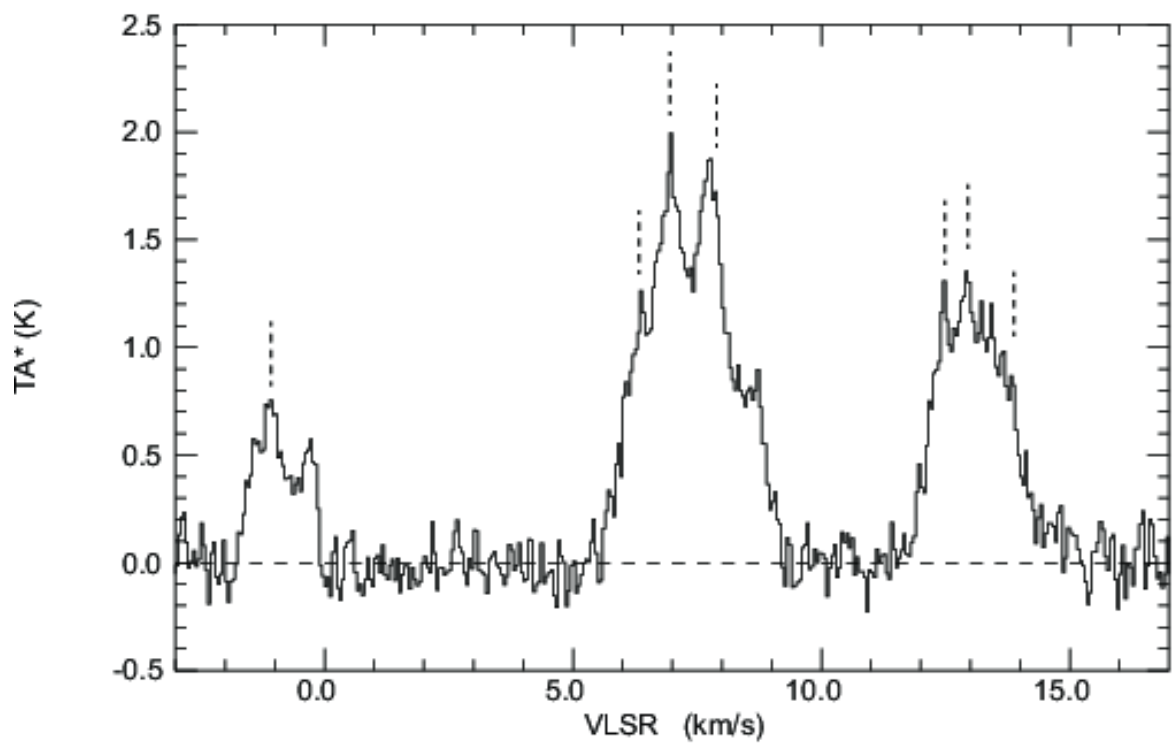}
\caption{N$_2$H$^+$ ($J=1-0$) line profile averaged in the 
$30''\times 30''$ area centered on the position 
(R.A. [J2000], Decl. [J2000]) $=$ (18:30:4.4, $-$02:04:23.5).
The velocity resolution is set to 0.05 km s$^{-1}$.  
The reference frequency is set to the rest frequency of the main component
of $F_1F \rightarrow F'_1 F' = 23 \rightarrow 12$, 93.173777 GHz,
to make the plot.}
\label{fig:lineprofile}
\end{figure}

The N$_2$H$^+$ velocity channel maps are presented 
in Figure \ref{fig:n2h+channel}, which shows that 
the main filamentary structure is running from south to north.
The filament has a large-scale velocity gradient of 
about 0.5 km s$^{-1}$ pc$^{-1}$ along its axis.
The filament contains several velocity components, 
implying the existence of the internal structures.
Figure \ref{fig:lineprofile} illustrates an example of the N$_2$H$^+$
line profile showing two velocity components, where the rest frequency
of the strongest hyperfine component is set to the velocity center.
The profile presented is averaged in the $30'' \times 30''$ area
centered on the position (R.A. [J2000], Decl. [J2000]) $=$ (18:30:04.4,
$-$02:04:23.5) and the velocity resolution is set to 0.05 km s$^{-1}$. 
The isolated line at $V_{\rm LSR} \sim -1$ km s$^{-1}$ shows two velocity
components with about 1 km s$^{-1}$ separation.
Such structure qualitatively resembles those of the B213 filament 
in Taurus \citep{hacar13}.
At the velocity of 7.0 km s$^{-1}$,
the central clump contains two filamentary structures (shown with
the dashed curves).
At the velosity of 7.8 $-$ 7.9 km s$^{-1}$, the arc-like filamentary
structure (shown with the dashed curve) 
can be recognized in the north-east of the central clump.

\section{N$_2$H$^+$ Line Analysis}
\label{sec:fitting}

The N$_2$H$^+$ ($J=1-0$) consists of seven hyperfine components over 4.6
MHz. Following \citet{caselli02} and \citet{difrancesco04}, 
we derive the total line optical depth ($\tau_{\rm tot}$), 
line width ($\Delta V$), line-of-sight velocity ($V_{\rm LSR}$), 
excitation temperature ($T_{\rm ex}$), 
and column density ($N_{\rm N_2H^+}$)
by fitting all seven hyperfine components 
with multiple Gaussian components
under the assumption that all the seven hyperfine lines have 
a single line width and excitation temperature.
The rest frequencies and intrinsic relative intensities of the 
hyperfine components are taken from \citet{caselli95} and
\citet{tine00}, respectively.
For the hyperfine fitting, we reconstructed the data cube of 
the intensity with a spatial grid size of $15''$ and 
velocity resolutoin of $0.1$ km s$^{-1}$ by applying 
the convolutional scheme with a spheroidal function, so that
the data rms noise level is reduced to $\Delta T_A^* \sim 76$ mK 
and the hyperfine fitting procedure is more reliable.
The effective angular resolution of the data is $37''$,
comparable to the AzTEC 1.1 mm and Herschel images.
For the following analysis, we use the hyperfine fitting data at all the pixels 
above the 2$\sigma$ levels for the fitted peak antenna temperature of the 
N$_2$H$^+$ weakest hyperfine component.

Table \ref{tab:n2h+} lists the mean, rms, maximum, and minimum values
of $\Delta V$, $V_{\rm LSR}$, $T_{\rm ex}$, and $\tau_{\rm tot}$ 
obtained from the hyperfine fitting.
Here, we compute these values by dividing the observed area 
into three parts: north (Decl. [J2000] $\ge$ $-02$:02:02.4), 
center ($-02$:02:02.4 $\ge$ Decl. [J2000] $\ge$ $-02$:05:32.5), and 
south (Decl. [J2000] $\le$ $-02$:05:32.5), because
these three parts appear to be separated
in the velocity channel maps presented in Figure \ref{fig:n2h+channel}.
The northern, central, and southern parts contain
the northern, central, and southern clumps, respectively
(see Figure \ref{fig:aztec}b).

In the following, we briefly describe the spatial distributions 
of these physical quantities obtained by the hyperfine fitting.

\subsection{Column Densities and Abundances}

We estimate the H$_2$ column densities by applying 
the spectral energy distribution (SED) fitting using the Herschel 
160, 250, 350, and 500 $\mu$m data.
The detail of the analysis will be described in 
Sugitani et al. (2013, in preparation). For the SED fitting, we used
the gas-to-dust ratio of 100 and the dust opacity per unit mass
of $\kappa _\nu = 0.005$ (1.3 mm/$\lambda$)$^\beta$ 
cm$^2$ g$^{-1}$ with $\beta=2$.
These parameters are the same as those adopted by \citet{konyves10}.

\begin{figure}[h]
\epsscale{0.3}
\plotone{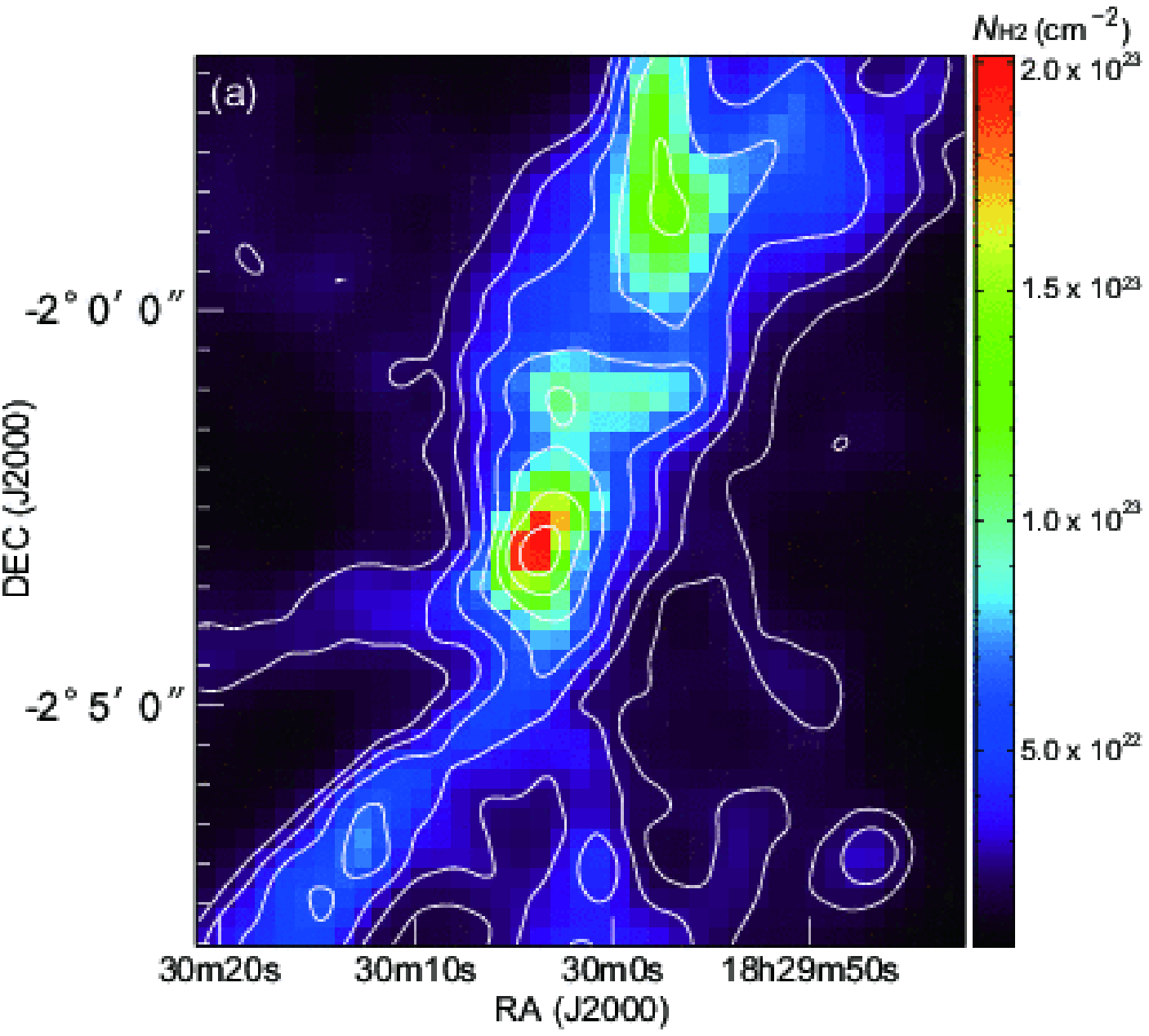}
\plotone{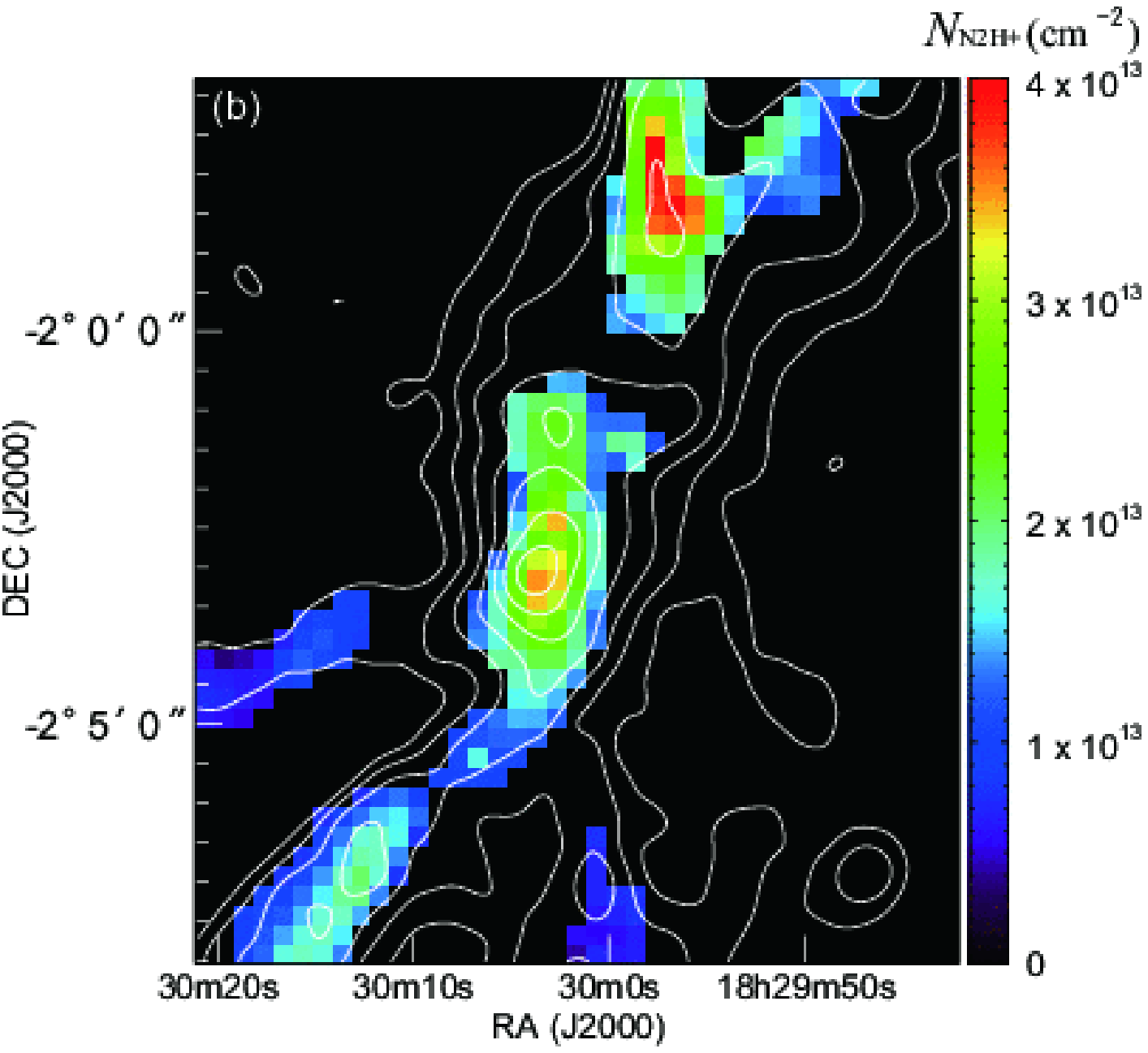}
\plotone{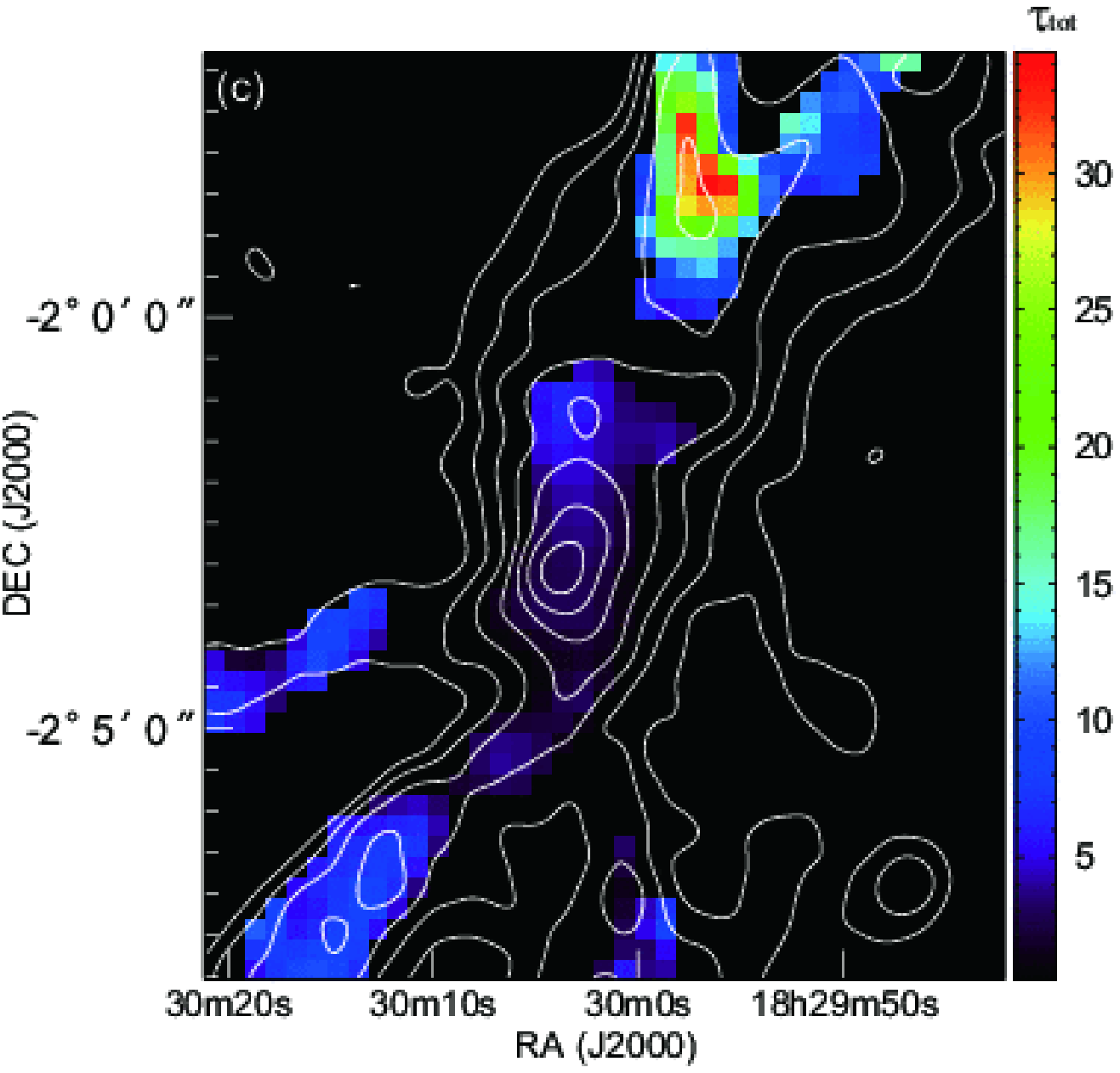}
\plotone{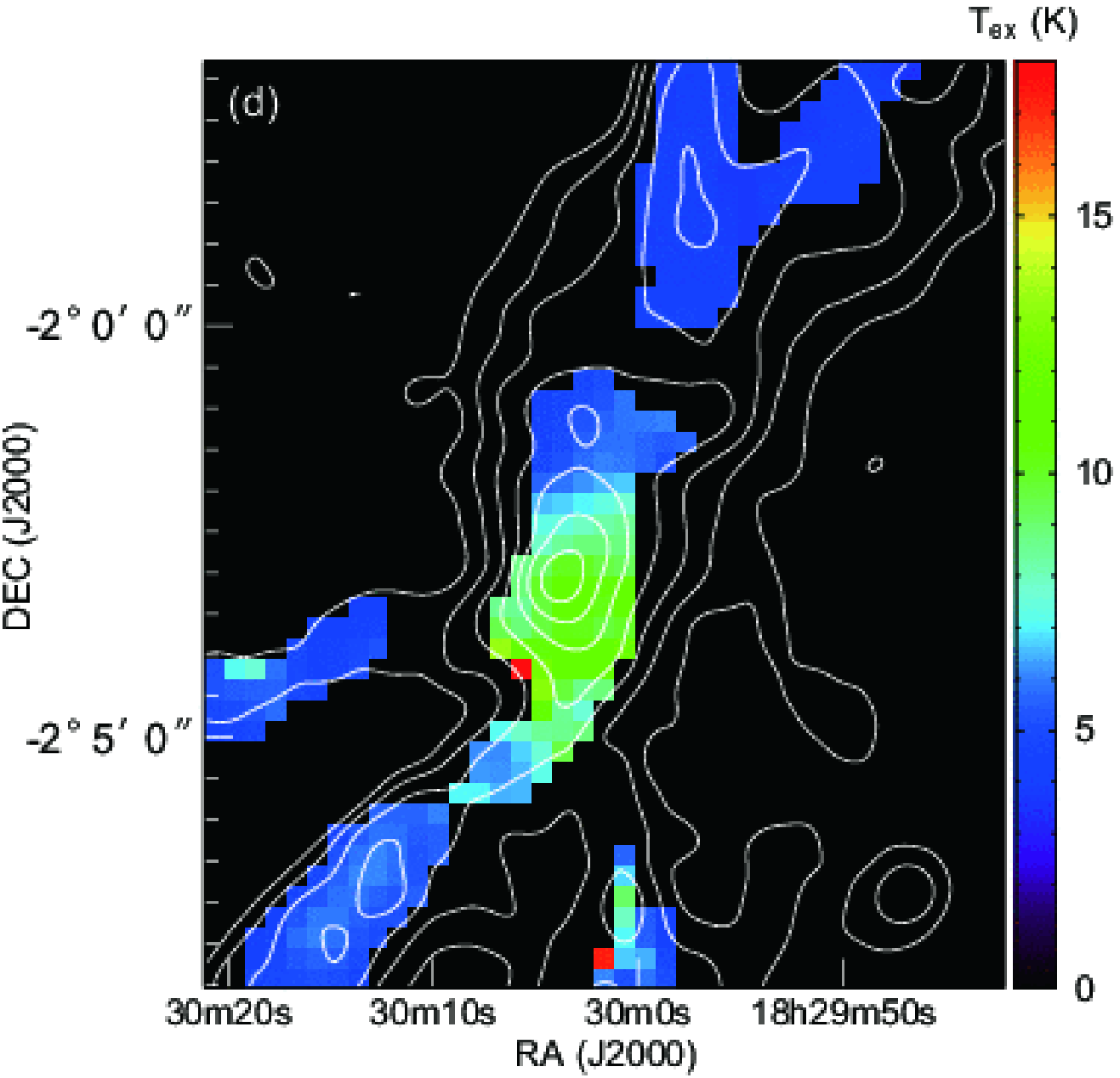}
\plotone{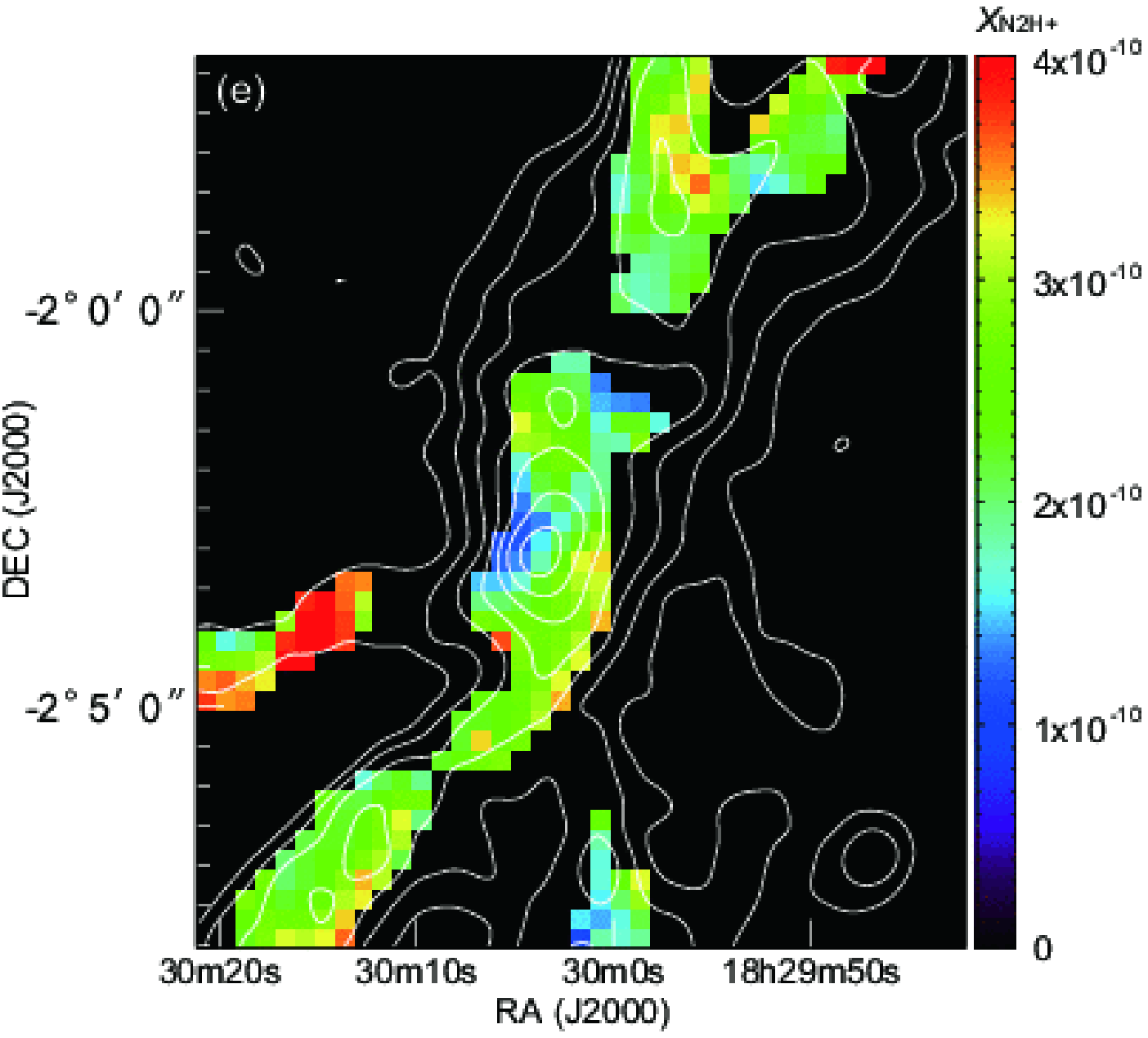}
\plotone{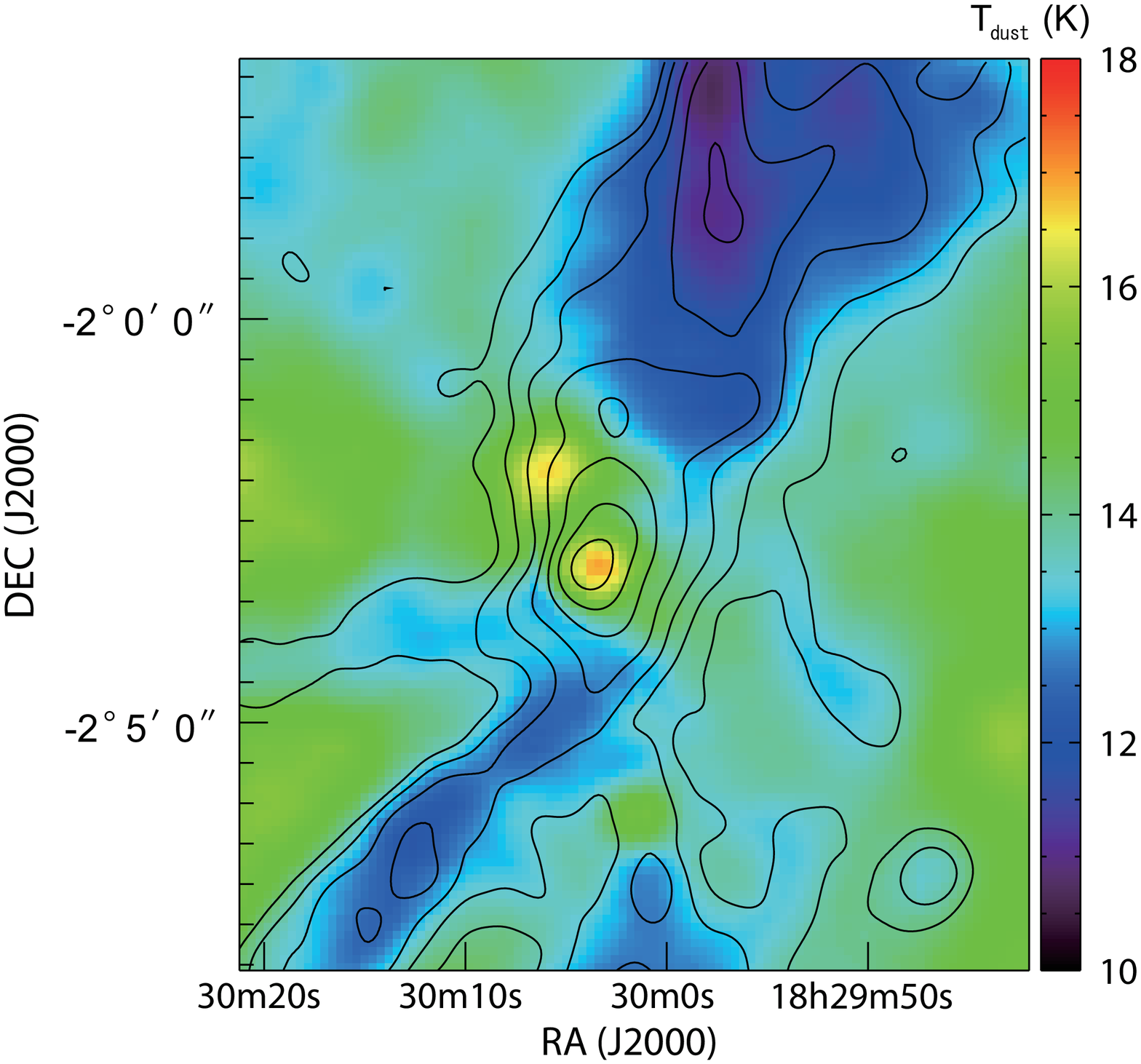}
\caption{(a) H$_2$ column density distribution 
($N_{\rm H_2}$) obtained 
from the SED fitting of the Herschel data.
(b) N$_2$H$^+$ column density distribution ($N_{\rm N_2H^+}$) obtained from 
the hyperfine fitting of the N$_2$H$^+$ data cube. 
(c) total optical depth distribution obtained from 
the hyperfine fitting of the N$_2$H$^+$ data cube.
(d) excitation temperature distribution obtained from 
the hyperfine fitting of the N$_2$H$^+$ data cube toward
Serpens South.
(e) N$_2$H$^+$ fractional abundance distribution 
($X_{\rm N_2H^+} = N_{\rm N_2H^+}/N_{\rm {H_2}}$) toward
Serpens South.
(f) Dust temperature distribution ($T_{\rm dust}$) obtained 
from the SED fitting of the Herschel data.
For all the panels, the contour lines indicate the 
the 1.1 mm dust continuum emission distribution. 
The contours start from $-$1.0 with an interval of 0.25 
in a logarithmic scale. 
}  
\label{fig:n2hpfit}
\end{figure}

Figure \ref{fig:n2hpfit}a shows the H$_2$ column density 
distribution in Serpens South obtained from the Herschel SED fitting. 
To compare $N_{H_2}$ with the N$_2$H$^+$ data, the H$_2$ column density data 
were remapped into the grid 
with the same pixels as that of the N$_2$H$^+$ data.
The estimated column density is essentially the same as that estimated by 
\citet{andre10} because 
the mass per unit length, line mass, is almost equal to their estimate of 
about 400 M$_\odot$ pc$^{-1}$ at a distance of 415 pc.
The N$_2$H$^+$ column densities are calculated by using the formula 
given in \citet{caselli02}. We used the optically-thin (thick)
formula when the optical depth of the strongest line is smaller 
(larger) than unity, to take into account the effect of the optical 
depth. The obtained column density distribution of N$_2$H$^+$ is presented
in Figure \ref{fig:n2hpfit}b.
Our hyperfine analysis reveals that the V-shape clump which has
relatively weak emission in the velocity integrated intensity map 
has a large column density peak. This is due to the fact that 
the optical depths of the N$_2$H$^+$ lines presented in 
Figure \ref{fig:n2hpfit}c are largest toward the V-shape clump.
For comparison, we also show the spatial distribution of the
excitation temperature $T_{\rm ex}$ in Figure \ref{fig:n2hpfit}d.
The excitation temperature tends to be higher toward the central
cluster-forming clump. In the northern clump, the excitation 
temperature tends to be as low as about 4 K, in spite of the 
high column density.  We note that at the position of 
the dust continuum emission peak (CSO-N), high-transition molecular 
lines such as CS ($J=5-4$), HCO$^+$ ($J=3-2$),
H$^{13}$CO$^+$ ($J=3-2$), and HCN ($J=3-2$) were not detected
with our CSO observations. These lines are not likely to be sufficiently 
excited due to either low temperatures or low densities, or both.

Figure \ref{fig:n2hpfit}e shows the N$_2$H$^+$ fractional abundance
across Serpens South. The mean fractional abundance 
of this region is estimated to be $X_{\rm N_2H^+}
\simeq 2.5 \times 10^{-10}$, which is similar to the 
values derived for other dark clouds;
e.g., $X_{\rm N_2H^+} \approx 10^{-10}$. 
For example, \citet{difrancesco04} and \citet{friesen10} 
estimated the N$_2$H$^+$ fractional abundances 
of $1.3\times 10^{-10}$ and $5.6\times 10^{-10}$
for the Ophiuchus A and B2 regions, respectively.
Figure \ref{fig:n2hpfit}e shows that the N$_2$H$^+$ fractional abundance varies 
by a factor of a few with the lower values 
of around $1.5\times 10^{-10}$ at the 1.1 mm continuum peak,
the northern part of the central clump, and the south-west 
filament, and the higher values of 
$(4 - 5) \times 10^{-10}$ at positions in the eastern and 
southern parts of the central clump and the northern V-shape
clump.  
Table \ref{tab:n2h+2} lists the mean, rms, minimum, and maximum of 
the N$_2$H$^+$ fractional abundances presented in Figure 
\ref{fig:n2hpfit}e.

\begin{figure}[h]
\plotone{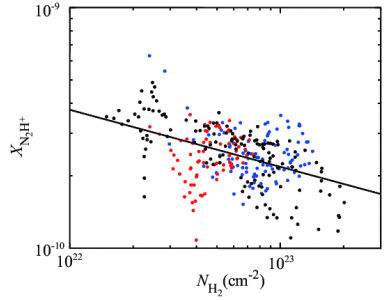}
\caption{Fractional abundance of N$_2$H$^+$ relative to H$_2$
against the H$_2$ column density in Serpens South. 
The H$_2$ column densities are derived by the SED fitting of the 
Herschel data. The solid line shows the best-fit power-law
of $X_{\rm N_2H^+} = 2.2 \times 10^{-10} 
(N_{\rm H_2}/10^{23} {\rm cm}^{-2})^{-0.237}$.
The blue, red, and black dots indicate the values at the pixels 
located in the  northern, central, and southern areas, respectively.
The northern, central, and southern areas are defined as 
follows: north (Decl. [J2000] $\ge$ $-02$:02:02.4), 
center ($-02$:02:02.4 $\ge$ Decl. [J2000] $\ge$ $-02$:05:32.5), and 
south (Decl. [J2000] $\le$ $-02$:05:32.5).
}  
\label{fig:abundance}
\end{figure}

In Figure \ref{fig:abundance}, the fractional abundances
$X_{\rm N_2H^+}$ are presented as a function of the H$_2$ column density.
Figure \ref{fig:abundance} indicates that 
the N$_2$H$^+$ fractional abundance decreases with 
increasing H$_2$ column density.
For comparison, we plot the best-fit power-law function that is given by 
\begin{equation}
\log X_{\rm N_2H^+} = (-9.66\pm 0.60) + (-0.237\pm 0.026)
\log (N_{\rm H_2}/10^{23} {\rm cm}^{-2}) \ ,
\end{equation}
with the correlation coefficient of ${\cal R}=0.49$. 
Although the correlation is weak, the dependence
of $X_{\rm N_2H^+}$ on $N_{\rm H_2}$ is consistent with 
a theoretical consideration of the ionization fraction 
in molecular clouds, as also pointed out by \citet{maruta10}. 
In molecular clouds, the ionization fraction is determined 
roughly by the balance between the ionization rate by cosmic rays 
and the recombination rate of ion and electron \citep{elmegreen79}.
Here, the ionization and recombination rates are proportional to
$n_n$ and $n_i n_e$, respectively, 
where $n_n$, $n_i$, and $n_e$ ($\approx n_i$) are the densities 
of the neutral gas, positive ion, and electron, respectively.
As a result, the ionization fraction decreases with increasing 
neutral density $n_n$, as $n_i/n_n \propto n_n^{-1/2}$.
The ionization fraction is expected to be proportional to 
$N_{\rm H_2} ^{-1/4}$ because $N_{\rm H_2} \propto n_n^{1/2}$
in a self-gravitating cloud.


\subsection{Centroid Velocities and Line Widths}

\begin{figure}[h]
\epsscale{0.4}
\plotone{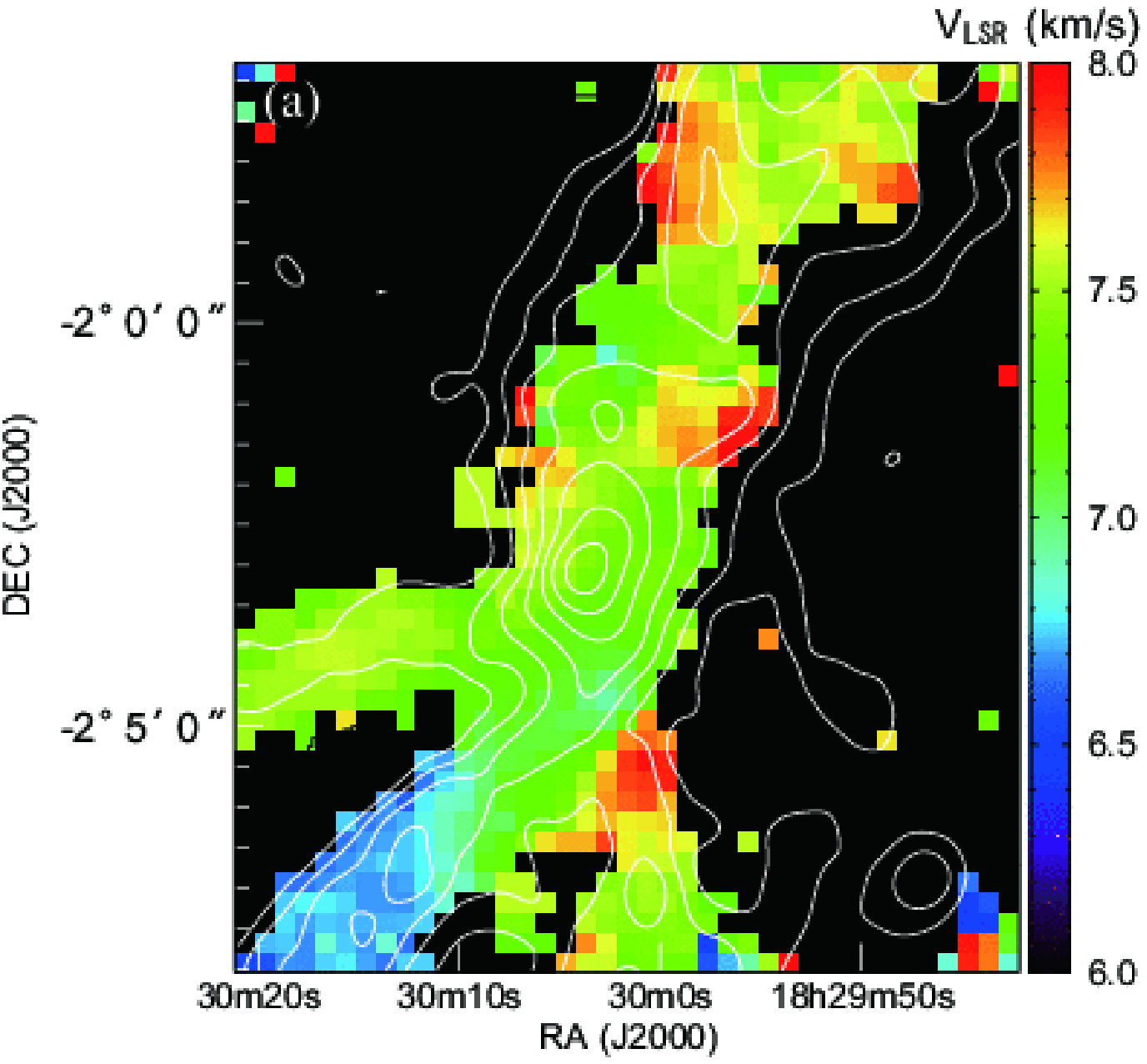}
\plotone{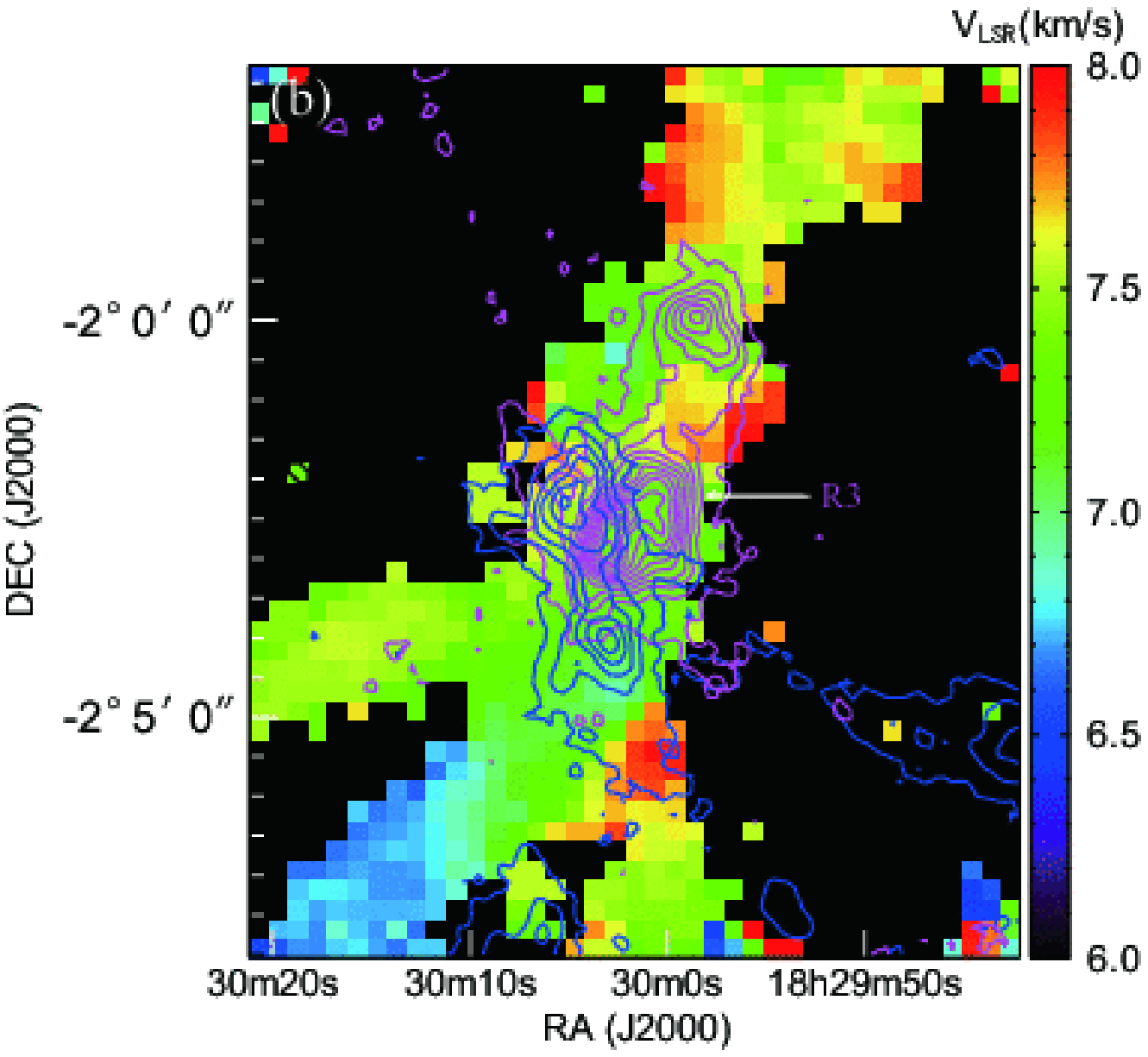}
\plotone{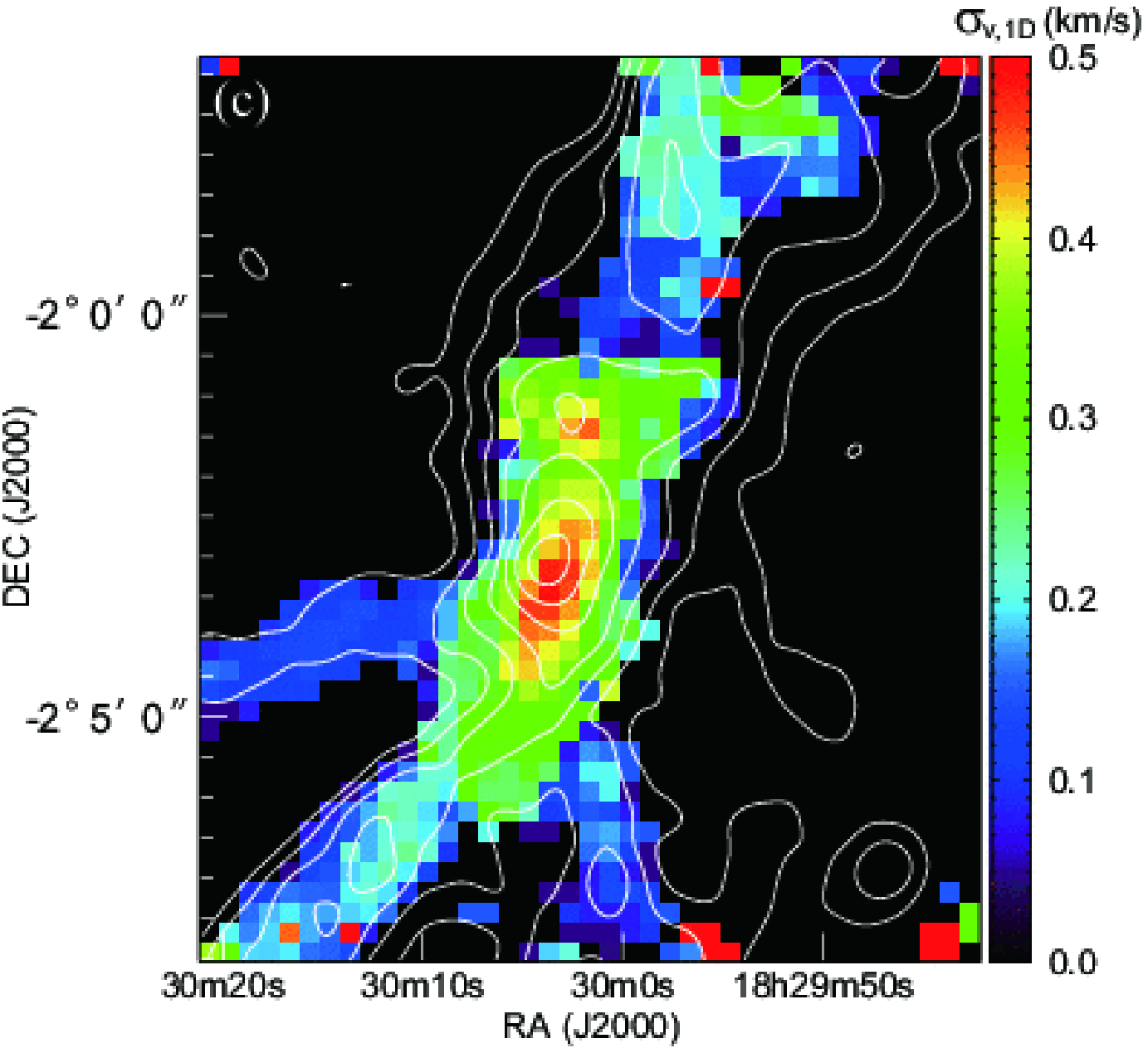}
\plotone{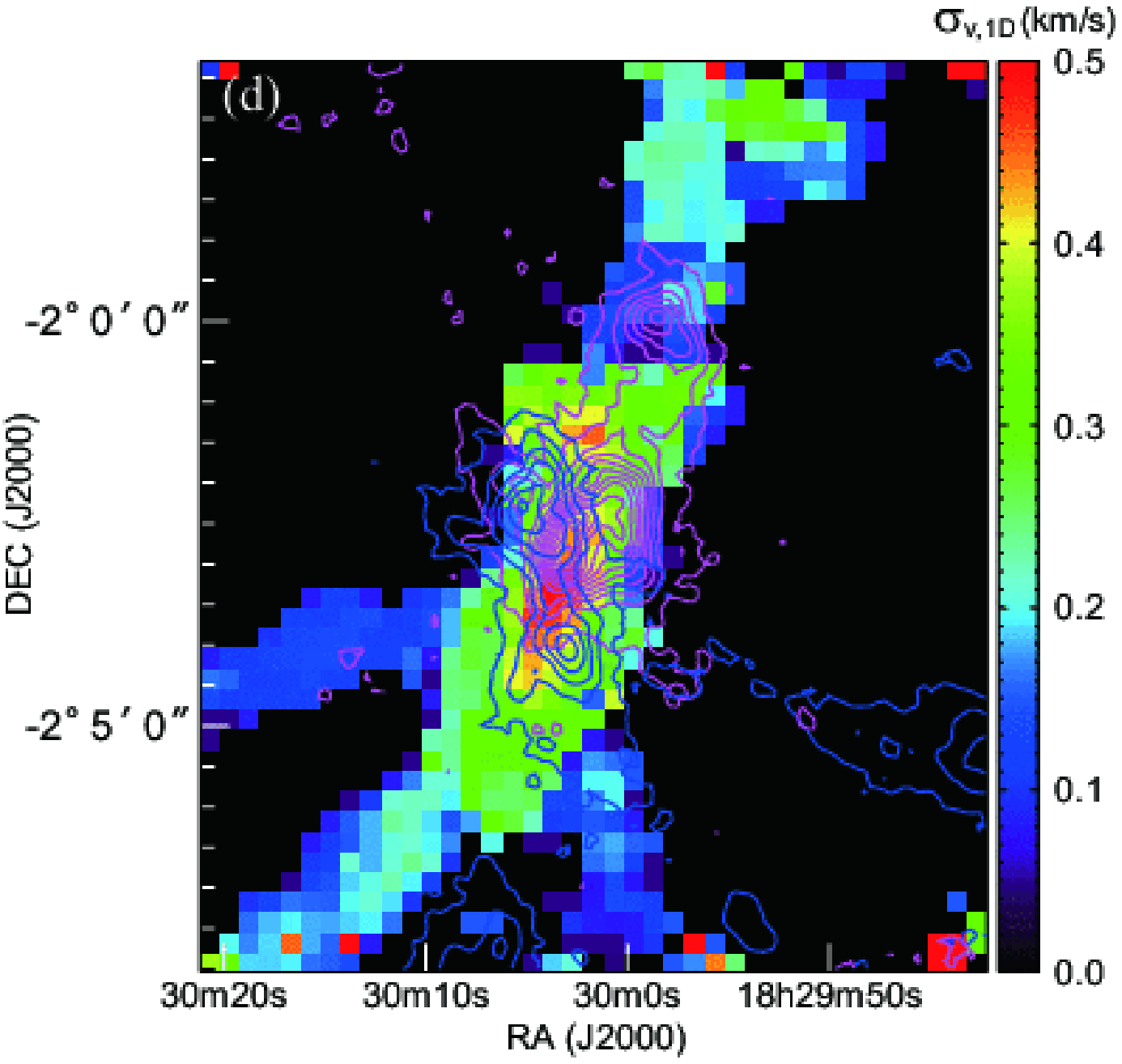}
\caption{(a) the centroid velocity distribution 
toward Serpens South.
(b) same as panel (a) but with the CO ($J=3-2$) outflow lobes overlaid.
(c) the distribution of the one-dimensional velocity dispersion
toward Serpens South.
(d) same as panel (c) but with the CO ($J=3-2$) outflow lobes overlaid.
The centroid velocity and 1D velocity dispersion are calculated in the 
range from 6 km s$^{-1}$ to 9.5 km s$^-1$ by integrating all the
 channels with the intensity larger than 0.3 K ($\simeq 4 \sigma$).
In panels (b) and (d), the blue contours represent blueshifted $^{12}$CO 
gas and magenta contours represent redshifted $^{12}$CO gas. 
The blue and magenta contour levels go up in 6 K km s$^{-1}$ steps, 
starting from 3 K km s$^{-1}$. The integration ranges are $-$9.75 
to 3.75 km s$^{-1}$ for blueshifted gas and 11.25 to 29.25 km s$^{-1}$ 
for redshifted gas. See \citet{nakamura11} in more detail.
We note that the isothermal sound speed is about 0.23 km s$^{-1}$ 
at the temperature of 15 K. Therefore, the pixels having green, 
yellow, or red color are roughly supersonic, the pixels having cyan 
are transonic, and the pixels having blue are subsonic.
}  
\label{fig:n2h+velocity}
\end{figure}

Figure \ref{fig:n2h+velocity}a shows a map of the centroid velocity 
(the first moment map), measured by the isolated hyperfine component
($F_1F \rightarrow F'_1 F' = 01 \rightarrow 12$), 
the second weakest hyperfine component, where
we calculated the centroid velocity by integrating the 
antenna temperatures of the isolated hyperfine component
for all the pixels above the 4$\sigma$ noise level.
A prominent characteristic of the centroid velocity 
distribution is a steep velocity change 
of 0.2 $-$ 0.3 km s$^{-1}$ near the southern part
of the central clump at 
(R.A. [J2000], Decl.[J2000]) $\sim$ (18:30:05, $-02$:05:00).
There is also a velocity jump of about 0.5 km s$^{-1}$
at the intersection between the main filament and 
the south-west filament of
(R.A. [J2000], Decl. [J2000]) $\sim$ (18:30:00, $-02$:05:00).
In the north-west part of the central clump,
the centroid velocity takes its maximum at 7.8 km s$^{-1}$,
which is shifted about 0.5 km s$^{-1}$
from that of the central clump (about 7.2 km s$^{-1}$). 
The area with this large centroid velocity in the central
clump is in good agreement with the position of the 
prominent redshifted outflow lobe discovered by 
\citet{nakamura11}, labeled with ``R3'', 
implying that the dense gas is interacting with 
the redshifted outflow lobe R3 (see Figure \ref{fig:n2h+velocity}b).

Figure \ref{fig:n2h+velocity}c shows a map of the one-dimensional 
(1D) velocity dispersion, $\sigma_{\rm v, 1D} (=\Delta V/\sqrt{8\ln 2})$,
(the second moment map) measured by the isolated hyperfine component.
The 1D velocity dispersion tends to be larger at the central clump
where active cluster formation is ongoing. 
In the central clump the velocity dispersion takes its maximum 
of about 0.5 km s$^{-1}$. The area having a large velocity dispersion
appears to overlap well with the blueshifted and redshifted outflow 
components blowing out of the central clump 
(see Figure \ref{fig:n2h+velocity}d).
In the other areas the velocity dispersion tends to be narrow 
with a mean of about 0.2 $-$ 0.3 km s$^{-1}$, 
which indicates that the internal motions are transonic.
In particular, the northern and southern clumps have
the narrow velocity dispersions of 0.25 km s$^{-1}$ and 0.2 km s$^{-1}$,
respectively. 
The spatial distribution of the velocity dispersion suggests
that the star formation activity has increased the internal
turbulent motions in this region.

\section{Dynamical State of Main Filament}
\label{sec:filament}

Here, we discuss the dynamical stability of the main filament.
An isothermal filament is in dynamical equilibrium when
the line mass is equal to the critical value 
of 
\begin{equation}
l_{\rm cr} = 2 \sigma_{v}^2/G \ , 
\end{equation}
where $\sigma_v$ denotes the three-dimensional 
velocity dispersion including the thermal contribution
\citep{stodolkiewicz63,ostriker64}. 
As mentioned in the previous section, the main filament has a line mass
of about 400 $M_\odot$ pc$^{-1}$ \citep[see also][]{andre10}. 
From the N$_2$H$^+$ hyperfine fitting, the mean line width 
is evaluated to be 0.81 km s$^{-1}$, corresponding to 
$\sigma_v \simeq 0.62$ km s$^{-1}$.
Thus, the critical line mass of the Serpens South main filament 
is roughly estimated to be $l_{\rm cr} \sim$ 180 $M_\odot$ pc$^{-1}$, 
under the assumption that the internal turbulence 
provides the isotropic pressure to support the filament
against the gravity.
Therefore, the line mass of the main filament is more than twice
the critical value of 180 $M_\odot$ pc$^{-1}$, suggesting that 
the filament is very unstable to radial collapse in the absence 
of additional support.
However, as mentioned in the next Section, 
rapid global collapse has not been detected in this region. 
Thus, the filament is likely to
be supported by additional forces.

A promissing support mechanism is the magnetic field. 
\citet{nakamura93} discussed that in the persence of longitudinal 
or helical magnetic field, a filamentary cloud  can
reach its dynamical equilibrium state even if its 
line mass exceeds $l_{\rm cr}$ \citep[see also][]{fiege00}. 
A rotation around the filament axis also provides the radial suppport
\citep{matsumoto94}.
However, there is no evidence of the rotation for the Serpens South filament.
According to \citet{nakamura93}, a longitudinal magnetic field 
can enhance the gravitational fragmentation with help of
the magnetic buoyancy instability, i.e., the Parker instability, and
the most unstable wavelength relative to the cloud diameter 
becomes shorter than that with no magnetic field.
The longitudinal component, if exists, should have its strength
of 
\begin{equation}
B_{\parallel} \simeq \sqrt{8\pi \sigma_{v}^2 \rho} \sim 200 
\left(\sigma_v \over 0.62 \ {\rm km \ s}^{-1} \right)
\left(\rho \over 10^5 \ {\rm cm}^{-3} \right)^{1/2} \ {\rm \mu G} \ ,
\end{equation}
to support the filament against the radial collapse
\citep{nakamura93}.
However, there is no clear evidence of the presence of the longitudinal
magnetic field in the dense regions.

In contrast, recent near-infrared polarimetric observations have revealed 
that the main filament is penetrated 
by the globally-ordered magnetic field that can provide the 
additional support against the gravity \citep{sugitani11}.
Therefore, the magnetic field may be roughly perpendicular to 
the filament.
A magnetic field perpendicular to the filament axis can also provide
the significant dynamical support against gravity if the field strength
is close to the critical value for magnetic support.
If the Serpens South filament was penetrated by the nearly 
critical magnetic field, the fragmentation is likely to have been 
triggered by the ambipolar diffusion since 
the filament appears to have already fragmented into several dense
clumps. 
However, since the near-infrared observations tend to trace the 
intermediate-density envelopes ($\sim 10^4$ cm$^{-3}$) of the clouds, further 
polarimetric observations such as submillimeter continuum polarimetry
are needed to 
uncover the magnetic field structures in the densest part.
We will discuss the magnetic support in more detail 
in Section \ref{subsec:bfield}.

\section{Dynamical State of Dense Clump}
\label{sec:virial}

\subsection{Clump Virial Analysis}

From the dust continuum map, we can recognize
several local peaks along the main filament, as already shown in 
Section \ref{subsec:1.1mmresults}.
In other words, the filament apparently fragments
into several pieces. 
Here, we define the subparsec-size substructures having 
the local peaks as the clumps.
We identified three dense clumps along the main filament: 
northern (V-shape), central (cluster-forming),
and southern clumps.
The clumps are defined as the regions enclosed by 
a contour of $5\times 10^{22}$ cm$^{-2}$ on the H$_2$ column density map
in each (northern, central, or southern) area presented 
in Fig. \ref{fig:n2hpfit}.
We interpret these clumps as sub-units of star or cluster formation
in this region. 
The threshold column density is somewhat arbitrary. However,
our main conclusion given in the present paper are not influenced
by the actual value of the threshold column density 
in the range between $\sim 5\times 10^{22}$
cm$^{-2}$ and $\sim 1\times 10^{23}$ cm$^{-2}$.
The clumps also contain substructures
that are recognized in the velocity-channel maps of N$_2$H$^+$
presented in Figure \ref{fig:n2h+channel}. 
We identify the internal structures of the clumps as the cores
in the next section.
Here, we perform the virial analysis using 
the Herschel and N$_2$H$^+$ data to investigate 
the dynamical states of the clumps.

Following \citet{nakamura11}, we adopt the virial equation 
of a spherical gas cloud that is given by
\begin{equation}
\frac{1}{2} \frac{\partial ^2 I}{\partial t^2} = 2U+W \ ,
\end{equation}
where $I$ is the moment of inertia, $U$ is the internal kinetic energy,
and $W$ is the gravitational energy.
In the above equation, the surface pressure term is omitted.
A clump is in virial equilibrium when $2U+W=0$.
The internal kinetic energy term $U$ and the gravitational energy
term $W$ are expressed as follows:
\begin{equation}
U=\frac{3M\Delta V^2}{16\ln 2}
\end{equation} 
and
\begin{equation}
W=-a\frac{GM^2}{R} 
\left[1-\left(\frac{\Phi}{\Phi_{\rm cr}}\right)^2\right]  \ , 
\end{equation} 
where the values $\Phi$ is the magnetic flux penetrating the clump 
and $\Phi_{\rm cr}$ is its critical value above which the clump
is fully supported by the the magnetic field 
against the self-gravity \citep{nakano98}.
In the discussion of this subsection, we assume $\Phi=0$, 
which means that the clumps are not supported by the magnetic forces.
We will discuss the contribution of the magnetic field to the cloud supoprt
in Section \ref{subsec:bfield}.
The value $a$ is a dimensionless parameter of order unity 
which measures how much the mass distribution deviates 
from a sphere with the profile of $\rho \propto r^{-2}$
\citep{bertoldi92}. For example, for a uniform sphere and a centrally-condensed 
sphere with $\rho \propto r^{-2}$, the dimensionless 
parameter $a$ is given by 3/5 and 1, respectively. 
For the identified clumps, we assume that the effects of the
nonspherical mass distribution is negligible.
This is a reasonable assumption because the aspect ratios of 
the clumps identified are not so far from unity and in such cases, 
the value of $a$ is close to unity \citep{bertoldi92}.
Below, we assume $a$ to be unity because each clump has a more or less 
centrally-condensed density distribution.

Table \ref{tab:clump} summarizes the physical quantities of the three clumps.
Here, we estimate the masses and radii from the Herschel column density
map. The clump radius is determined by taking the area enclosed by the 
threshold H$_2$ column density of $5\times 10^{22}$ cm$^{-2}$ 
and computing the radius of the circle 
required to reproduce the area.
The clump line widths are estimated by taking the intensity-weighted
average of the one-dimensional velocity dispersion obtained from 
hyperfine fitting of the N$_2$H$^+$ data.
The mass of the northern clump is estimated to be about 200 $M_\odot$,
which is comparable to that of the central cluster-forming clump.
In spite of the relatively large mass, only several YSOs are 
identified in the northern clump. This may imply that the northern 
clump is likely to be the site of future cluster formation, 
or star formation may be precluded by some additional clump support.
The southern clump has about 40 $M_\odot$ and is less massive than
the other two.
The line width of the northern clump is evaluated to be about 
0.7 km s$^{-1}$, which is as small as that of the typical molecular 
cloud core in quiescent regions. The southern clump, with which 
only a few YSOs appear to be associated, also has a 
similar line width. On the other hand, the central 
{\it cluster-forming} clump has a larger line width of about 
1.3 km s$^{-1}$. 
This may indicate that the star formation activity 
enhances the internal motions in the central clump.

For all the clumps, the terms $U$ and $W$ are estimated to be 
$U \approx 40 - 110$ M$_\odot$ km$^2$ s$^{-2}$ and 
$W \approx - 40 \sim -1000$ M$_\odot$ km$^2$ s$^{-2}$, respectively, 
which yields  the virial ratio, the ratio between $2U$ and $W$, of 0.1 $-$ 0.3.
For all the clumps, the virial ratios are estimated to 
be well below the unity, indicating that the 
internal turbulent motions play only a minor role in the clump support.
This conclusion is inconsistent with 
the previous estimation by \citet{nakamura11}, who
concluded that the Serpens South clump as a whole is 
close to quasi-virial equilibrium. 
The main difference between the two estimations comes
from the different line widths. \citet{nakamura11}
obtained the line width of the central clump using 
HCO$^+$ ($J=4-3$). However, this line tends to be optically thick 
toward the central clump, which may lead to the overestimation 
of the line width.  We believe that the present 
estimation is more reasonable because N$_2$H$^+$ ($J=1-0$) tends to be
more optically thin than HCO$^+$ ($J=4-3$) in the densest part.  
This cloud as a whole seems 
to be extremely gravitationally unstable 
in the absence of the additional support mechanisms
such as the magnetic support.

We note that the virial ratios of the clumps are not sensitive to the 
threshold column density that defines the clump boundary ($N_{\rm cr}$).
For example, for the threhold density of 
$N_{\rm cr} \lesssim 1\times 10^{23}$ cm$^{-2}$, 
the virial ratios of the northern and central clumps are almost 
the same as those in Table \ref{tab:clump}.
For the southern clump whose peak column density is $7\times 10^{22}$ 
cm$^{-2}$, the virial ratio is almost the same as that in Table 
 \ref{tab:clump} as long as $N_{\rm cr} \lesssim 6\times 10^{22}$ cm$^{-2}$.

The clumps and cores having the small virial ratios have recently been
reported by a couple of groups. For example, \citet{dili12} carried out
NH$_3$ observations toward OMC-2 and OMC-3 and found that the clumps in
this area have small virial ratios. 
\citet{pillai11} performed the virial analysis of the 
IRDC clumps and found that the IRDC clumps often have
low virial ratios (see also Pillai et al. 2013, in prep.).

\subsection{Global Infall}

As shown above, the cloud as a whole seems not to be 
supported by the internal turbulent motions, and therefore
the global infall motions are expected.
In fact, the profile maps of the optically thick lines 
often show blueshifted part stronger than the redshifted part,
i.e., blue-skewed \citep[see][]{nakamura11}. 
Such blue-skewed profiles are probably 
indicative of global infall motions.

\begin{figure}[h]
\epsscale{0.8}
\plotone{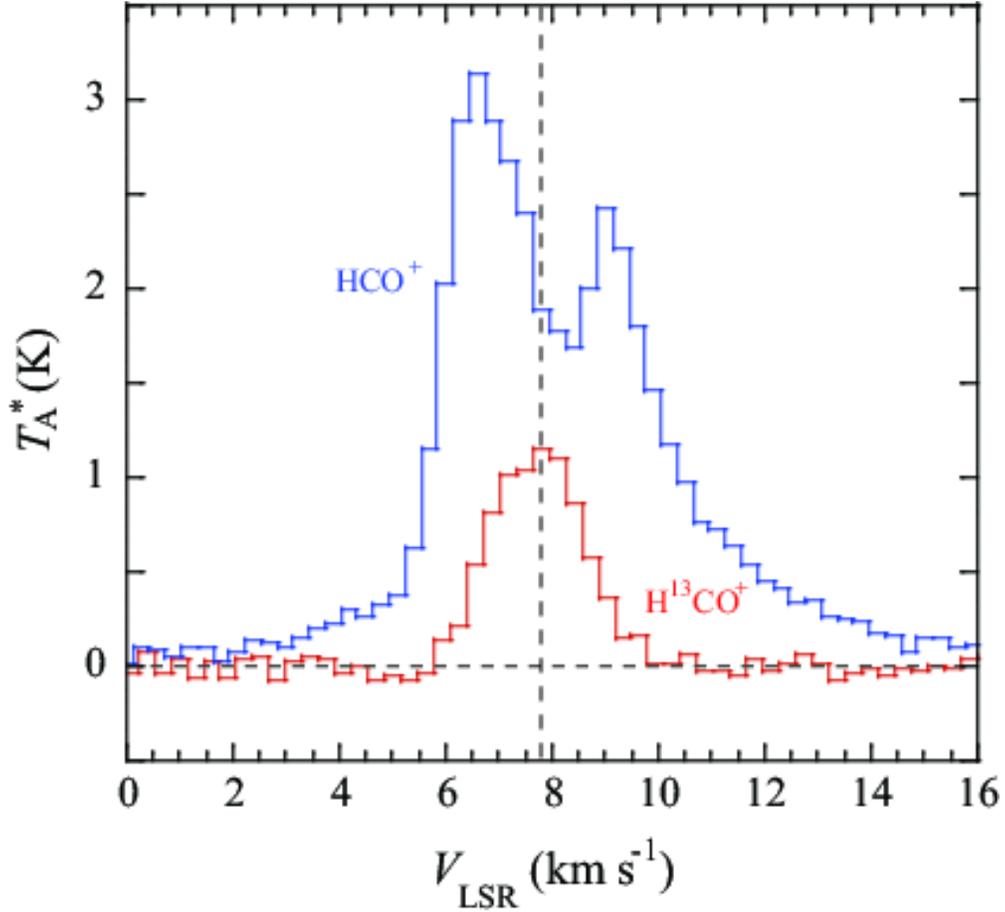}
\caption{HCO$^+$ ($J=3-2$, {\it blue}) and 
H$^{13}$CO$^+$ ($J=3-2$, {\it red}) line profiles
 toward the Serpens South clump at the position of 
(R.A. [J2000], Dec. [J2000]) $=$ (18:30:3.80, $-$2:03:03.9).
The FWHM beam size of the CSO 10.4-m telescope was 30$''$.
The optically-thick HCO$^+$ ($J=3-2$) line shows a clear blue-skewed
profile, i.e., infall signature, 
with prominent blueshifted and redshifted wings that are
probably caused by the powerful molecular outflows.
On the other hand, the optically-thin H$^{13}$CO$^+$ ($J=3-2$) line
shows a single-peak profile.
The dashed line indicates the value of $V_{\rm LSR}$ at 
the H$^{13}$CO$^+$ ($J=3-2$) emission peak.
}  
\label{fig:infall}
\end{figure}

In Figure \ref{fig:infall}, we present a HCO$^+$ ($J=3-2$) line profile 
detected toward CSO-C (in the Serpens South clump), whose coordinate is 
(R.A. [J2000], Decl, [J2000]) = (18:30:03.80, $-$02:03:03.9).
At CSO-C, the HCO$^+$ ($J=3-2$) line is likely to be
optically-thick. For comparison, an optically-thin line
H$^{13}$CO$^+$ ($J=3-2$) is presented in Figure \ref{fig:infall}.
The HCO$^+$ ($J=3-2$) line shows a clear blue-skewed profile, whereas
the H$^{13}$CO$^+$ ($J=3-2$) line shows a single-peak profile.
This indicates that the densest part has an infall motion.

Applying a simple model of \citet{myers96}, we evaluated 
the infall speed of 0.024 km s$^{-1}$, which is subsonic.
If the central clump as a whole is contracting
with this subsonic infall speed, the mass infall rate is estiamted to be
$\dot{M}= 4\pi R_{\rm clump}^2 \rho V_{\rm infall} \sim 
9\times 10^{-5} M_\odot$ yr$^{-1}$, where $\rho$ is set to the 
mean density in the clump, $\rho \sim 5\times 10^{-18}$ g cm$^{-3}$. 
This implies that 
the mass of the central clump will be doubled within 2.5 Myr.
The estimated mass infall rate is in good agreement with
that obtained by \citet{kirk13} within a factor of a few.

The estimated infall speed is, however, too slow compared to the 
free-fall velocity of $V\sim \sqrt{GM/R} \sim 1-2$ km s$^{-1}$,
corresponding to the FWHM line widths of 2 $-$ 4 km s$^{-1}$.
The clump FWHM line widths of 0.5 $-$ 1.3 km s$^{-1}$
are smaller than this free-fall velocity.
Hence, the additional support mechanism should be needed.

\subsection{Magnetic Support}
\label{subsec:bfield}

As discussed above, the internal turbulent motions 
play only a minor role in the clump support in this region.
However, there is no clear evidence showing the free-fall 
global infall motions 
of order of a few km s$^{-1}$. The line widths
along the line-of-sight are at most 0.5$-$ 1 km s$^{-1}$.
This implies that the additional support mechanism 
is needed. 
The most promising mechanism of the additional support
is the magnetic support. Here, we discuss whether 
the magnetic field plays an important role in the clump
support in this region.

\citet{sugitani11} carried out the near-infrared (JHKs) 
polarimetric observations toward Serpens South
and found that the Serpens South filament is 
penetrated perpendicularly to the large-scale ordered 
magnetic field. The global magnetic field also 
appears to be curved in the southern part of the main filament. 
Such morphology is consistent with the idea that the global 
magnetic field is distorted by gravitational contraction 
along the main filament toward the central cluster-forming clump.
Applying the Chandrasekhar-Fermi method,
they estimated the magnetic field strength of 
about 80$\mu$G at the filament envelope with the 
density of $10^4$ cm$^{-3}$, where we rescaled the 
distance to 415 pc and used the smaller line width
estimated from our N$_2$H$^+$ data.  
Assuming an inclination angle of $45^\circ$, the 
magnetic field strength is roughly estimated to be
about 120 $\mu$G.

The mean densities of the clumps can be estimated to be
about $10^5$ cm$^{-3}$ [$=M_{\rm cl}/(4\pi R_{\rm cl}^3/3)$]. 
Assuming that the magnetic field strength is
proportional to the square root of the density, 
$B\propto \rho^{1/2}$, the strength of the magnetic field associated
with the clumps is evaluated to be about 400 $\mu$G.
If the clumps have the field strengths of about 400 $\mu$G, the
mass-to-magnetic-flux ratios are estimated to be around 
unity, close to critical. 
In other words, the magnetic support is expected to 
play a crucial role in the clump dynamics and evolution.

According to our estimation, the parent cloud was presumably 
close to magnetically-critical and either 
the ambipolar diffusion or mass accretion along the magnetic field 
lines may have triggerd the slow global infall and eventually 
the active cluster formation that is observed in the central clump 
\citep{nakamura10,bailey12}.

\subsection{Clump-Mass-Size Relation}

\begin{figure}[h]
\epsscale{0.8}
\plotone{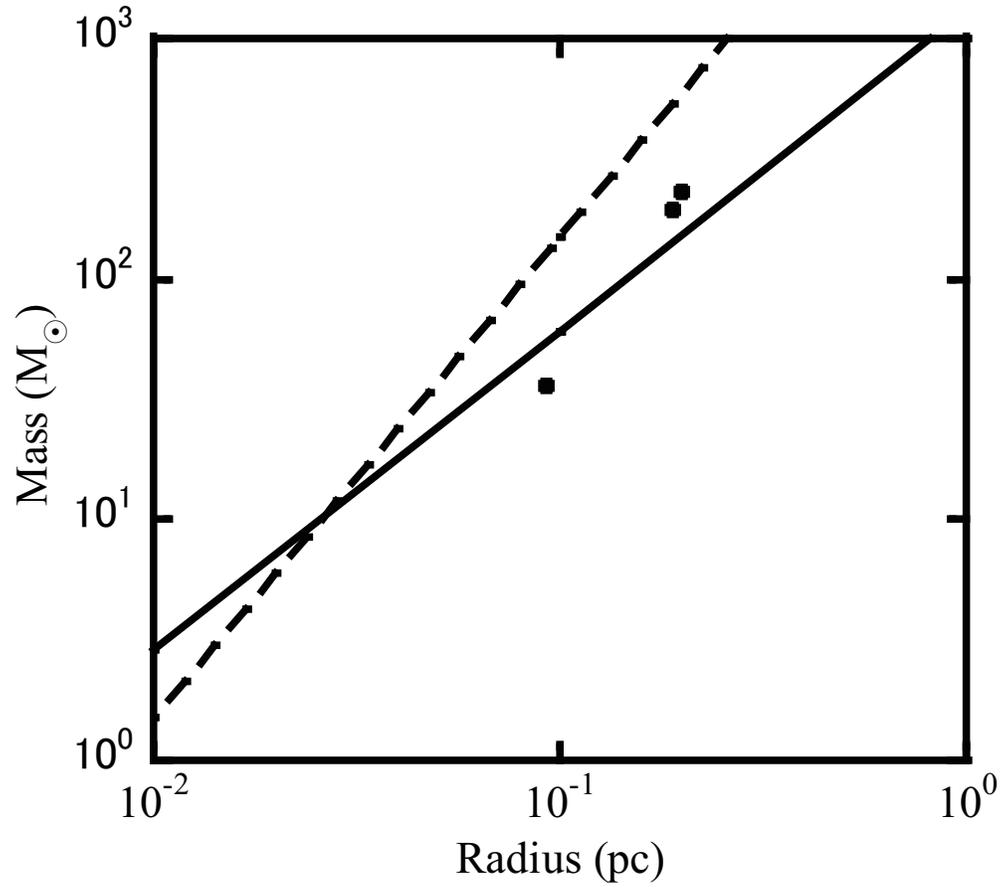}
\caption{Clump mass-radius relation for the three Serpens South clumps.
The solid and dashed lines indicate the criteria for the massive star 
formation by \citet{kauffmann10} and \citet{krumholz08}, respectively.
}  
\label{fig:clumpmass-radius}
\end{figure}

Recently, \citet{kauffmann10a} and \citet{kauffmann10} 
analyzed the mass and radius of cloud
substructures in several nearby molecular clouds over a wide range of 
spatial scales ($0.05 \lesssim r/{\rm pc} \lesssim 3$) and 
found that the clumps forming massive stars ($M\gtrsim 10M_\odot$) 
have masses larger than $M \gtrsim 1300 M_\odot (r/{\rm pc})^{1.33}$,
where we rescaled Kauffmann \& Pillai's relation by a factor of 1.5 
to take into account the difference in the adopted dust parameters.
On the other hand, \citet{krumholz08} proposed the threshold 
column density of 1 g cm$^{-2}$ for massive star formation.
In Figure \ref{fig:clumpmass-radius} 
we plot the three Serpens South clumps
(northern, central, and southern clumps) on the 
mass-radius diagram with Kauffmann \& Pillai's relation and 
Krumholz \& McKee's criteria for massive star formation.
The northern and central clumps are located above the critical 
mass-radius relation but below the Krumholz criteria.
Thus, we expect that these two clumps are borderline clumps 
for forming massive stars.

Interestingly, all the Serpens South clumps have column densities
comparable to about $10^{23}$ cm$^{-2}$.
Such clumps are the densest ones found 
among nearby star-forming molecular clouds whose distances are 
less than about 400 pc, except in the Orion molecular cloud,
the nearest massive star forming region.
In the Orion molecular cloud,
a number of clumps have the column densities similar to
or larger than the Serpens South clumps 
\citep[see Figure 1 of ][]{kauffmann10a}.
In fact, the clumps containing the most 
massive core found for small radii in the Perseus and Ophiuchus
molecular clouds have the column densities comparable to 
the Serpens South clumps. Therefore, we expect that 
the physical conditions of the Serpens South clumps resemble
at least those of the nearby cluster-forming clumps where 
only low-mass and intermediate-mass stars are forming (e.g., NGC 1333).

The central clump already harbors a cluster of protostars,
whereas the northern and southern clumps show no active 
star formation.
Since the northern clump has a mass and radius comparable 
to the central clump, the northern clump is a promissing 
candidate of {\it pre-protocluster} clump.

\subsection{Scenario of Clump Formation and Evolution}

\begin{figure}[h]
\epsscale{0.8}
\plotone{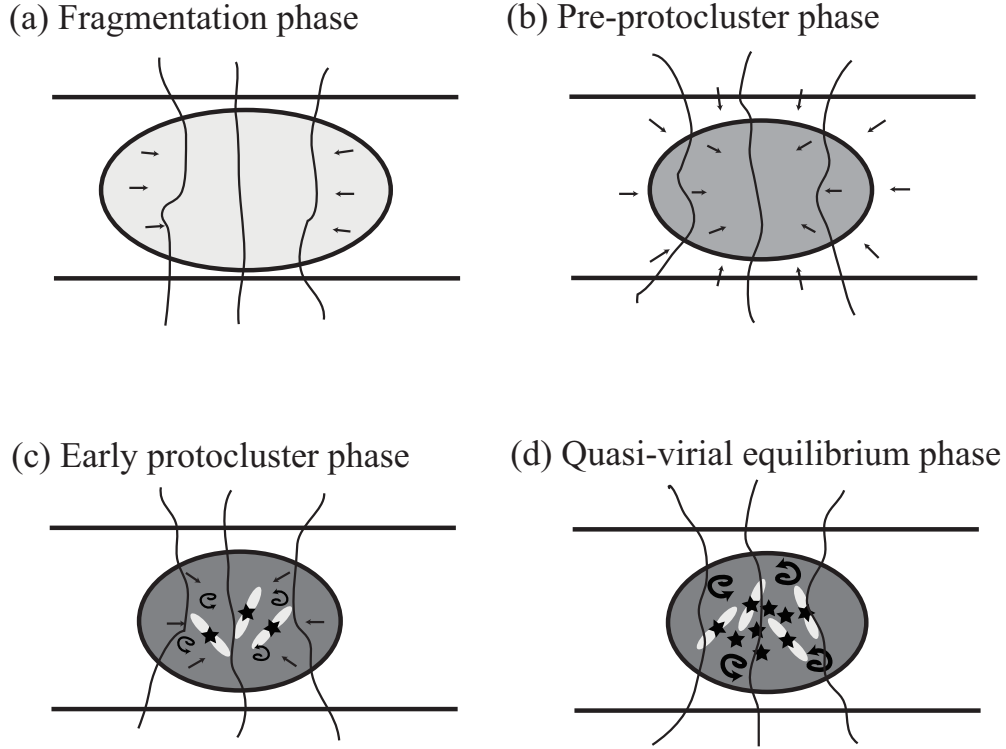}
\caption{Magnetically-regulated cluster formation in
Serpens South.
(a) When the initial filament is supported by the strong magnetic 
field, the fragmentation is driven by the ambipolar diffusion
to form dense clumps.
(b) a dense clump grows in mass due to the mass infall 
along and/or across the magnetic field lines.
Or, the merging of the clumps along the filament axis may 
play a role in increasing the clump mass. Although the accretion or merging 
supplies turbulent motions in the clump, the internal motions 
are relatively small. The virial ratio of such a clump may be 
only a few tenth unless the magnetic support is not taken into account.
This stage is refered to as ``pre-protocluster'' stage.
(c) active cluster formation is initiated in the clump.
(d) protostellar outflows inject additional momentum in the ambient material 
and the clump as a whole is supported by the 
outflow-driven turbulence. 
}  
\label{fig:scenario}
\end{figure}

We propose the following 
scenario of the clump formation and evolution in Serpens South.
Figure \ref{fig:scenario} shows the conceptual diagram of the scenario.

(a) Fragmentation phase
 
Dense clumps form by fragmentation of a filamentary cloud 
that was close to magnetically critical. The fragmentation 
is likely to be driven by the ambipolar diffusion.
In the absence of the ambipolar diffusion, the magnetically-critical 
filaments cannot fragment into clumps. 
Without the ambipolar diffusion, even in the
mildly-magnetically-supercritical filaments, it is difficult to
obtain the clumps whose separations are several times as large as 
the filament diameter, as seen in the dust continuum image,
because the clump separations would become much larger.

(b) Pre-protocluster phase

The clumps grow in mass by gas accumulation either
along or across the magnetic field lines, or both.
The merging of the fragments along the filament axis may 
also play a role in increasing the clump mass.
This stage corresponds to the one immediately before
active cluster formation is initiated. Thus, we refer to this stage
as the pre-protocluster stage.
The pre-protocluster clumps are the magnetically-supported clumps 
of cold, quiescent, dense gas. 

(c) Early protocluster phase 

When the density becomes high enough, active cluster formation is
initiated and protostellar outflows start injecting supersonic turbulence
in the clumps. Hence, the line widths of the clumps 
become larger.

(d) Quasi-virial equilibrium phase

Eventually, the clumps reach quasi-virial equilibrium due to 
the supersonic turbulence driven by protostellar outflows
\citep[see also][]{sugitani10}.

The southern, northern, and central clumps are presumably in 
Stages (a), (b) and (c), respectively.

We note that the physical properties of the pre-protocluster clumps
identified in Serpens South are different from those expected 
from the accretion-driven turbulence scenario
\citep{klessen10,enrique10}, for which the supersonic turbulence
is maintained by the accretion flows onto the dense clumps.
The accretion-driven turbulence scenario considers that 
star formation proceeds dynamically, and predicts 
that the turbulent energy balances with the gravitational energy 
in the dense clumps in formation.  
In contrast, the dense clumps in Serpens South appear more quiescent,
in particular prior to the cluster formation.
In Serpens South, stellar feedback is likely to be more responsible 
for driving turbulent motions in the dense clumps.

On the other hand, it remains unclear how the main filament was formed. 
The collision between the multiple filaments may have played a role 
because in the channel maps, several velocity-coherent components can 
be recognized. Further investigation is needed to understand 
how the Serpens South filament was formed.

\section{Physical Properties of Dense Cores}
\label{sec:core}

As shown in Section \ref{sec:results}, the N$_2$H$^+$
velocity channel maps show that the clumps contain 
substructures, or dense cores. Here, we identify the dense cores
from the N$_2$H$^+$ data cube and estimate the physical 
properties of the dense cores.

We note that the core identification depends strongly on the 
angular and velocity resolutions. Thus, the identified cores 
may not be the actual units that are separated dynamically 
from the background media. However, we attempt to identify 
the substructures recognized in the current data
as dense cores, following the previous studies.

\subsection{Identification of Dense Cores}

We apply the clumpfind to identify dense cores 
from the N$_2$H$^+$ data cube \citep{williams94}. The conditions 
adopted for the core identification 
are essentially the same as described in \citet{ikeda07}. 
The threshold and stepsize of the core identification are 
set to the 2$\sigma$ noise level.
A core identified must contain more than two continuous velocity channels
and each channel must contain more than 3 pixels whose 
intensities are larger than the 3$\sigma$ noise levels.
In addition, the core must have pixels connected to both 
the velocity and space domains
(see Ikeda et al. 2007 for further details).
Most of the N$_2$H$^+$ ($J=1-0$) hyperfine components are  
blended significantly over much of the area observed. 
Therefore, we used only the isolated component 
($F_1F \rightarrow F'_1 F' = 01 \rightarrow 12$) to identify the cores.  
A minimum mass of the core identified by this procedure 
is evaluated to about 0.17 M$_\odot$ (corresponding 
to about 60 $\sigma$ levels)
for $T_{\rm ex} \approx 7$ K and 
$X_{\rm N_2H^+} \approx 2.5 \times 10^{-10}$. 
In total, we identify 70 dense cores.

\subsection{Derivation of Physical Quantities}

We determine the physical properties of the cores are determined 
by using the definitions described in Section 3.3 of \citet{ikeda07}. 
The position and local standard of rest (LSR) velocity 
of a core are set to the values at the pixel having the largest
antenna temperature, $T_{\rm A, peak}^*$. 
The core radius, $R_{\rm core}$, is calculated as the radius of 
a circle having the same area projected on the position-position plane
as the identified core, by using Equation 2 of \citet{ikeda07}.
The aspect ratio of the core is calculated by
the two-dimensional Gaussian fitting to its total integrated 
intensity distribution. 
The FWHM line width, $dV_{\rm core}$, is corrected 
for the velocity resolution of the spectrometers ($=0.025$ km s$^{-1}$). 
The Local Thermodynamic Equilibrium (LTE) mass is estimated under 
the assumption that all the hyperfine 
components are optically thin (see Ikeda et al. 2007 for further details). 
As mentioned in the previous section, 
the optically-thin assumption is reasonable in most of the area because 
the mean total optical depth derived from the hyperfine fitting is 
$\tau \sim 4-6 $ except for the northern part.
For simplicity, the N$_2$H$^+$ fractional abundance relative to H$_2$ 
is assumed to be spatially constant and equal to the mean fractional abundance 
of the observed area, $2.5 \times 10^{-10}$. 
The excitation temperatures are also set to be spatially constant at 
4, 7, and 6 K (the mean values listed in Table \ref{tab:n2h+})
for the northern, central, and southern regions, respectively. 

We note that the physical properties of the identified cores are 
partly affected by the core identification method and its adopted
parameters (see Section 4.1 of \citet{maruta10} for further discussion).

\subsection{Core Properties}

In Table \ref{tab:serpscore}, we show 
the physical quantities of the 70 dense cores 
identified from the N$_2$H$^+$ data cube.  
In Table \ref{tab:serpscore2}, we also list the minimum, 
maximum, and mean values of $R_{\rm core}$, $dv_{\rm core}$,
$M_{\rm LTE}$, $M_{\rm vir}/M_{\rm LTE}$, $\bar{n}$, and aspect ratio.
Here, we compute these values by dividing the observed area 
into the three parts as in Table \ref{tab:n2h+}:  
north, center, and south.

\begin{figure}[h]
\epsscale{0.4}
\plotone{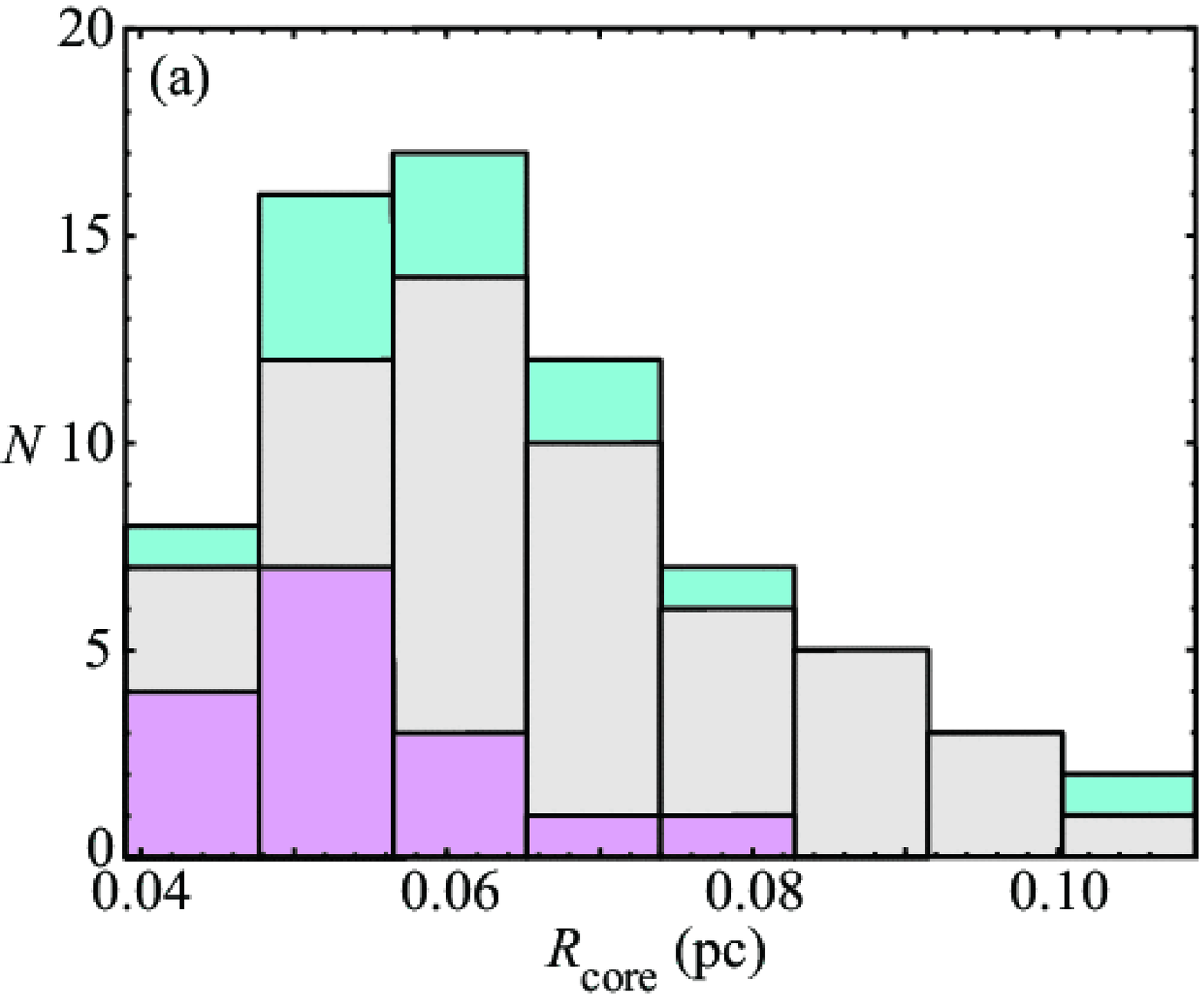}
\plotone{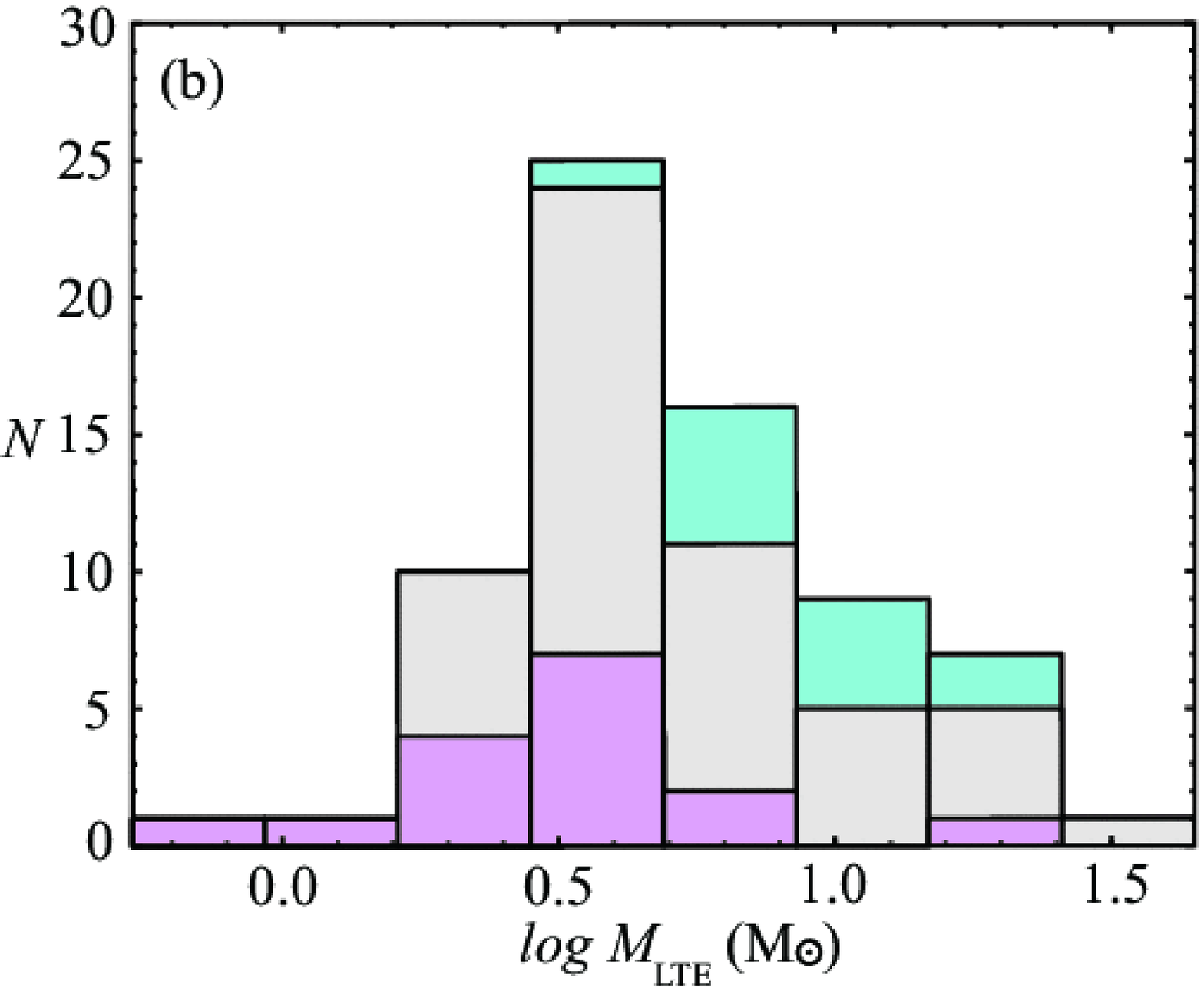}
\plotone{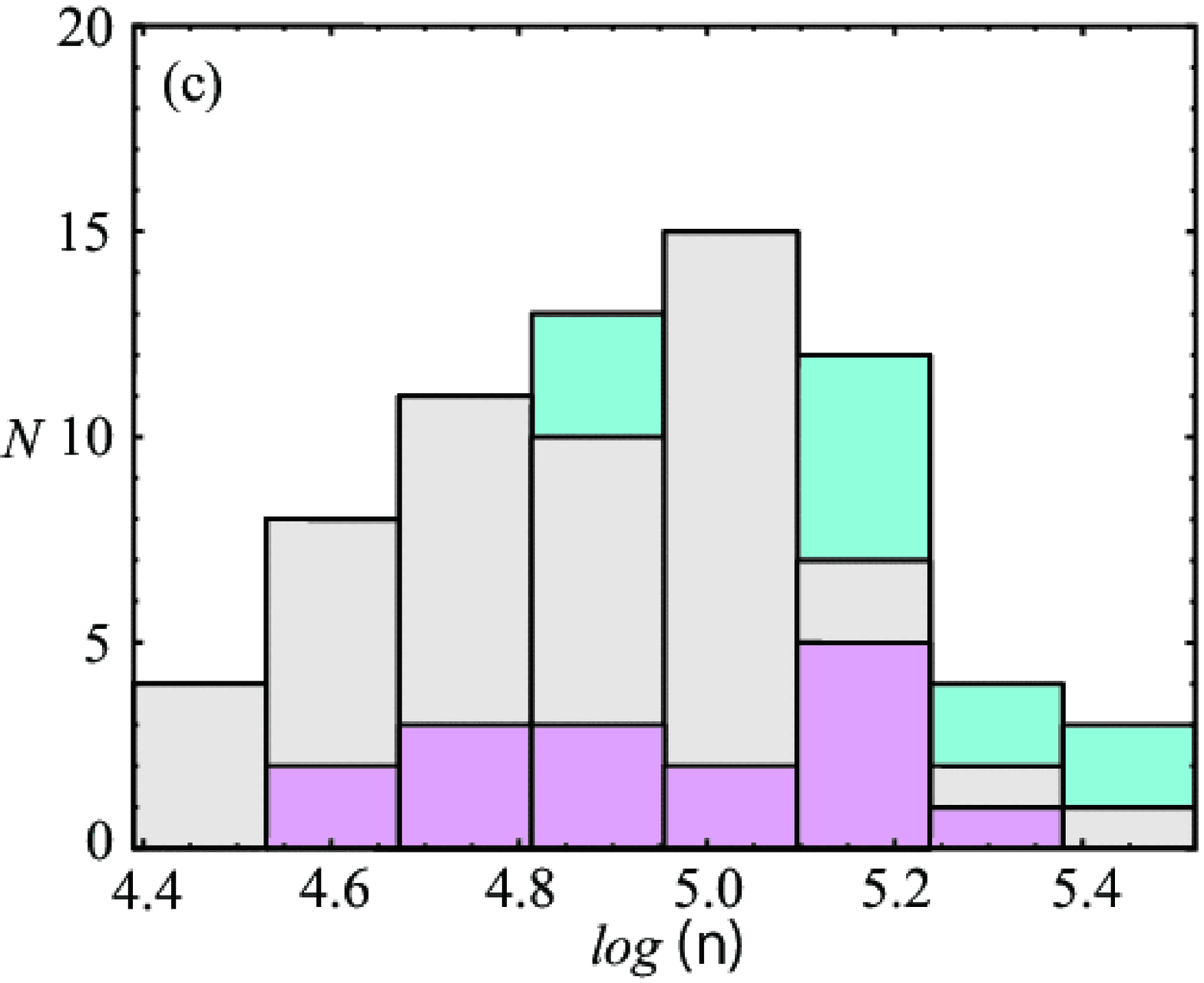}
\plotone{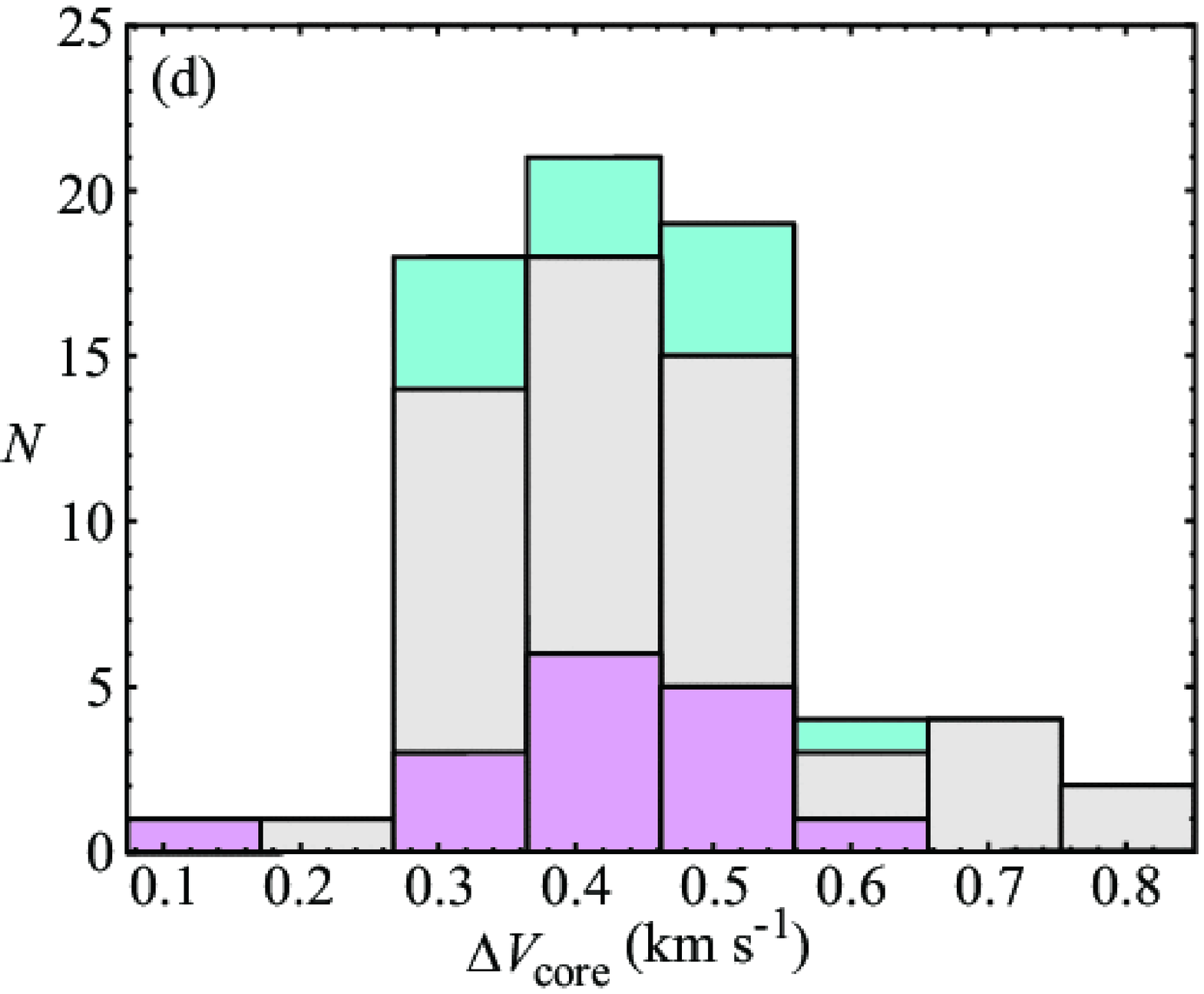}
\plotone{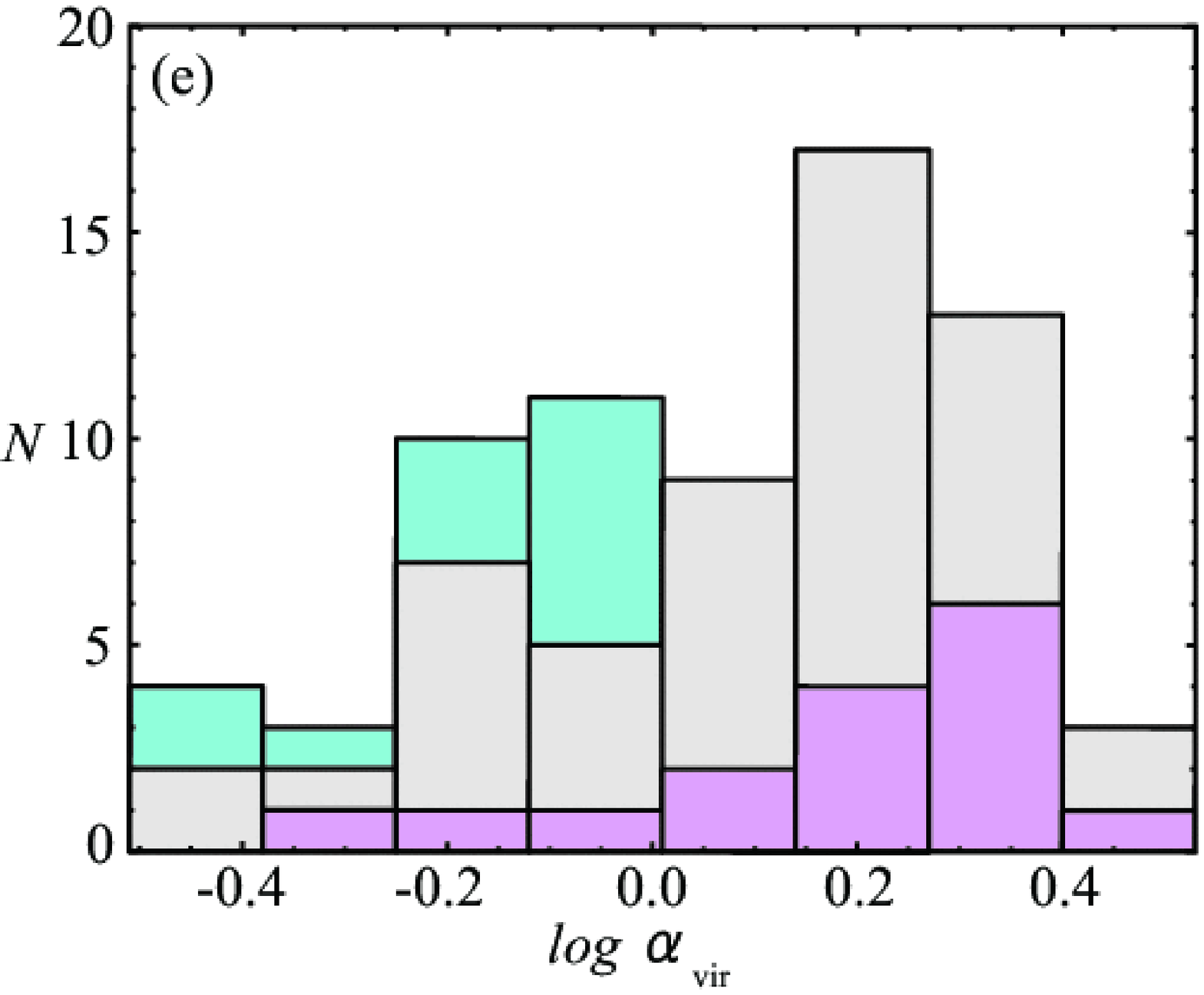}
\plotone{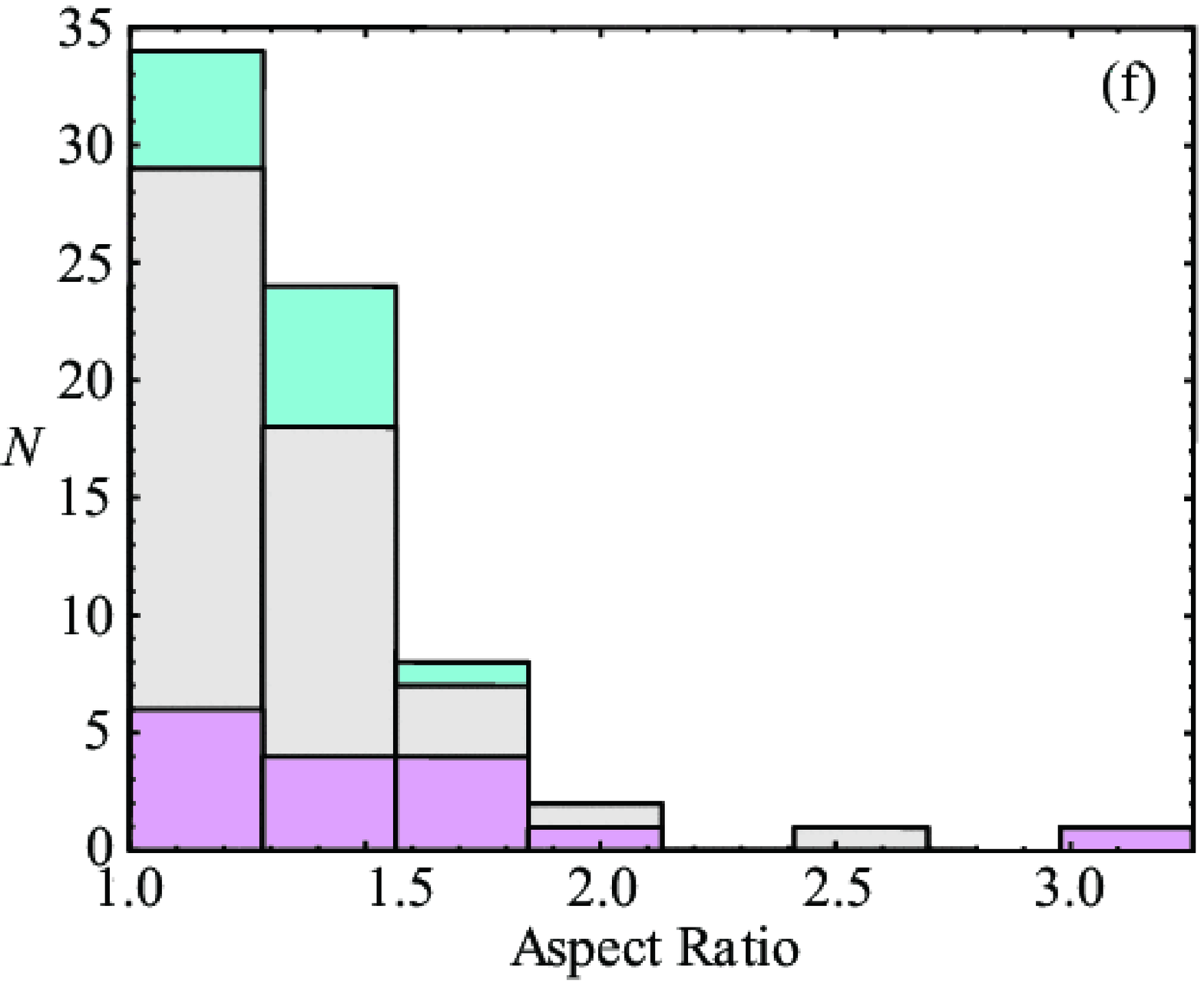}
\caption{Histograms of (a) radius ($R_{\rm core}$), (b) LTE mass
 ($M_{\rm LTE}$), (c) mean density ($\bar{n}$), 
(d) FWHM line width ($\Delta V_{\rm core}$), 
(e) virial ratio ($\alpha_{\rm vir}$), 
and (f) aspect ratio of the  N$_2$H$^+$ cores.
The cyan, grey, and magenta  histograms indicate the cores located in 
the northern, central, and southern areas, respectively.
}  
\label{fig:various histogram}
\end{figure}

Figures \ref{fig:various histogram}a through \ref{fig:various histogram}f
indicate the histograms of the radius ($R_{\rm core}$), 
LTE mass ($M_{\rm LTE}$), mean density ($\bar{n}$), 
line width ($\Delta V_{\rm core}$), virial ratio 
($\alpha_{\rm vir}$), and aspect ratio of the N$_2$H$^+$ cores.
In Figures \ref{fig:various histogram}a through 
\ref{fig:various histogram}, the cyan, grey, and magenta histograms 
show the cores located in the northern, central, and 
southern areas, respectively.
The distribution of the core radius takes its maximum at about 0.06 pc, 
ranging from 0.04 to 0.11 pc.  
The peak of its distribution is close to the mean of 0.065 pc
(see Figure \ref{fig:various histogram}a).  
The maximum and minimum radii of the N$_2$H$^+$ cores
are very close to the mean core radius, indicating that  
the distribution of the core radius is very narrow.
The cores in the southern part tend to be somewhat smaller than
those in the other parts.
On the other hand, the histogram of the LTE mass shows
a much broader distribution than that of the core radius. 
The distribution of the LTE mass takes its maximum at about 4 M$_\odot$ 
ranging from 0.8 to 40 M$_\odot$ (see Figure \ref{fig:various histogram}b).
The distribution of the mean density of the N$_2$H$^+$ cores 
has a single peak at about $1\times10^5$ cm$^{-3}$ that
is close to the mean value (see Figure \ref{fig:various histogram}c).
The distribution of the core mean density is also broad,
ranging from $0.25 \times 10^5 $ to $3.2 \times 10^5$ cm$^{-3}$
(see Figure \ref{fig:various histogram}c).
The cores located in the northern part tend to have larger
LTE masses and larger mean densities, compared to the cores in the other parts.

Figure \ref{fig:various histogram}d shows the 
distribution of the FWHM line width 
of the N$_2$H$^+$ cores.
The distribution of the line width has a single 
peak at around a mean of 0.45 km s$^{-1}$, ranging 
from 0.078 to 0.85 km s$^{-1}$.
Its distribution also appears to be concentrated 
in the range of 0.3 to 0.5 km s$^{-1}$.
Some cores in the central part have large line widths.
Applying the procedure described by \citet{myers91}, 
we classify the identified cores into the following two groups: 
gthermal corehand gturbulent core,h on the basis of the critical 
line width, $dV_{\rm cr}$. 
The critical line width is defined as follows:
\begin{equation}
dV_{\rm cr}=\left[8\ln2 \, k_B T 
\left({1\over \mu m_H}+{1\over m_{\rm obs}}\right)\right]^{1/2} \ , 
\label{eq:velocity width}
\end{equation}
where $k_B$ is the Boltzmann constant, 
$\mu=2.33$ is the mean molecular weight, $m_H$ is the mass of 
a hydrogen atom, and $m_{\rm obs}$ is the mass of a observed molecule,
i.e., N$_2$H$^+$ in this paper.
The critical line width of the N$_2$H$^+$ cores is evaluated 
to be 0.55 km s$^{-1}$ when the gas temperature of $T=14$ K is adopted.
We note that $T=14$ K is close to the dust temperature obtained from the
SED fitting of the Herschel data.
Among the 70 identified N$_2$H$^+$ cores, most of the cores 
(about 76 \%) are categorized as the thermal cores.
Even for the cores classified as the turbulent cores, most of them
have transonic nonthermal motions.
In other words, the cores in the Serpens South IRDC are relatively 
quiescent.

Figure \ref{fig:various histogram}e shows the distribution of the virial
ratios. First, most of the cores have the virial ratios not so far from
unity. In addition, the virial ratios appear to depend weakly on area. 
For example, in the northern part, almost all the cores have the virial
ratios smaller than unity. In contrast, in the central and southern
parts, the majority of the cores have the virial ratios larger than unity.

Figure \ref{fig:various histogram}f shows the distribution of the aspect
ratios of the cores. 
The majority of the cores have the aspect ratios close to unity.
The cores in the southern part tend to be somewhat more elongated than
those in the other parts.

\subsubsection{Line Width-Radius Relation}

\begin{figure}[h]
\epsscale{0.45}
\plotone{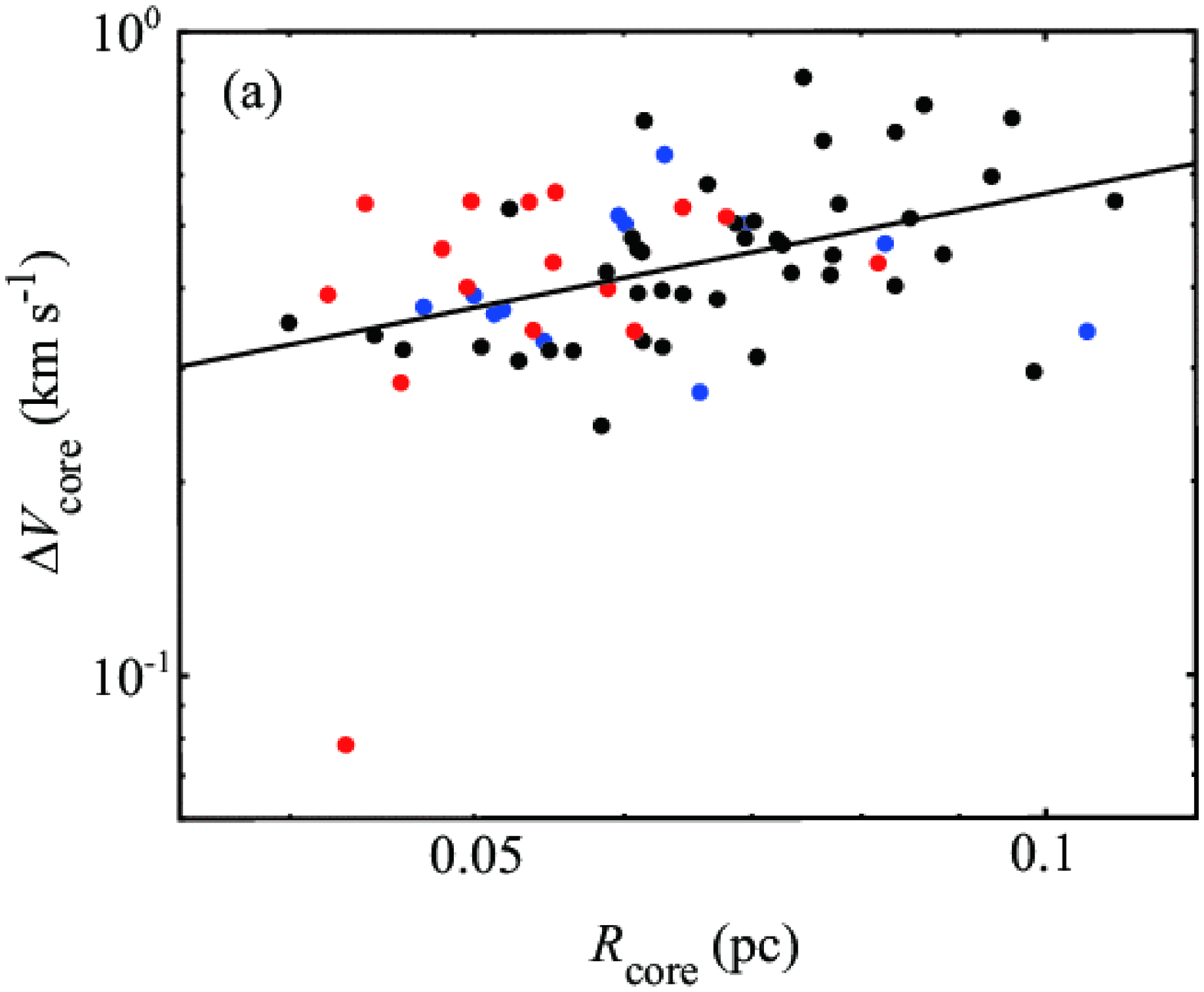}
\plotone{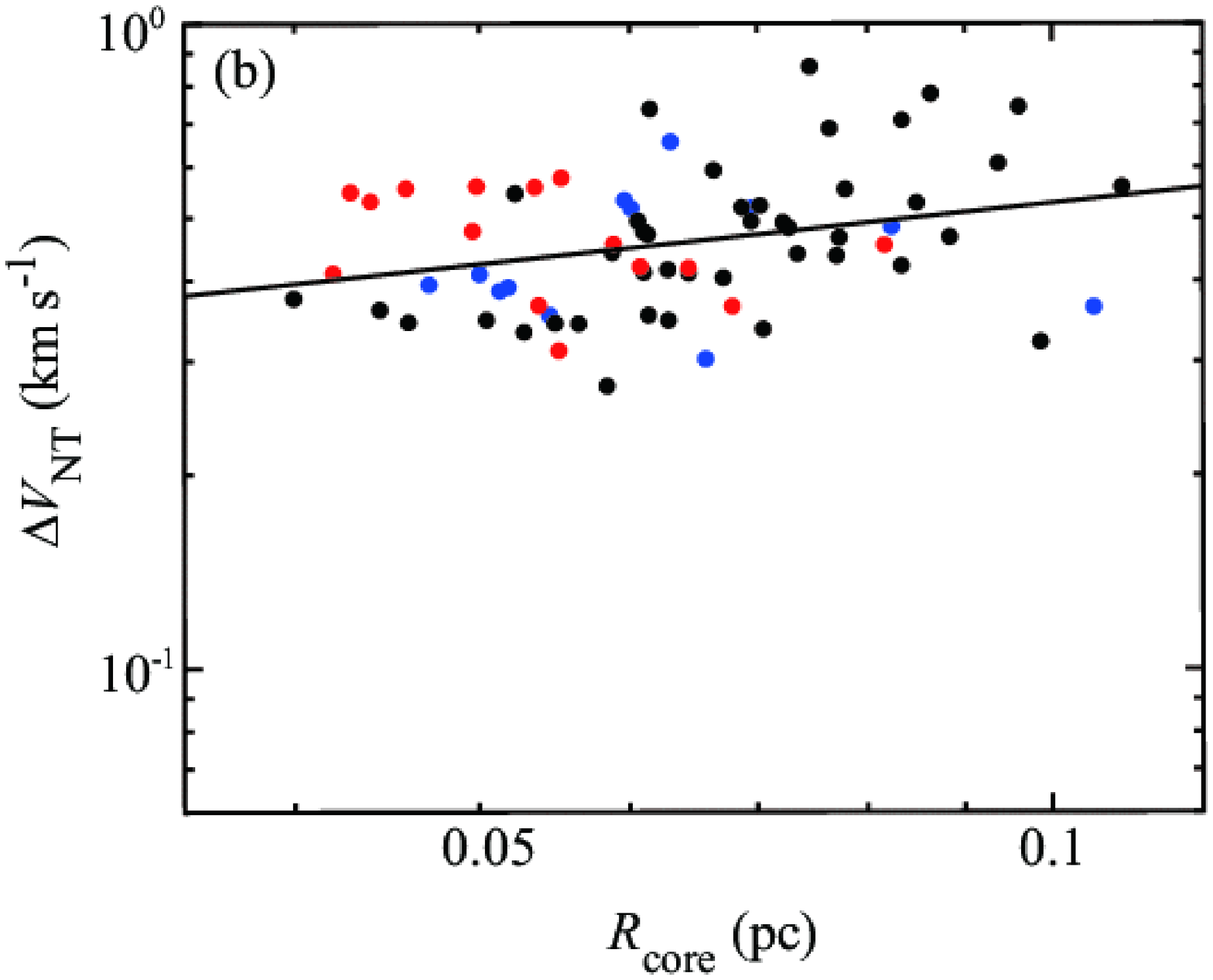}
\plotone{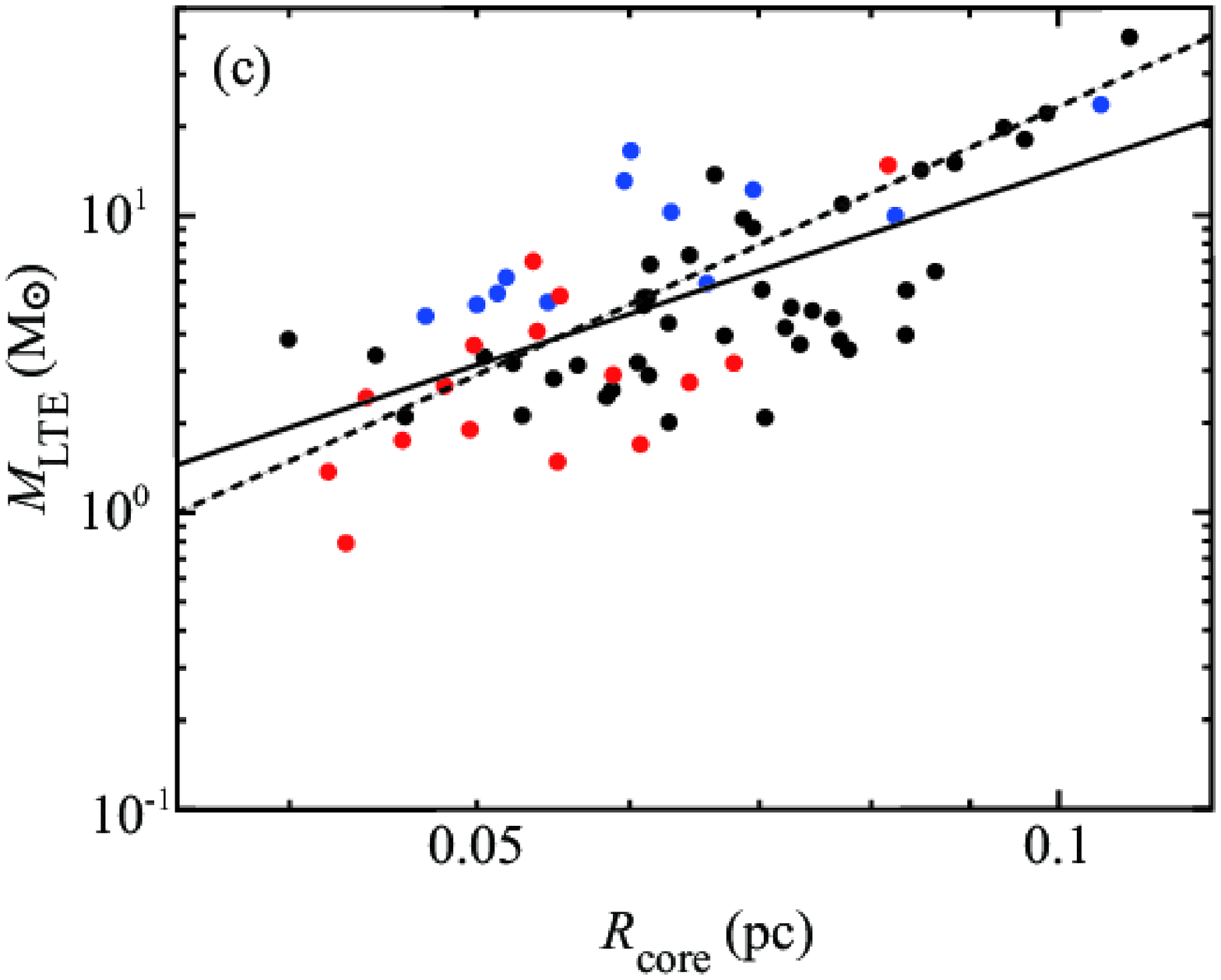}
\plotone{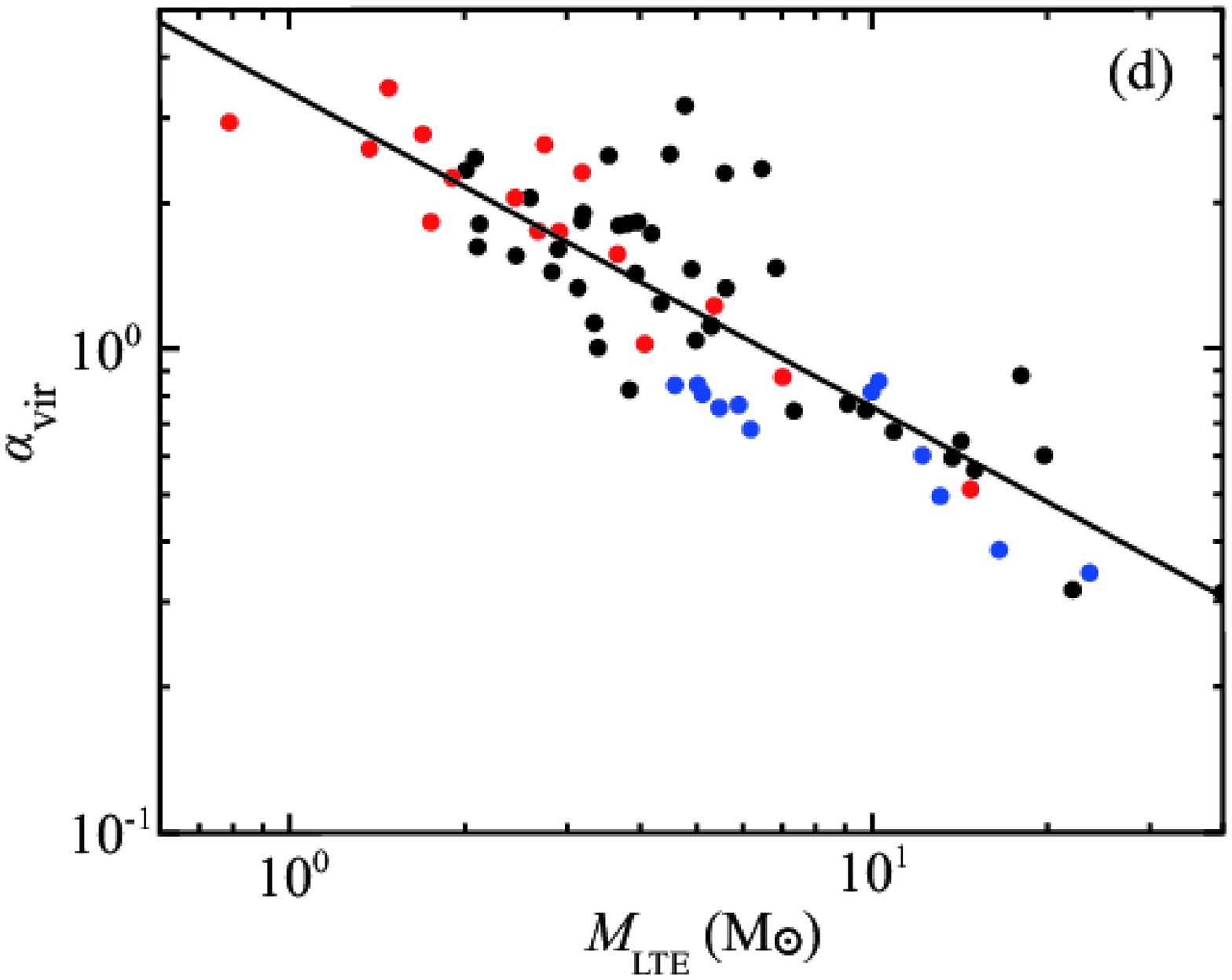}
\caption{(a) Line-width-radius relation, (b) nonthermal line-width-radius
 relatoin, (c) mass-radius relation, and 
(d) virial-ratio-mass relation of the N$_2$H$^+$ cores. 
The blue, red, and black circles are the same as those of Figure 
\ref{fig:abundance}.
}  
\label{fig:velocity-size}
\end{figure}

Figure \ref{fig:velocity-size}a shows th line-width-radius 
relation of the 70 dense cores identified from the 
N$_2$H$^+$ data cube.
For comparison, we show in Figure \ref{fig:velocity-size}a
the best-fit power-law that is given by 
\begin{equation}
\log \left(\Delta V_{\rm core} \over \ {\rm km \ s}^{-1}\right) = 
(0.335 \pm 0.194) + (0.587\pm 0.161)
\log \left({{R_{\rm core} \over {\rm pc}}}\right) \ \ ,
\end{equation}
where the correlation coefficient is  ${\cal R}=0.40$. 
The line-width-radius relation of the identified cores does not 
show a clear power-law like the so-called \citet{larson81}'s law. 
The weak dependence on the line widths 
of the core radius is also pointed out by
recent studies of other star-forming regions such as 
in Orion A \citep{ikeda07}, $\rho$ Oph \citep{maruta10}, and
massive-star-forming regions \citep{sanchez-monge13}.
The weak dependence of the line-width-radius relation 
may suggest that the identified cores are in 
various dynamical states as discussed below.

Figure \ref{fig:velocity-size}a indicates that the nonthermal 
components of the line widths appear to be important 
in the identified cores.
To quantify the contribution of the nonthermal components
in the line widths of the cores, we plot
in Figure \ref{fig:velocity-size}b 
the nonthermal line width 
$\Delta V_{\rm NT} [\equiv (\Delta V_{\rm core}^2-8\ln 2 \,
k_B T/m_{\rm obs})^{1/2}]$ as a function of core radius.
For comparison, we also plot in Figure \ref{fig:velocity-size}b 
the best-fit power-law that is given by
\begin{equation}
\log \left(\Delta V_{\rm NT} \over {\rm km \ s}^{-1}\right)= 
(0.037\pm 0.150) + (0.315\pm 0.125)
\log \left({R_{\rm core} \over {\rm pc}}\right)
\end{equation}
with ${\cal R}=0.36$.
Again, the correlation between the nonthermal line width and core radius
is weak.

The weak dependence of the line width on the core radius is 
consistent with the results of the recent numerical simulations 
of turbulent, magnetized molecular clouds.
The numerical simulations show that the cores formed in 
turbulent media are not in perfect virial equilibrium, but
in various dynamical states that are influenced by 
the ambient environments \citep{nakamura08,nakamura11b,offner08}.


\subsubsection{Mass-Radius Relation}

In Figure \ref{fig:velocity-size}c, we present the LTE mass 
as a function of core radius.
For comparison, we plot with a solid line 
the best-fit power-law that is given by 
\begin{equation}
\log \left(M_{\rm LTE} \over M_\odot \right)=
(3.32\pm 0.35) + (2.17\pm 0.29) \log
\left({R_{\rm core} \over {\rm pc}}\right) \ ,
\end{equation}
where ${\cal R}=0.69$. 
The relation of 
$M_{\rm LTE}=4\pi \mu m_H \bar{n} R_{\rm core}^3 /3$ is also 
shown with a dashed line in Figure \ref{fig:velocity-size}c, where
the mean density is set to $\bar{n}=9.6\times 10^4$ cm$^{-3}$,
which is close to the critical density of N$_2$H$^+$ ($J=1-0$), 
$n_{\rm cr}\simeq 1.5\times 10^5$ cm$^{-3}$.

\subsubsection{Virial Ratio-LTE-Mass Relation}
\label{subsec:virial analysis}

The importance of the self-gravity in a core 
is often assessed using the virial ratio, the ratio
between the virial mass and core mass
($\alpha_{\rm vir} = M_{\rm vir}/M_{\rm LTE}$).
In Figure \ref{fig:velocity-size}d, we present
the virial ratio as a function of the LTE mass, where 
we adopt a dimensionless parameter of $a=1$, corresponding to
a uniform sphere. For comparison, we show 
in Figure \ref{fig:velocity-size}d the best-fit power-law 
function that is given by
\begin{equation}
\log \alpha_{\rm vir} =  (0.53 \pm 0.04) + (-0.651\pm 0.050)
\log \left({M_{\rm LTE} \over M_\odot}\right) \ ,
\end{equation}
with the correlation coefficient of ${\cal R}=0.85$.
The virial ratio ranges from 0.3 to 3.5, having 
a mean of 1.4. 
Figure \ref{fig:velocity-size}d indicates that the cores 
with larger LTE masses tend to have smaller virial ratios.
The majority of the cores have virial ratios
smaller than 2, indicating that the total energy (the gravitational 
plus internal kinetic energies) of a core is negative. 
Therefore, the self-gravity of the individual cores appears 
to be important for the majority of the cores.

However, we note that the physical properties of the 
cores identified by the above procedure 
(and any other core identification schemes) depend 
strongly on the telescope beam size because 
the structures smaller than the beam size 
cannot be spatially resolved \citep[see][]{maruta10}.
According to \citet{maruta10}, the cores identified 
from the data having the higher spatial resolution tend to be
gravitational unbound and the external pressure due to 
the ambient turbulence plays an 
important role in the formation and evolution of the cores
\citep[see also Figure 10 of][]{nakamura11b}. 
It remains uncertain whether a core identified here 
is a basic structure that is separated dynamically 
from the ambient turbulent media. 
In fact, recently, \citet{nakamura12} revealed that the two 
starless cores which was previously identified as single cores 
using the single-dish telescopes, contain several 
substellar-mass condensations whose masses are larger than 
or at least comparable to the critical Bonner-Ebert mass.
The substellar-mass condensations might be real basic units of star 
formation in molecular clouds.
The data with higher angular resolution on the basis of 
the interferometric observations will be necessary to
address the issue of what is the basic units of star formation 
in star-forming molecular clouds.

\section{Conclusion}
\label{sec:conclusion}

We summarize the main results of the present paper as follows.

1. From the Herschel and 1.1 mm dust continuum data, we identified 
three dense clumps along the main filament: 
northern (V-shap), central (cluster-forming), 
and southern clumps.
The clump masses are estimated to be around 40 $-$ 230 $M_\odot$.

2. Applying a hyperfine fitting to the N$_2$H$^+$ ($J=1-0$) data, 
we derived the spatial distributions of the N$_2$H$^+$ column density, 
excitation temperature, fractional abundance, optical depth, and 
line width in Serpens South. We found that the cluster-forming clump has 
larger line widths of about 1 km s$^{-1}$, whereas the regions
with no signs of active star formation have smaller line widths of 
about 0.6 km s$^{-1}$. The fractional abundance of N$_2$H$^+$ is 
estimated to be $X_{\rm N_2H^+} \simeq 2.5 \times 10^{-10}$ over 
the whole area and tends to decrease with increasing column
density.  
We interpret that the overall dependence 
of $X_{\rm N_2H^+}$ on the column density reflects the ionization degree 
in the dense molecular gas, which 
is determined by the balance between the cosmic-ray ionization 
and the recombination.

3. Applying the virial analysis, we found that all the three clumps 
have small virial ratios of 0.1 $-$ 0.3.
This indicates that the internal turbulent motions play only 
a minor role in the clump support.
The northern clump has a mass and radius comparable to the central
cluster-forming clump, although there is no sign of active 
cluster formation. Thus, it is a likely candidate of {\it
pre-protocluster clump}, 
where active cluster formation will be in the future.

4. Although the clumps show the observational signs of the 
global infall motions, the infall speed appears to be 
much slower than the free-fall velocities of a few km s$^{-1}$. 
We propose that the slow global infall is due to the fact
that the Serpens South filament is supported 
by the large-scale ordered magnetic field that was discovered
by \citet{sugitani11}. According to our estimation, the parent
cloud was magnetically critical 
and either the ambipolar diffusion or mass accretion along 
the magnetic field lines may have triggerd the global
infall and the active cluster formation that is observed 
in the central clump.

5. The physical properties of the pre-protocluster clump are 
(1) the column density higher than $10^{23}$ cm$^{-2}$, 
(2) a lower temperature compared to the ambient gas, 
(3) a small internal velocity dispersion,
(4) a small virial ratio, and 
(5) the magnetic field that is close to the critical.
These properties contradict the accretion-driven turbulence scenario, 
which predicts that in the pre-protocluster clump, 
the turbulent energy should balance with the gravitational energy due 
to the momentum injection driven by the global accretion flow.

6. Applying the clumpfind to the 3D data cube of the N$_2$H$^+$
emission taken with the Nobeyama 45-m telescope, 
we identified 70 dense cores. The cores with larger LTE masses 
tend to have smaller virial ratios. 
The majority of the cores have virial ratios
smaller than 2, indicating that for most of the cores, the total energy
(the gravitational plus internal kinetic energies) is negative.
Therefore, the self-gravity appears to be important 
for most of the cores.
A caveat is that the physical properties of the identified cores 
depend strongly on the telescope beam size because 
the substructures smaller than the telescope beam size 
($\approx 20''$, corresponding to 0.04 pc) cannot be resolved.
Higher spatial resolution and higher sensitivity observations will 
be necessary to uncover the basic units of star formation.

\acknowledgements

This work is supported in part by a Grant-in-Aid for Scientific 
Research of Japan (20540228).
We thank Thushara Pillai, Jens Kauffmann, Alvaro Hacar, Philippe
Andr\'e, and Huei-Ru Chen for valuable comments and suggestions. 
We are grateful to the staffs at the Nobeyama Radio Observatory
(NRO) for both operating the 45-m and helping us with the data 
reduction. NRO is a branch of the National Astronomical 
Observatory, National Institutes of Natural Sciences, Japan.
We also thank the staffs at the Caltech Submillimeter Observatory
for giving us an opportunity to use the telescope.

\clearpage

\begin{deluxetable}{lllll}
\tabletypesize{\scriptsize}
\tablecolumns{5}
\tablecaption{N$_2$H$^+$ Line Characteristics in Serpens South}
\tablewidth{9cm}
\tablehead{\colhead{Physical Quantities} &\colhead{Mean\tablenotemark{a}} 
&\colhead{$\sigma$\tablenotemark{b}} 
&\colhead{Min\tablenotemark{a}} 
&\colhead{Max\tablenotemark{a}} 
}
\startdata
whole area  & &  & & \\ \hline
$V_{\rm LSR}$ (km s$^{-1}$)  & 7.34 $\pm$ 0.06 & 0.31 & 6.70 $\pm$ 0.06& 7.76 $\pm$ 0.07 \\
$\Delta V$ (km s$^{-1}$)     & 0.81 $\pm$ 0.08& 0.33 & 0.32 $\pm$ 0.08& 1.48  $\pm$ 0.06\\
$T_{\rm ex}$  (K)            & 6.0 $\pm$ 1.6 & 2.3 & 3.6 $\pm$ 0.84& 18.6  $\pm$ 5.75\\
$\tau_{\rm tot}$             & 7.6 $\pm$ 2.9& 6.4 & 0.5 $\pm$ 0.18& 34.4 $\pm$ 10.2\\ \hline
north  & &  & & \\ \hline
$V_{\rm LSR}$ (km s$^{-1}$)  & 7.59 $\pm$ 0.06& 0.09 & 7.29 $\pm$ 0.06& 7.76  $\pm$ 0.07\\
$\Delta V$ (km s$^{-1}$)     & 0.65 $\pm$ 0.08& 0.11 & 0.44 $\pm$ 0.10& 0.96  $\pm$ 0.10\\
$T_{\rm ex}$  (K)            & 4.1 $\pm$ 0.89& 0.2 & 3.6 $\pm$ 0.84& 4.6  $\pm$ 1.05\\
$\tau_{\rm tot}$             & 14.4 $\pm$ 5.31& 7.4 & 5.4 $\pm$ 1.49& 34.4 $\pm$ 10.2\\ \hline
center  & &  & & \\ \hline
$V_{\rm LSR}$ (km s$^{-1}$)  & 7.37 $\pm$ 0.06& 0.16 & 7.00 $\pm$ 0.06& 7.66 $\pm$ 0.06\\
$\Delta V$ (km s$^{-1}$)     & 1.05 $\pm$ 0.08& 0.35 & 0.32 $\pm$ 0.08& 1.48 $\pm$ 0.06\\
$T_{\rm ex}$  (K)            & 7.4 $\pm$ 1.95& 2.5 & 4.1 $\pm$ 0.89& 18.6 $\pm$ 5.75\\
$\tau_{\rm tot}$             & 3.8 $\pm$ 1.37& 1.9 & 0.5 $\pm$ 0.18& 9.6 $\pm$ 3.49\\ \hline
south  & &  & & \\ \hline
$V_{\rm LSR}$ (km s$^{-1}$)  & 6.95 $\pm$ 0.06& 0.30 & 6.70 $\pm$ 0.06& 7.63 $\pm$ 0.08\\
$\Delta V$ (km s$^{-1}$)     & 0.59 $\pm$ 0.08& 0.16 & 0.37 $\pm$ 0.07& 1.07 $\pm$ 0.09\\
$T_{\rm ex}$  (K)            & 5.7 $\pm$ 1.78& 1.6 & 4.0 $\pm$ 1.03& 17.3 $\pm$ 20.8\\
$\tau_{\rm tot}$             & 6.0 $\pm$ 2.51& 2.3 & 0.6 $\pm$ 0.87& 10.7 $\pm$ 7.03 
\enddata
\tablenotetext{a}{With standard deviation of the error of the hyperfine
 fitting including the flux calibration error of about $\pm$ 20 \%
and the spectrometer's velocity resolution of 0.05 km s$^{-1}$}
\tablenotetext{b}{standard deviation of the fitted data points from the
 mean values}
\tablecomments{The whole observed area is divided into three 
subregions: (1) north (Decl. $\ge -2~00~2.4$ ), 
(2) center ($-2~05~32.5 \le$ Decl. $\le -2~00~2.4$ ), 
and (3) south (Decl. $\le -2~05~32.5$ ).}
\label{tab:n2h+}
\end{deluxetable}

\begin{deluxetable}{lllll}
\tabletypesize{\scriptsize}
\tablecolumns{5}
\tablecaption{Column Densities and N$_2$H$^+$ Fractional 
Abundances in Serpens South}
\tablewidth{9cm}
\tablehead{\colhead{Physical Quantities} &\colhead{Mean\tablenotemark{a}} 
&\colhead{$\sigma$\tablenotemark{b}} 
&\colhead{Min\tablenotemark{a}} 
&\colhead{Max\tablenotemark{a}} 
}
\startdata
whole area  &  &  & & \\ \hline
$N_{\rm H_2}$ ($10^{22}$ cm$^{-2}$)  & ~6.9 $\pm$ 1.4& 1.5 & 1.5 $\pm$ 0.3& 20.1  $\pm$ 4.0\\
$N_{\rm N_2H^+}$ ($10^{13}$ cm$^{-2}$)   & ~~\,1.7 $\pm$ 0.5& ~\,0.8 & ~\,0.4 $\pm$ 0.1& ~\,~\,4.0  $\pm$ 1.0\\
$X_{\rm N_2H^+}$ ($10^{-10}$)            & ~~\,2.5 $\pm$ 1.4& ~\,0.7 & ~\,1.1 $\pm$ 0.4& ~\,~\,6.3  $\pm$ 4.2\\ \hline
north  &  &  & & \\ \hline
$N_{\rm H_2}$ ($10^{22}$ cm$^{-2}$)  & ~8.3 $\pm$ 1.7& 1.2 & 2.4 $\pm$ 0.5& 14.2  $\pm$ 2.8\\
$N_{\rm N_2H^+}$ ($10^{13}$ cm$^{-2}$)   & ~~\,2.0 $\pm$ 0.6& ~\,0.8 & ~\,0.8 $\pm$ 0.4& ~\,~\,4.0  $\pm$ 1.0\\
$X_{\rm N_2H^+}$ ($10^{-10}$)            & ~~\,2.5 $\pm$ 1.1& ~\,0.7 & ~\,1.5 $\pm$ 0.6& ~\,~\,6.3  $\pm$ 4.2\\ \hline
center  &  &  & & \\ \hline
$N_{\rm H_2}$ ($10^{22}$ cm$^{-2}$)  & ~7.2 $\pm$ 1.4& 1.7 & 1.5 $\pm$ 0.3& 20.1  $\pm$ 4.0\\
$N_{\rm N_2H^+}$ ($10^{13}$ cm$^{-2}$)   & ~~\,1.6 $\pm$ 0.5& ~\,0.7 & ~\,0.5 $\pm$ 0.2& ~\,~\,3.5  $\pm$ 0.7\\
$X_{\rm N_2H^+}$ ($10^{-10}$)            & ~~\,2.6 $\pm$ 1.7& ~\,0.8 & ~\,1.1 $\pm$ 0.4& ~\,~\,4.9  $\pm$ 2.0\\ \hline
south  &  &  & & \\ \hline
$N_{\rm H_2}$ ($10^{22}$ cm$^{-2}$)  & ~4.8 $\pm$ 1.0& ~\,0.5 & 2.4 $\pm$ 0.5& ~\,7.4  $\pm$ 1.5\\
$N_{\rm N_2H^+}$ ($10^{13}$ cm$^{-2}$)   & ~~\,1.2 $\pm$ 0.5& ~\,0.4 & ~\,0.4 $\pm$ 0.1& ~\,~\,2.0  $\pm$ 0.5\\
$X_{\rm N_2H^+}$ ($10^{-10}$)            & ~~\,2.4 $\pm$ 1.5& ~\,0.5 & ~\,1.1 $\pm$ 0.4& ~\,~\,3.6  $\pm$ 1.6 
\enddata
\tablecomments{The definitions of the subareas are the same as 
those of Table \ref{tab:n2h+}.}
\tablenotetext{a}{With standard deviation of the error of the hyperfine
fitting including the flux calibration error of about $\pm$ 20 \%
and the spectrometer's velocity resolution of 0.05 km s$^{-1}$}
\tablenotetext{b}{standard deviation of the fitted data points from the
 mean values}
\label{tab:n2h+2}
\end{deluxetable}

\begin{deluxetable}{llll}
\tabletypesize{\scriptsize}
\tablecolumns{4}
\tablecaption{Physical Quantities of Clumps in Serpens South}
\tablewidth{9cm}
\tablehead{\colhead{Physical Quantities} &\colhead{Serps N} 
&\colhead{Serps C} 
&\colhead{Serps S} 
}
\startdata
Mass ($M_\odot$)  & 193 & 232 & 36   \\
Radius (pc) & 0.189 & 0.200 & 0.093 \\
$T_{\rm dust} (K)^a $ & 11.3 & 14.0 & 12.7 \\
$\Delta V$ (km s$^{-1}$) & 0.665 & 1.25 & 0.580 \\
$U$ ($M_\odot$ km$^2$ s$^{-2}$) & 36 & 114 & 5.7 \\
$W$ ($M_\odot$ km$^2$ s$^{-2}$) & $-$721 & $-$1000 & $-$43.6 \\
$\alpha_{\rm vir}$ & 0.10 & 0.23 & 0.26 \\
$\Gamma \ ^b$ & 1.5 & 1.6 & 1.1 
\enddata
\tablenotetext{a}{Column-density weighted dust temperature derved from
 the SED fitting of the Herschel data.}
\tablenotetext{b}{the mass-to-magnetic flux ratio normalized to the
 critical value $2\pi G^{1/2}$.}
\label{tab:clump}
\end{deluxetable}

\begin{deluxetable}{lllccccccccccc}
\rotate
\tablecolumns{14}
\tabletypesize{\footnotesize}
\tablewidth{0pt}
\tablecaption{Properties of the N$_2$H$^+$ cores in Serpens South
\label{tab:serpscore}}
\tablehead{
\colhead{ID} & 
\colhead{R.A.} & 
\colhead{Decl.} & 
\colhead{$V_{\rm LSR}$} & 
\colhead{$T^*_{A, \rm peak}$\tablenotemark{a}} & 
\colhead{$R_{\rm core}$} & 
\colhead{$R_{\rm core}$} & 
\colhead{Aspect} &
\colhead{$\Delta v_{\rm core}$} & 
\colhead{$M_{\rm LTE}$} & 
\colhead{$M_{\rm vir}$} & 
\colhead{$\alpha_{\rm vir}$} &
\colhead{$\bar{n}$} & 
\colhead{Area$^b$} \\
\colhead{} & 
\colhead{(J2000.0)} & 
\colhead{(J2000.0)} & 
\colhead{(km s$^{-1}$)} & 
\colhead{(K)} & 
\colhead{(arcsec)} & 
\colhead{(pc)} &
\colhead{ Ratio}  & 
\colhead{(km s$^{-1}$)} & 
\colhead{($M_\odot$)} & 
\colhead{($M_\odot$)} & 
\colhead{} &
\colhead{($10^5$ cm$^{-3}$)}  &
\colhead{}
}
\startdata
1 & 18 29 44.9 & -02 05 33.0 & 8.87 & 0.65 & 27.4 & 0.0551 & 3.26 & 0.438 & 1.48 & 5.12 & 3.45 & 0.37 & s \\
2 & 18 29 47.1 & -02 00 46.6 & 7.51 & 0.61 & 35.0 & 0.0705 & 1.12 & 0.312 & 2.09 & 5.16 & 2.47 & 0.25 & c \\
3 & 18 29 49.5 & -02 05 09.4 & 7.62 & 0.56 & 31.3 & 0.0629 & 1.05 & 0.324 & 2.01 & 4.70 & 2.33 & 0.34 & c \\
4 & 18 29 49.6 & -01 58 08.0 & 7.85 & 0.76 & 25.7 & 0.0518 & 1.29 & 0.370 & 6.19 & 4.22 & 0.68 & 1.84 & n \\
5 & 18 29 52.3 & -01 57 46.9 & 7.88 & 0.71 & 27.1 & 0.0545 & 1.41 & 0.331 & 5.12 & 4.12 & 0.80 & 1.31 & n \\
6 & 18 29 54.2 & -02 00 48.2 & 7.43 & 0.61 & 38.0 & 0.0764 & 1.03 & 0.677 & 4.50 & 11.34 & 2.52 & 0.42 & c \\
7 & 18 29 55.2 & -01 58 39.8 & 7.51 & 0.94 & 52.3 & 0.1052 & 1.22 & 0.342 & 23.62 & 8.12 & 0.34 & 0.84 & n \\
8 & 18 29 55.7 & -01 59 29.6 & 7.72 & 0.63 & 25.5 & 0.0513 & 1.47 & 0.364 & 5.47 & 4.13 & 0.75 & 1.68 & n \\
9 & 18 29 55.9 & -01 58 34.9 & 7.93 & 0.61 & 32.7 & 0.0658 & 1.23 & 0.275 & 5.90 & 4.51 & 0.76 & 0.86 & n \\
10 & 18 29 56.4 & -02 01 21.4 & 8.02 & 0.62 & 31.2 & 0.0628 & 1.27 & 0.396 & 4.34 & 5.38 & 1.24 & 0.72 & c \\
11 & 18 29 56.6 & -01 57 51.8 & 7.62 & 0.92 & 29.9 & 0.0601 & 1.77 & 0.502 & 16.53 & 6.34 & 0.38 & 3.15 & n \\
12 & 18 29 56.8 & -01 59 58.8 & 7.74 & 0.61 & 23.4 & 0.0471 & 1.45 & 0.374 & 4.59 & 3.86 & 0.84 & 1.82 & n \\
13 & 18 29 57.5 & -01 59 44.8 & 7.22 & 0.67 & 24.9 & 0.0500 & 1.45 & 0.389 & 5.02 & 4.23 & 0.84 & 1.66 & n \\
14 & 18 29 57.6 & -02 00 40.3 & 7.30 & 0.63 & 30.6 & 0.0615 & 1.18 & 0.727 & 6.85 & 10.04 & 1.47 & 1.22 & c \\
15 & 18 29 57.8 & -02 01 44.3 & 7.17 & 0.60 & 37.1 & 0.0746 & 1.81 & 0.849 & 4.78 & 15.16 & 3.17 & 0.48 & c \\
16 & 18 29 58.3 & -02 06 27.1 & 7.71 & 0.61 & 33.8 & 0.0679 & 1.82 & 0.515 & 3.18 & 7.35 & 2.31 & 0.42 & s \\
17 & 18 29 58.6 & -02 01 15.2 & 7.65 & 0.89 & 32.0 & 0.0644 & 1.24 & 0.391 & 7.35 & 5.46 & 0.74 & 1.14 & c \\
18 & 18 29 58.8 & -02 01 50.0 & 7.98 & 0.67 & 26.2 & 0.0528 & 2.45 & 0.308 & 2.12 & 3.84 & 1.81 & 0.60 & c \\
19 & 18 29 58.8 & -01 59 05.0 & 7.54 & 0.66 & 34.6 & 0.0695 & 1.01 & 0.503 & 12.21 & 7.34 & 0.60 & 1.50 & n \\
20 & 18 29 59.5 & -02 03 33.7 & 7.53 & 0.65 & 42.9 & 0.0864 & 1.28 & 0.768 & 6.48 & 15.21 & 2.35 & 0.42 & c \\
21 & 18 29 59.5 & -01 58 13.0 & 7.80 & 0.78 & 31.3 & 0.0630 & 1.05 & 0.644 & 10.28 & 8.79 & 0.86 & 1.70 & n \\
22 & 18 29 59.6 & -02 00 51.6 & 8.04 & 0.64 & 30.5 & 0.0614 & 1.16 & 0.330 & 2.90 & 4.64 & 1.60 & 0.52 & c \\
23 & 18 29 60.0 & -02 05 43.0 & 7.75 & 0.66 & 29.2 & 0.0588 & 1.46 & 0.399 & 2.91 & 5.06 & 1.74 & 0.59 & s \\
24 & 18 30 00.2 & -01 59 55.6 & 7.37 & 0.91 & 29.6 & 0.0596 & 1.13 & 0.518 & 13.10 & 6.49 & 0.50 & 2.56 & n \\
25 & 18 30 00.7 & -02 02 19.9 & 7.04 & 0.97 & 47.7 & 0.0961 & 1.16 & 0.733 & 18.03 & 15.86 & 0.88 & 0.84 & c \\
26 & 18 30 00.8 & -02 06 15.3 & 7.43 & 0.65 & 22.7 & 0.0458 & 1.01 & 0.285 & 1.75 & 3.19 & 1.82 & 0.76 & s \\
27 & 18 30 01.0 & -02 07 11.9 & 7.91 & 0.69 & 32.0 & 0.0644 & 1.12 & 0.533 & 2.74 & 7.23 & 2.64 & 0.42 & s \\
28 & 18 30 01.2 & -02 01 22.6 & 7.96 & 0.91 & 19.9 & 0.0400 & 1.01 & 0.353 & 3.84 & 3.15 & 0.82 & 2.49 & c \\
29 & 18 30 01.3 & -02 04 21.3 & 7.70 & 0.59 & 38.3 & 0.0771 & 1.22 & 0.418 & 3.81 & 6.89 & 1.81 & 0.34 & c \\
30 & 18 30 01.3 & -02 04 18.0 & 6.54 & 0.66 & 38.7 & 0.0779 & 1.13 & 0.540 & 3.54 & 8.85 & 2.50 & 0.31 & c \\
31 & 18 30 01.5 & -02 01 26.1 & 7.49 & 0.98 & 22.0 & 0.0444 & 1.38 & 0.338 & 3.39 & 3.40 & 1.00 & 1.61 & c \\
32 & 18 30 01.6 & -02 07 18.3 & 5.30 & 0.69 & 21.3 & 0.0428 & 2.02 & 0.078 & 0.79 & 2.31 & 2.93 & 0.42 & s \\
33 & 18 30 01.8 & -02 05 29.5 & 7.87 & 0.70 & 36.1 & 0.0727 & 1.23 & 0.466 & 4.91 & 7.14 & 1.46 & 0.53 & c \\
34 & 18 30 02.0 & -02 01 00.6 & 7.04 & 0.95 & 38.4 & 0.0773 & 1.00 & 0.449 & 10.90 & 7.34 & 0.67 & 0.98 & c \\
35 & 18 30 02.5 & -02 02 12.9 & 7.90 & 1.04 & 42.2 & 0.0849 & 1.07 & 0.513 & 14.22 & 9.14 & 0.64 & 0.96 & c \\
36 & 18 30 03.0 & -02 03 01.2 & 7.22 & 1.65 & 49.0 & 0.0986 & 1.29 & 0.296 & 22.11 & 7.01 & 0.32 & 0.95 & c \\
37 & 18 30 03.1 & -02 03 07.4 & 6.75 & 1.21 & 43.9 & 0.0884 & 1.16 & 0.450 & 15.01 & 8.41 & 0.56 & 0.90 & c \\
38 & 18 30 03.1 & -02 04 41.6 & 7.07 & 1.39 & 46.6 & 0.0937 & 1.29 & 0.595 & 19.75 & 11.87 & 0.60 & 0.99 & c \\
39 & 18 30 03.4 & -02 03 15.0 & 7.71 & 1.48 & 54.1 & 0.1088 & 1.30 & 0.545 & 39.86 & 12.49 & 0.31 & 1.28 & c \\
40 & 18 30 03.5 & -02 00 45.1 & 7.58 & 0.65 & 28.0 & 0.0564 & 1.42 & 0.319 & 3.13 & 4.18 & 1.33 & 0.72 & c \\
41 & 18 30 03.7 & -02 06 12.3 & 7.62 & 0.75 & 24.8 & 0.0499 & 1.28 & 0.545 & 3.66 & 5.73 & 1.57 & 1.22 & s \\
42 & 18 30 03.8 & -01 59 54.4 & 7.19 & 0.61 & 40.9 & 0.0823 & 1.47 & 0.468 & 9.98 & 8.12 & 0.81 & 0.74 & n \\
43 & 18 30 04.0 & -02 03 58.2 & 6.88 & 1.20 & 33.0 & 0.0664 & 2.01 & 0.579 & 13.74 & 8.16 & 0.59 & 1.94 & c \\
44 & 18 30 04.3 & -02 01 45.5 & 7.49 & 0.90 & 34.2 & 0.0687 & 1.38 & 0.503 & 9.76 & 7.27 & 0.74 & 1.24 & c \\
45 & 18 30 05.0 & -02 05 01.8 & 6.68 & 0.68 & 33.4 & 0.0672 & 1.30 & 0.384 & 3.94 & 5.62 & 1.43 & 0.54 & c \\
46 & 18 30 05.3 & -02 05 27.5 & 7.41 & 0.60 & 30.5 & 0.0613 & 1.08 & 0.455 & 5.30 & 5.88 & 1.11 & 0.95 & c \\
47 & 18 30 05.4 & -02 01 06.8 & 6.80 & 0.59 & 41.5 & 0.0834 & 1.18 & 0.697 & 5.59 & 12.88 & 2.30 & 0.40 & c \\
48 & 18 30 06.1 & -02 06 12.5 & 7.56 & 0.61 & 21.8 & 0.0439 & 1.02 & 0.541 & 2.44 & 5.00 & 2.05 & 1.20 & s \\
49 & 18 30 06.2 & -02 04 44.1 & 7.54 & 0.68 & 30.3 & 0.0610 & 1.24 & 0.460 & 5.29 & 5.91 & 1.12 & 0.96 & c \\
50 & 18 30 06.8 & -02 05 20.5 & 6.93 & 0.94 & 34.5 & 0.0695 & 1.33 & 0.477 & 9.09 & 6.98 & 0.77 & 1.12 & c \\
51 & 18 30 07.0 & -02 07 04.4 & 7.51 & 0.82 & 20.8 & 0.0419 & 1.29 & 0.391 & 1.37 & 3.55 & 2.59 & 0.77 & s \\
52 & 18 30 07.1 & -02 03 15.8 & 7.58 & 0.70 & 34.9 & 0.0703 & 1.41 & 0.507 & 5.62 & 7.48 & 1.33 & 0.67 & c \\
53 & 18 30 07.2 & -02 01 05.6 & 7.76 & 0.61 & 36.5 & 0.0735 & 1.27 & 0.422 & 3.68 & 6.61 & 1.79 & 0.38 & c \\
54 & 18 30 07.8 & -02 06 26.2 & 7.14 & 0.59 & 24.7 & 0.0496 & 1.59 & 0.401 & 1.90 & 4.29 & 2.25 & 0.65 & s \\
55 & 18 30 07.9 & -02 04 13.1 & 6.86 & 0.64 & 29.0 & 0.0584 & 1.09 & 0.244 & 2.45 & 3.81 & 1.55 & 0.51 & c \\
56 & 18 30 08.1 & -02 04 14.0 & 7.42 & 0.74 & 30.3 & 0.0610 & 1.42 & 0.392 & 4.99 & 5.18 & 1.04 & 0.91 & c \\
57 & 18 30 08.1 & -02 03 36.8 & 7.20 & 0.67 & 27.2 & 0.0548 & 1.64 & 0.320 & 2.83 & 4.06 & 1.44 & 0.71 & c \\
58 & 18 30 08.2 & -02 05 53.1 & 7.29 & 0.59 & 23.9 & 0.0481 & 1.84 & 0.460 & 2.67 & 4.67 & 1.75 & 0.99 & s \\
59 & 18 30 08.7 & -02 05 14.3 & 7.30 & 0.64 & 22.8 & 0.0459 & 1.14 & 0.321 & 2.11 & 3.41 & 1.62 & 0.90 & c \\
60 & 18 30 09.7 & -02 05 44.3 & 6.73 & 0.92 & 26.7 & 0.0538 & 1.69 & 0.344 & 4.08 & 4.17 & 1.02 & 1.09 & s \\
61 & 18 30 09.9 & -02 02 06.4 & 7.57 & 0.60 & 41.4 & 0.0834 & 1.30 & 0.403 & 3.96 & 7.22 & 1.82 & 0.28 & c \\
62 & 18 30 10.8 & -02 03 42.8 & 7.45 & 0.69 & 25.1 & 0.0505 & 1.58 & 0.324 & 3.34 & 3.77 & 1.13 & 1.08 & c \\
63 & 18 30 11.0 & -02 06 12.5 & 6.90 & 1.04 & 26.6 & 0.0535 & 1.56 & 0.544 & 7.02 & 6.13 & 0.87 & 1.90 & s \\
64 & 18 30 11.4 & -02 04 25.6 & 7.07 & 0.61 & 29.2 & 0.0587 & 1.10 & 0.423 & 2.59 & 5.30 & 2.05 & 0.53 & c \\
65 & 18 30 11.5 & -02 06 53.4 & 6.91 & 0.90 & 27.4 & 0.0552 & 1.42 & 0.563 & 5.37 & 6.57 & 1.22 & 1.32 & s \\
66 & 18 30 13.3 & -02 04 17.7 & 7.50 & 0.66 & 26.0 & 0.0522 & 1.09 & 0.531 & 3.18 & 5.84 & 1.84 & 0.92 & c \\
67 & 18 30 13.4 & -02 07 00.7 & 6.17 & 0.59 & 30.2 & 0.0608 & 1.11 & 0.343 & 1.70 & 4.70 & 2.77 & 0.31 & s \\
68 & 18 30 13.5 & -02 03 23.5 & 7.48 & 0.59 & 35.9 & 0.0722 & 1.32 & 0.476 & 4.19 & 7.23 & 1.73 & 0.46 & c \\
69 & 18 30 13.7 & -02 06 41.9 & 6.70 & 1.29 & 40.6 & 0.0817 & 1.08 & 0.436 & 14.77 & 7.56 & 0.51 & 1.12 & s \\
70 & 18 30 15.9 & -02 04 43.6 & 7.49 & 0.66 & 30.1 & 0.0606 & 1.41 & 0.478 & 3.20 & 6.10 & 1.91 & 0.59 & c \\
\enddata
\tablecomments{Units of right ascension are hours, minutes, and seconds, 
and units of declination are degrees, arcminutes, and arcseconds.
Taking into account the calibration error of the observed intensities 
of about $\pm$20 \%, the derived LTE masses, virial ratios, and mean
 densities also have an error of at least about $\pm$20 \%.}
\tablenotetext{a}{$T_A^*$ denotes the peak antenna temperature of the
 isolated hyperfine component.}
\tablenotetext{b}{The symbols n, c, and s mean, respectively, the
 northern, central, and southern areas where the cores are located.}
\end{deluxetable}

\begin{deluxetable}{llll}
\tablecolumns{3}
\tablecaption{Summary of the Physical Properties 
of the N$_2$H$^+$ cores\label{tab:serpscore2}}
\tablewidth{\columnwidth}
\tablehead{\colhead{Property} & \colhead{Minimum} & \colhead{Maximum}
 &\colhead{Mean\tablenotemark{a}}} 
\startdata
whole area  &  &  & \\ \hline
$R_{\rm core}$ (pc) & 0.0400  & ~\,0.1088  & 0.0649 $\pm$ 0.0154 \\
$dv_{\rm core}$ (km s$^{-1}$) &0.078  & ~\,0.849  & 0.448 $\pm$ 0.132 \\
$M_{\rm LTE}$ (M$_\odot$) &0.79  & 39.86  & 6.81 $\pm$ 6.49 \\
$M_{\rm vir}/M_{\rm LTE}$  &0.313  & ~\,3.45  & 1.42 $\pm$ 0.76 \\
$\bar{n}$ ($\times 10^5 $cm$^{-3}$) &0.25  & ~\,3.15  & 0.96 $\pm$ 0.58 \\
Aspect Ratio & 1.00  & ~\,3.26  & 1.36 $\pm$ 0.36 \\ \hline
north  &  &  & \\ \hline
$R_{\rm core}$ (pc) & 0.0471  & ~\,0.1052  & 0.0634 $\pm$ 0.0158 \\
$dv_{\rm core}$ (km s$^{-1}$) &0.275  & ~\,0.644  & 0.423 $\pm$ 0.100 \\
$M_{\rm LTE}$ (M$_\odot$) &4.59  & 23.62  & 9.83 $\pm$ 5.57 \\
$M_{\rm vir}/M_{\rm LTE}$  &0.344  & ~\,0.86  & 0.68 $\pm$ 0.18 \\
$\bar{n}$ ($\times 10^5 $cm$^{-3}$) &0.74  & ~\,3.15  & 1.64 $\pm$ 0.67 \\
Aspect Ratio & 1.01  & ~\,1.77  & 1.33 $\pm$ 0.21 \\ \hline
center  &  &  & \\ \hline
$R_{\rm core}$ (pc) & 0.0400  & ~\,0.1088  & 0.0693 $\pm$ 0.0150 \\
$dv_{\rm core}$ (km s$^{-1}$) &0.244  & ~\,0.849  & 0.464 $\pm$ 0.141 \\
$M_{\rm LTE}$ (M$_\odot$) &2.01  & 39.86  & 7.16 $\pm$ 7.12 \\
$M_{\rm vir}/M_{\rm LTE}$  &0.313  & ~\,3.17  & 1.43 $\pm$ 0.67 \\
$\bar{n}$ ($\times 10^5 $cm$^{-3}$) &0.25  & ~\,2.49  & 0.81 $\pm$ 0.45 \\
Aspect Ratio &1.00  & ~\,2.45  & 1.30 $\pm$ 0.27 \\ \hline
south  &  &  & \\ \hline
$R_{\rm core}$ (pc) & 0.0419  & ~\,0.0817  & 0.0546 $\pm$ 0.0101 \\
$dv_{\rm core}$ (km s$^{-1}$) &0.078  & ~\,0.563  & 0.426 $\pm$ 0.122 \\
$M_{\rm LTE}$ (M$_\odot$) &0.79  & 14.77  & 3.61 $\pm$ 3.27 \\
$M_{\rm vir}/M_{\rm LTE}$  &0.512  & ~\,3.45  & 1.97 $\pm$ 0.79 \\
$\bar{n}$ ($\times 10^5 $cm$^{-3}$) &0.31  & ~\,1.90  & 0.85 $\pm$ 0.43 \\
Aspect Ratio &1.01  & ~\,3.26  & 1.54 $\pm$ 0.54 
\enddata
\tablenotetext{a}{With standard deviation}
\end{deluxetable}

\clearpage

\end{document}